\documentclass[a4paper,11pt]{article}
\pdfoutput=1
\usepackage{jheppub}
\usepackage{xcolor}
\colorlet{mdtRed}{red!50!black}
 \usepackage[]{hyperref}
\hypersetup{colorlinks=true,
linkcolor=gray,
urlcolor=mdtRed
}
\usepackage{tikz}
\usepackage{fancybox}
\usepackage{float}
\usepackage{amsfonts}
\usepackage{amsmath}
\usepackage{amsbsy}
\usepackage{graphicx}
\usepackage{amssymb}
\usepackage{amsthm}
\usepackage{bm}
\usepackage{comment}
\usepackage{epsfig}
\usepackage{latexsym}
\usepackage{pdflscape}
\usepackage{color}
\usepackage{slashed}

\makeatletter
\def\@fpheader{\relax}
\makeatother

\usepackage{orcidlink}

\newcommand{\beq}{\begin{eqnarray}}
\newcommand{\eeq}{\end{eqnarray}}
\newcommand{\bea}{\begin{eqnarray}}
\newcommand{\eea}{\end{eqnarray}}
\newcommand{\be}{\begin{equation}}
\newcommand{\ee}{\end{equation}}
\newcommand{\bq}{\begin{equation}}
\newcommand{\eq}{\end{equation}}

\newcommand{\half}{\frac{1}{2}}
\newcommand{\nn}{\nonumber}

\newcommand{\sump}[0]{\sum_{(h,g)}\!{\raise 4pt \hbox{$'$}}\,}

\def\le{\left}
\def\ri{\right}

\def\half{\frac12}

\def\m{\mu}
\def\n{\nu}

\def\6{\partial}

\def\tr{{\rm Tr}}

\def\le{\left}
\def\ri{\right}
\def\6{\partial}

\def\hri#1#2{\href{http://arxiv.org/abs/#1}{[arXiv:#1]#2}}
\def\hre#1#2{\href{http://arxiv.org/abs/#2/#1}{[arXiv:#1/#2]}}

\def\hrj#1#2{\href{https://doi.org/#1}{#2}}

\title{A microscopic Normal Matrix Model for $(A)dS_2$}

\author[a]{Panos Betzios\orcidlink{0000-0002-5350-9404},}
\affiliation[a]{\href{https://www.ugent.be/we/physics-astronomy/en}{Department of Physics and Astronomy},
Ghent University, \\ Krijgslaan, 281-S9, 9000 Gent, Belgium}

\emailAdd{panos.betzios@ugent.be}

\abstract{We describe the duality between the gravitating $c=1$ (compact) Sine-Gordon model and a normal matrix model. From a two-dimensional quantum gravity perspective and due to the periodic nature of the potential, this model admits both anti-de Sitter and de-Sitter saddles, similarly to simpler models of Sine-Dilaton gravity, as well as more complicated interpolating ``wineglass wormhole'' geometries. From a string theory perspective the Euclidean de-Sitter (genus zero) saddles are related to the presence of a classical entropic contribution associated to the target space geometry. The gravitating Sine-Gordon model corresponds to a well defined CFT by construction and the eigenvalues of the dual normal matrix model are supported in a compact region of the complex plane. The duality with the normal matrix model is operationally defined even for a finite, but sufficiently large matrix size $N$, depending on the precise observable to be determined. We define and study a ``microscopic'' version of the large-N limit that allows us to recover non-perturbative results for all physical observables.}

\vspace{10cm}

\dedicated{Dedicated to the memory of Umut G\"ursoy}

\begin{document}
\maketitle
\flushbottom

\section{Introduction}

Constructing a precise and complete model of quantum gravity in spacetimes with positive cosmological constant ($\Lambda >0$), such as de-Sitter space, has been a longstanding unsolved problem. This is in stark contrast with spacetimes having anti-de Sitter asymptotics ($\Lambda < 0$). A well known difficulty is constructing quantum observables in cosmological spacetimes that do not contain any asymptotic region where gravitational fluctuations can be neglected and where gauge invariant ``detectors'' can be placed. Observers are surrounded by cosmological horizons shielding from them any (potential) asymptotic portions of the spacetime, and what is even worse, in closed cosmologies the associated Cauchy surfaces are compact, leading to spacetimes having no such asymptotic regions. 

These challenges already appear in the simplified two-dimensional case of one time and one space dimension and for this reason, there has been a revived interest in very simple models of two dimensional de-Sitter space such as JT gravity with a positive cosmological constant~\cite{Cotler:2024xzz}, or the more general model of Sine-Dilaton~\cite{Blommaert:2024ydx,Blommaert:2024whf,Blommaert:2025avl} gravity. Sine-Dilaton gravity has also the advantage that it can descend from a microscopic model - the double scaled SYK model, which has also been argued to be related to the physics of two-dimensional de-Sitter space from a different perspective in~\cite{Susskind:2022bia,Rahman:2022jsf,Narovlansky:2023lfz}. Another interesting avenue opens, by embedding Sine-Dilaton gravity to the complex-Liouville string~\cite{Collier:2024kmo,Collier:2025pbm,Collier:2024lys}, which provides a possible non-perturbative completion of the effective Sine-Dilaton models through its dual matrix model description. Other recent semi-classical approaches to understand two dimensional cosmological spacetimes can be found in~\cite{Anninos:2021ene,Anninos:2024fty,Anninos:2024iwf} and references therein\footnote{We should also mention the possibility that when the cosmological constant is replaced by a field dependent potential, it could allow de-Sitter space to be embedded inside AdS, as the scalar field runs. This idea has both lower dimensional~\cite{Anninos:2017hhn,Anninos:2022hqo} as well as higher dimensional incarnations~\cite{Betzios:2024oli,Betzios:2024iay}. Some solutions that we study in section~\ref{ComplexLiouville} do have this property.}.

Nevertheless, it is fair to say that there is no general consensus on whether any of these models satisfies all the desiderata 
of a complete microscopic model of de-Sitter space (even at low dimensions). Some of them contain no locally propagating degrees of freedom and exhibit a non-normalizable Hartle-Hawking type of state, an infinite Euclidean path integral on $S^2$, leading to an infinite de-Sitter entropy according to the Gibbons-Hawking prescription. Others on the other hand fall short in handling non-trivial topologies and have no clear UV completion, while the ones that could potentially solve this issue rely on theories that fundamentally have a complex central charge and many calculations and physical results have to be understood through various complexifications and analytic continuations of parameters and basic building blocks. 

It remains therefore of utmost importance, to expand even further our arsenal of tools and candidate models of lower dimensional de-Sitter space, since they are our best hope of grappling with the associated difficult and deep cosmological questions. 

In this work, we explore from this perspective \emph{a duality between
the gravitating $c=1$ compact Sine-Gordon (SG) model
and a Normal Matrix Model (NMM)}.  The flat space version of the compact two dimensional Sine-Gordon model we are interested in, is defined by the (Euclidean) action
\be\label{SGmodel}
S_{SG} = \int d^2 x \, \left( \partial_\m X \partial^\m X + t_0 \cos \frac{X}{R} \right) \, , \qquad X \sim X + 2 \pi R \, .
\ee
Using the principle of general covariance, we can couple the two-dimensional Sine-Gordon model to gravity, resulting in an action $S_{SG}[{g} ; X]$, where ${g}$ is the two dimensional metric~\cite{Moore:1992ga}. Classically, the theory $S_{SG}[{g} ; X]$ in two dimensions can always be understood as being a conformal field theory (even if the original theory in flat space is not a CFT). To understand why this is so, one should recall that the stress energy tensor is defined by ${\delta S / \delta {g}^{\m \n}} = T_{\m \n}$.
Since the Liouville mode corresponds to the conformal mode/overall
scale of the metric, seen by decomposing ${g}_{\m \n} = e^{2  \phi}  \hat{g}_{\m \n}$,
one finds that demanding that the classical Liouville equation of motion is satisfied (in the background metric $\hat{g}_{\m \n}$), leads to the trace of the energy momentum tensor being zero: ${T}^{\m}_{\mu}=0$. Coupling a two dimensional field theory to gravity can therefore be used to define a classical conformal field theory~\cite{Ginsparg:1993is}. 

Quantum mechanically, we would like to understand
(Euclidean at this point) quantum gravitational path integrals of the form
\be
\langle \mathcal{O}_1 \, \dots \, \mathcal{O}_n \rangle
= \frac{1}{Z} \, \int \frac{\mathcal{D} g \, \mathcal{D} X}{\text{Vol}({\it Diff\/})}
\,\mathcal{O}_1 \, \dots \, \mathcal{O}_n  \, e^{- \kappa_0 \int d^2 x \sqrt{g} R - \mu_0 A - S[g,X]} \, ,
\ee
where $\mathcal{O}_i$ are general covariant operators.
By passing to a conformal gauge $g_{\m \n} = e^{2  \phi} \hat{g}_{\m \n} $, the matter/gravity
system can be written as a system of matter coupled to Liouville theory. The passage to conformal gauge introduces Faddeev-Popov reparametrization ghosts and the role of quantum effects from the path integral measure, is subsumed in a renormalization of the parameters of the classical matter/Liouville system\footnote{There is an additional renormalization due to rendering the measure invariant under parallel translation in $\phi$ - space viz. $|\delta \phi|^2 = \int d^2 x \sqrt{\hat{g}} (\delta \phi)^2$}, see \cite{Ginsparg:1993is,Nakayama:2004vk} for reviews on this. The final renormalized action $S_{ren}[\hat{g} ; \phi, X]$ should then be treated as a \emph{quantum effective action}, that can describe an exact CFT at the quantum level. This exact CFT is placed on the effective metric $\hat{g}_{\m \n}$, that in two dimensions can be treated as being classical and fixed (since all the quantum dynamics of the conformal mode are subsumed in the path integral over the Liouville field $\phi$) - there are no other local gravitational degrees of freedom remaining\footnote{This is again an arbitrary choice of frame, and it is typical in the literature to fix $\hat{g}_{\m \n} = \delta_{\m \n}$}.

As we describe in section~\ref{sec:SineGordon}, the $c=1$ gravitational Sine-Gordon model, as all other $c=1$ models in two dimensions, exhibits appreciable quantum effects and hence the parameters/coefficients of the renormalized effective action differ by $O(1)$ terms with respect to the  ``naive classical CFT" action one would have obtained without incorporating the quantum effects and transformations in the path integral measure\footnote{For example the typical parameters $Q = b + 1/b$ in Liouville theory are $b=1 , \, Q=2$ when $c=1$, and hence there is no large or small $b$ expansion.}. Nevertheless, it is still possible to study such a model in great detail using exact CFT techniques, or especially  its dual matrix model description (as we do later on in sections~\ref{MMduals} and~\ref{sec:selfdual}). One might ponder at this point, how relevant is the study of the resulting effective ``quantum geometries'', especially if we wish to think of such a model as a toy version of putative higher dimensional models of de-Sitter space. To this, we believe that this property of the $c=1$ models is actually an interesting feature, since it has long been argued that in expanding spacetimes such as de-Sitter, effects due to quantum loops and IR divergences can become increasingly important at late times, invalidating or modifying appreciably naive semi-classical expectations for physical observables~\cite{Antoniadis:1985pj,Tsamis:1996qq,Polyakov:2007mm,Betzios:2020nry,Anninos:2024fty}. The difference/simplification that exists in simple two dimensional models is that the metric has only a single (conformal) degree of freedom and one can incorporate all these quantum effects using a simple local (Liouville) effective action.
An advantage of our model, is that \emph{it contains propagating dynamical degrees of freedom} - the $c=1$ matter boson and hence it contains richer physics in comparison to de-Sitter JT, or Sine-Dilaton gravity, whilst at the same time \emph{having an in-principle non perturbative definition through its normal matrix model dual}.

In section~\ref{SolutionsEOM}, we show that the effective action of the gravitating SG model admits both de-Sitter and anti-de Sitter saddles, due to the presence of periodic minima and maxima in the cosine potential for the matter field $X$\footnote{In fact one can even realize slow-roll inflation in this model (in Lorentzian signature), assuming that the cosine potential in~\eqref{SGmodel} is slowly varying i.e. $R \gg 1$.}. While being as yet unable to analytically solve the relevant saddle point equations in complete generality, in section~\ref{ComplexLiouville} we numerically construct interpolating geometries between AdS and dS, similar to the ``wineglass'' wormholes of~\cite{Betzios:2024oli,Betzios:2024iay} and the ``centaur'' geometries of~\cite{Anninos:2017hhn,Anninos:2022hqo}.  In section~\ref{MinisuperspaceWDW}, we analyze the minisuperspace Wheeler DeWitt (WDW) equation and show how it encapsulates the AdS and cosmological type of physics of the afforementioned solutions.

In section~\ref{genuszerofree} we improve upon the classical analysis of the effective action, using exact CFT and matrix model techniques - these allow for an exact evaluation of the genus zero ($S^2$) free energy and entropy of the model, that are both finite and non-zero. This leads to an additional property of our two-dimensional model that makes its analysis extremely interesting - one can always adopt a string theory perspective, where the two dimensional Liouville CFT (with SG matter in our case) is interpreted as the worldsheet CFT
of a (Liouville) string theory. This leads to the following important observation: \emph{The presence of Euclidean de-Sitter ($S^2$) saddles on the worldsheet can be linked to the presence of a classical entropic contribution associated to a target space geometry}. 
The gravitating compact SG model we consider here and its T-dual version\footnote{This is a version, where one turns on winding/vortex modes on the worldsheet, instead of Tachyons/vertex modes, see section~\ref{sec:SineGordon} for more details.}, describe two-dimensional target space manifolds with a compact direction, having a thermodynamic interpretation containing a classical entropic piece~\cite{Alexandrov:2002pz}. In the T-dual model, this classical entropic piece can be thought of as being related to the presence of a classical horizon in the target space geometry, pinpointing to a connection between a ``worldsheet'' de-Sitter space and a ``target space'' black hole\footnote{This is consistent with the physical picture proposed in~\cite{Susskind:1994sm}. Here due to the string scale size of the target space geometry, there is a regime of parameters where the black hole can also be thought of as a condensate of gravitating long strings that carry the entropy, see~\cite{Betzios:2022pji} and references within for more details.}. One would expect that this relation generalizes even in higher dimensional examples of string theory where in the presence of target space black hole backgrounds the worldsheet (or a higher dimensional brane) action admits a classical Euclidean de-Sitter background solution\footnote{The work~\cite{Betzios:2020zaj} makes such a suggestion for higher dimensional brane(world)s, but from a bottom up perspective.}.

In section~\ref{sec:Chiral}, we briefly review the matrix quantum mechanics model (MQM) dual to $c=1$ Liouville string theory using the chiral formalism of~\cite{Alexandrov:2002fh,Alexandrov:2003ut}. A priori this model describes the case of uncompactified $X$, and if we wish to compactify it, we need to pass to a finite temperature version of the MQM model and chiral description. In the compactified case, there exists in fact a version of the duality (including the SG deformations) more suitable for our purposes  - involving a \emph{Normal Matrix Model} (NMM)~\cite{Alexandrov:2003qk,Mukherjee:2005aq}. We describe this model and its properties in section~\ref{sec:Normal}. The NMM crucially differs from the compactified MQM at the non-perturbative level, exhibiting two important advantages consistent with expectations for a microscopic holographic model of de-Sitter space. Upon diagonalization of the matrix, the eigenvalues are found to occupy a finite (compact) region in the complex plane. In addition when computing certain observables, the duality with the gravitational SG model can operationally be defined even at finite-N, as first explained in~\cite{Mukherjee:2006hz,Mukherjee:2005aq}. We review and expand upon the relevant arguments in section~\ref{sec:Normal}. Our analysis clarifies that there exist two versions of performing a large-N limit that are of physical interest:
A usual ``macroscopic'' 't Hooft (saddle point) type of limit, where the eigenvalues form a
continuous compact fluid droplet on the complex plane with sharp edges (on which we expand upon in section~\ref{MacrolargeNlimit}), as well as a more ``microscopic'' version of the large-N limit, that invokes the original (unscaled) orthogonal polynomials and the associated Christoffel-Darboux (CD) kernel of the NMM. While the first limit can only reproduce expectation values of observables to leading order in the genus/topological expansion, the second limit is able to reproduce the all-genus expansion and can be used to non-perturbatively define the quantum gravity path integral, for all observables. It is important to emphasize at this point that \emph{this ``microscopic'' large-N limit is different than the usual double scaling limit of matrix models}, that focuses near their spectral edge. In particular it \emph{keeps the original orthogonal polynomials unscaled}. More details on the properties of normal matrix models, their orthogonal polynomials - CD kernels, and on how to perform the specific large-N limits are provided in Appendix~\ref{NormalMatrix}.

In section~\ref{sec:selfdual}, we analyze further the case when the compactification radius is $R=1$. This is sometimes called the ``self-dual'' radius (this terminology descends from the string theory perspective). In this case the model simplifies considerably and can be completely and analytically solved. In particular in section~\ref{MacrolargeNlimit} we analyze the ``macroscopic'' version of the large-N limit, computing observables
such as the expectation value of loop operators/Wheeler-DeWitt (WDW) wavefunctions ($\Psi(\ell) = \langle W(\ell) \rangle$) at genus zero in section~\ref{LoopMacro}.
In sections~\ref{MicroLargeN} and~\ref{LoopMicro} we perform the exact ``microscopic'' large-N analysis, that leads to non-perturbative results for the loop operators/WDW wavefunctions, that are found to exhibit two behaviours/phases - one that qualitatively resembles the genus zero non-normalizable modified Bessel wavefunctions for a small SG deformation parameter and another that exhibits a bounded oscillatory wavefunction with a power law decaying envelope at $\ell \rightarrow \infty$ indicating its normalizability, when the deformation parameter is large.
These results are consistent with the minisuperspace bulk results and indicate that the cosmological WDW wavefunction is in fact normalisable in this model. In sections~\ref{dosrealmu} and~\ref{nurealproperties}, we analyze the dual density of states, upon assuming that $\Psi(\ell) = \langle W(\ell) \rangle = \langle Z(\ell) \rangle = \int d E \langle \rho(E) \rangle e^{- \ell E}$, with the loop length having an interpetation as an inverse temperature $\beta \equiv \ell$ for the macroscopic (loop) boundary dual (if it exists). The results of these sections are quite interesting showing that, while in the AdS case/phase
of the model there can potentially exist an interpetation of an averaged boundary dual living on the macroscopic loop boundary, this is evidently not the case in the cosmological regime, whereby the density of states $\langle \rho(E) \rangle$ inevitably contains regions where it becomes negative\footnote{Something similar happens also for the complex Liouville string~\cite{Collier:2024kmo,Collier:2025pbm,Collier:2024lys}.}.
There are various perspectives that one can take upon this result, that we discuss further in sections~\ref{dosrealmu} and~\ref{nurealproperties}. The most conservative perspective is that this result is not surprising, because as we mentioned in the case of dS or a compact Cauchy surface cosmology, there is no true asymptotic boundary where gravity can be turned off in contrast with the AdS case. In this case the loop boundary (even if macroscopic) is always gravitating, in contrast with the AdS case. Nevertheless, the NMM is perfectly well defined in all cases and offers a direct non-perturbative description of the bulk physics.

We finally mention that the $R=1$ NMM, exhibits interesting connections with several other models that have appeared in the literature. It is directly related to topological strings and the Imbimbo-Mukhi-Kontsevich-Penner matrix model, as well as with the Gaussian matrix models that describe the $1/2$-BPS sector of $\mathcal{N}=4$ SYM~\cite{Imbimbo:1995np,Ghoshal:1995wm,Mukhi:2003sz,Gopakumar:2022djw,Gopakumar:2024jfq}. We briefly review these correspondences in the discussion section, together with further comments on future directions to explore. Throughout this work, we complement our analysis with various detailed appendices that provide relevant background material and derivations of formulae that are used in the main text.

\section{The gravitational Sine-Gordon model}
\label{sec:SineGordon}

The action of the $c=1$ Liouville theory (dual to gauged matrix quantum mechanics, see~\cite{Ginsparg:1993is,Klebanov:1991qa,Nakayama:2004vk,Martinec:2004td,Alexandrov:2003ut} for reviews) is
\bea\label{Liouvilleaction}
S_{c=1} = \frac{1}{4 \pi} \int_{\mathcal{M}} d^2 z \sqrt{\hat{g}} \left[ \hat{g}^{\m \n } \partial_\m \phi \partial_\n \phi + Q \hat{R} \phi + \mu (\phi) e^{2 b \phi} + \hat{g}^{\m \n} \partial_\m X \partial_\n X \right] \, . \nn \\
c_X = 1 \, , \quad c_L = 1 + 6 Q^2 = 25 \, , \quad Q = b+ \frac{1}{b} = 2 \, , \quad b=1 \, ,
\eea
where $\phi(z,\bar{z})$ is the Liouville field, $\mu$ is the cosmological constant\footnote{We do not include the factor of $4 \pi$ in the definition of $\mu$, since we are essentially using the physical $\mu_{KPZ}$ that appears also in the string theory genus expansion.} and $X(z,\bar{z})$ is the $c=1$ matter boson, which we consider to be compact with periodicity $X \sim X + 2 \pi R$. The parameters in the action~\eqref{Liouvilleaction} are fixed, so that it defines an exact CFT at the quantum level.

Notice also the parenthesis in the Lagrangian of~\eqref{Liouvilleaction}. The (``dressed'') cosmological constant operator is actually a linear combination of terms $\phi e^{2  \phi}$ and $e^{2  \phi}$~\cite{Polchinski:1990mf}.  Which term should be used, depends on the phase of the model and the ensemble of the dual matrix quantum mechanics model that one is working on (i.e. canonical vs. grand-canonical).  Here we shall analyze the canonical possibility ($\mu e^{2  \phi}$). See also~\cite{Klebanov:1994pv} for more details on other phases of the model with microscopic ``touching'' wormhole interactions. On a manifold with boundaries we add the following boundary term~\cite{Fateev:2000ik,Teschner:2000md}  
\be\label{boundaryaction} 
S_{\partial \mathcal{M}} = \int_{\partial \mathcal{M}} d u \hat{g}^{1/4} \left(\frac{K \phi}{ \pi} + \mu_B e^{\phi} \right)\, , 
\ee
with $K$ being the extrinsic curvature and the parameter $\mu_B$ corresponding to the boundary cosmological constant, which is sometimes parametrised by $\mu_B = \sqrt{\mu} \cosh \pi \sigma$, with $\sigma$ a dimensionless parameter. While the matter field being a free boson, can satisfy simple Dirichlet or Neumann boundary conditions, we find from \eqref{boundaryaction} that the analogous possibilities for the Liouville field are ($n$ is a unit normal vector the the boundary)
\be
\delta \phi |_{\partial M} = 0 \, ,  \quad \frac{ \partial  \phi}{\partial n} + 2 K + 2 \pi \mu_B  e^{ \phi} \, |_{\partial M} = 0 \, .
\ee
The same options will continue to hold for the gravitational Sine-Gordon model, that we shall now introduce.

The physical $(1,1)$ operators of the (compactified) $c=1$ Liouville CFT are the vertex/vortex operators that take the form\footnote{The vertex operators are also target space Tachyon operators from a string theory perspective.}
\bea\label{Liouvilleoperators}
{T}^-_{\pm n/ R} &=& e^{\pm i \frac{n}{R}(X_L + X_R) } e^{\left(2 - \frac{n}{R} \right) \phi} \, , \quad {T}^+_{\pm n/ R} = e^{\pm i \frac{n}{R}(X_L + X_R) } e^{\left(2 + \frac{n}{R} \right) \phi} \, , \nn \\
\mathcal{T}^-_{\pm n R} &=& e^{\pm i n R(X_L - X_R) } e^{\left(2 - n R \right) \phi} \, , \quad \mathcal{T}^+_{\pm n R} = e^{\pm i n R(X_L - X_R) } e^{\left(2 + n R \right) \phi}
\eea
When $\mu \neq 0$, we recover one linear combination for the possible dressing due to the presence of the Liouville wall that reflects the modes. In particular only the $(-)$ (non-normalisable) modes are considered as the physical asymptotic vertex/vortex operators\footnote{The $+$ operators can be thought of as changing the state of the theory~\cite{Betzios:2022pji}.}.

We now define a generalised gravitational $c=1$ (compact) Sine-Gordon model by deforming the action with two complex conjugate vertex modes
\be\label{SineGordon}
S_{SG} = S_{c=1} + S_{v} \, , \quad S_v =  \frac{1}{4 \pi} \int_{\mathcal{M}} d^2 z \sqrt{\hat{g}} \left(   t_n  \mathcal{T}^-_{+ n/R} + t_{-n} \mathcal{T}^-_{-n/R} \right) \, ,
\ee
the standard gravitational Sine-Gordon model corresponding to the case when only a single mode $t_n = t_{-n} = t$
is turned on~\cite{Moore:1992ga}. This deformation makes sense as long as $2 > n/R$, so that the perturbation of the Lagrangian does not blow up in the asymptotic weakly coupled region and is a relevant deformation\footnote{The (normalisable) operators with the $(+)$ dressing in eqn. \eqref{Liouvilleoperators} are always relevant and grow in the strongly coupled region $\phi \rightarrow \infty$.}. Under this assumption $S_{SG}$ is expected to lead to a well defined CFT\footnote{To our knowledge this has not been proven rigorously yet.}. We should also clarify at this point that, while the actions~\eqref{Liouvilleaction} and~\eqref{SineGordon} are classical CFTs under the replacement
$Q \rightarrow 1/b , \, \xi_n = 2 - |n|/R \rightarrow 2 b$, nevertheless when $c=1$ and $b=1$ quantum effects are important and have a large effect in the renormalization of the parameters that appear in the quantum effective Liouville action, as can be seen from the parameters of eqns.~\eqref{Liouvilleaction} and~\eqref{Liouvilleoperators} and our analysis in appendix~\ref{Weylindependence}.

A basic quantity that one can describe in both sides of the MQM/Liouville duality is the partition function/free energy containing general vertex perturbations 
\be\label{mainmatch}
 Z_{\text{gen. SG}}(t;\mu,R) = \left\langle  e^{ \sum_n t_n \left( \mathcal{T}^-_{n/R}+ \mathcal{T}^-_{- n/R}  \right) }   \right\rangle_{c}   = \mathcal{F}_{\text{target}}(t ; \mu R) = \log \mathcal{Z}_{\text{Matrix Model}}( t ; \mu , R ) \, .
\ee
This equality exhibits the usual relation between the 2d quantum gravity partition function of connected closed surfaces (or the ``worldsheet'') and the free energy on the target space (when adopting a string theory interpretation). In particular the connected correlators of the $c=1$ compact worldsheet CFT with general vertex perturbations are related to the partition function of a generalised Sine-Gordon model coupled to two dimensional quantum gravity ($Z_{\text{gen. SG}} $), that also has a free energy interpretation for the target space ``string-field theory'' action. This also corresponds to the free energy of the dual matrix model that captures the complete geometric genus expansion (as well as various non perturbative effects). In some cases, and in particular when the characteristic scale of the size of the 2d surfaces (expressed via $\alpha'$ in the string theory language) is small compared to the characteristic scales of the target space, one can approximately perform this computation using a low energy local target space effective action. In more general cases, as in our specific $c=1$ example, such a local target space effective action ceases to be trustworthy and one has to resort either to 2d quantum gravity (worldsheet) or matrix model techniques to analyze the physics in detail.

In the $c=1$ case, one version of the dual matrix model is Matrix Quantum Mechanics (MQM) in the presence of the appropriate Tachyon/vertex deformations, which we briefly discuss in section~\ref{sec:Chiral}. On the other hand, as we review in section~\ref{sec:Normal}, there exists a second Normal Matrix Model (NMM) description that can be used instead as the matrix model dual of the general gravitating Sine-Gordon models when $X$ is compact, that differs non-perturbatively from the compactified $c=1$ MQM. This NMM description is found to carry both technical and conceptual advantages, in relation to Cosmological/de-Sitter space physics, that we analyze in section~\ref{sec:Normal} and in much more detail in section~\ref{sec:selfdual}.

In the rest, we shall mostly focus in the case where we turn on a single vertex deformation, that corresponds to the usual Sine-Gordon model in its gravitating form. In this case there exist two parameters ($t_n, \, \mu$) on which physical quantities depend on, so the physical scales will be set on either by the SG interaction or the cosmological constant term respectively.  A KPZ-DDK scaling analysis of the action/free energy\footnote{This is performed by shifting the Liouville field $\phi \rightarrow \phi + \phi_0$, and checking how the various terms scale.} indicates that dimensionally
\be\label{KPZDDK}
 t_{\pm n} \sim \mu^{1- |n|/ 2 R} \, , \qquad g_s \sim 1/\mu  \, .
\ee
We can then define the dimensionless combinations 
$\zeta_n = t_n/\mu^{1-|n|/2 R} $. These parameters govern the relative importance between the terms in the action of the gravitating SG model. In particular for large $\zeta_n$, the corresponding vertex deformation becomes large and the string coupling and scale are set by $t_n$, while for small $\zeta_n$ the most relevant scale is $1/\mu$.
The gravitating SG model can exhibit phase transitions as one tunes $\zeta_n$\footnote{This was first observed in~\cite{Moore:1992ga} for the gravitating Sine-Gordon model and in~\cite{Betzios:2022pji} for the T-dual model consisting of vortex deformations.}. In the case that the scale of the theory is governed by the SG coupling $t_n$, it is possible to turn off the coupling $\mu$, so that the model runs in the IR to the so-called Sine-Liouville (SL) CFT. Our analysis of the saddle point equations in section~\ref{SolutionsEOM}, confirms this picture, since the de Sitter and cosmological classical solutions exist only in the phase/regime of relatively large deformation, that is at large enough $\zeta_n$\footnote{In the case of the T-dual vortex deformation this corresponds to the black hole phase~\cite{Betzios:2022pji}.}, or in the SL model where $\mu \rightarrow 0$. As we shall delve upon later this leads to a classical contribution to the entropy of the system, that has both a $2d$ quantum gravity and target space interpretation (if the $2d$ surface is interpreted as the worldsheet of a string theory).

\section{Solutions of the effective action}\label{SolutionsEOM}

In this section, we consider classical (saddle point) solutions of the (renormalised) effective action\footnote{In the sense of having taken into account the normalization shifts of the effective action parameters due to ghosts and the non-trivial measure.} of Liouville coupled to the gravitating SG model in the fixed background metric $\hat{g}_{\m \n}$. Of course this analysis does not completely take into account the quantum fluctuations of the Liouville field itself, since we do not integrate over it in the path integral, and in the later sections we consider both exact CFT and matrix model techniques to compute the exact contributions to the free energy and various observables, either at genus zero or even non-perturbatively in some specific cases (such as when the compactification radius is $R=1$).

We shall find though that the analysis of the renormalized effective action, produces results that are in good qualitative agreement with the exact results (at leading order in the genus expansion). Something analogous was also observed for the gravitational Sine-Gordon model in~\cite{Moore:1992ga}, and our analysis is clarifying and extending these former results. As an example of the latter, we are able to find new saddles of the resulting equations that interpolate between the de-Sitter and anti-de Sitter saddles, in section~\ref{ComplexLiouville}.

\subsection{Saddle point equations}\label{sec:sols1}

Let us now turn on to a description of the most simple classical solutions of the saddle point equations of motion (EOMs) of the gravitational SG model in the canonical ensemble/case\footnote{The other option with the Liouville potential $\mu \phi e^{2 \phi}$~\cite{Polchinski:1990mf,Klebanov:1994pv}, can be treated similarly and it would be interesting to explore it further in the future.} where the Liouville potential term is $\mu e^{2 \phi}$. We define $\xi_n = 2 - |n|/R$ and consider the equations stemming from the effective SG action
\bea\label{EOMsa}
-  \hat{\nabla}^2 \phi +  \hat{R} +  \mu e^{2 \phi} +  t \xi_n e^{\xi_n \phi} \cos \frac{n}{R} X = 0 \, , \nn \\
-  \hat{\nabla}^2 X -  t \frac{n}{R} e^{\xi_n \phi} \sin \frac{n}{R} X = 0 \, , \nn \\
2 ( \hat{\nabla}_\m \hat{\nabla}_\n \phi - \hat{g}_{\m \n} \hat{\nabla}^2 \phi ) - \partial_\m \phi \partial_\n \phi - \partial_\n X \partial_\m X + \frac{\hat{g}_{\m \n}}{2 } ( \mathcal{L}_{SG} - 2 \phi \hat{R} ) = - \hat{T}_{\m \n}^{class.} \, , \nn \\
\mathcal{L}_{SG} =  (\partial \phi)^2 + (\partial X)^2 + 2 \phi \hat{R} + \mu e^{2 \phi} + 2 t e^{\xi_n \phi} \cos \frac{n}{R} X \, .
\eea
The third equation defines the stress tensor treating the effective gravitational SG action as being purely classical, but in fact there is a renormalization of its exponential terms due to an appropriate regularization and renormalization of the $e^{a \phi}$ operators, (see appendix~\ref{regularizationexpoperators}). This leads to the following renormalized stress energy tensor in our example
\bea
\hat{T}^{(ren.)}_{\m \n} &=& 2 (   \hat{g}_{\m \n} \hat{\nabla}^2 \phi - \hat{\nabla}_\m \hat{\nabla}_\n \phi) + \partial_\m \phi \partial_\n \phi + \partial_\n X \partial_\m X - \nn \\
&-& \frac{\hat{g}_{\m \n}}{2 } \left( (\partial \phi)^2 + (\partial X)^2 + 2 \mu e^{2 \phi} + 2 \xi_n t e^{\xi_n \phi} \cos \frac{n}{R} X\right) \nn \\
\eea
One can check that this renormalized stress tensor is traceless on a flat background $\hat{g}_{\m \n} = \delta_{\m \n}$, upon using the first (Liouville field) equation, verifying that the model corresponds indeed to a CFT. On a more general background $\hat{g}_{\m \n}$, one needs to take into account also the presence of the conformal anomaly, so that $\hat{T}^{(ren.) m}_{\m} \sim  \hat{R} $ (the coefficient playing the role of an effective central charge).

Going to conformal gauge for the effective background metric: $\hat{g}_{\m \n} = e^{2 \sigma} \delta_{\m \n}$ so that $\hat{R} = - 2 e^{- 2 \sigma} \partial^2 \sigma \, $, the Liouville and matter equations reduce to
\bea\label{EOMsconformala}
- \partial^2 \phi - 2 \partial^2 \sigma + e^{2 \sigma} \left(  \mu e^{2 \phi} +  t \xi_n e^{\xi_n \phi} \cos \frac{n}{R} X \right) = 0 \, , \nn \\
- \partial^2 X - t e^{2 \sigma}\frac{n}{R} e^{\xi_n \phi} \sin \frac{n}{R} X = 0 \, . 
\eea
We observe that the equation of motion for $X$, is similar to the Sine-Gordon equation, but where now the scale factor $\sigma$ and $\phi$ dynamically change the magnitude of the cosine potential. In particular regions of large $\sigma, \,\phi$ increase its magnitude and lead to a trapping in its minima, while the opposite happens when $\sigma , \, \phi$ become smaller.

We shall now attempt to solve these equations in a gradual increase of generality. Solutions with constant $\phi$, are particularly important, since the metric has the same properties in any frame that involves it being rescaled by a $\phi$ dependent conformal factor.

\subsubsection{The undeformed $c=1$ Liouville theory }\label{undeformedLiouville} 

Let us first start by considering the undeformed $c=1$ Liouville theory by setting $t=0$. One then finds a simple solution when $X = X_0 \, , \, \phi = \phi_0$ are constants and
\be
\hat{R} = - 2 e^{- 2 \sigma} \partial^2 \sigma = - \mu e^{2 \phi_0} \, .
\ee
This describes a space of constant negative/positive curvature ($(A)dS_2$) for $\mu > 0 $ and $\mu < 0$ respectively\footnote{Rescaling by $\phi_0$ does not change this, since it is a constant.}. Normally the 
Liouville theory is defined for $\mu > 0$, since only then the potential is bounded from below (see though~\cite{CarneirodaCunha:2003mxy} for an analysis of the $\mu < 0 $ de-Sitter regime). The deformations $t \neq 0$ will cure this pathology, whilst retaining the dS solutions, since the cosine-potential is well defined and bounded from below. Notice also that the topology of the $EAdS_2$ space could also be that of a cylinder/two boundary wormhole for a hyperbolic conjugacy class of $SL(2,R)$~\cite{Ginsparg:1993is}. The background field $\sigma$ though, will diverge on each of the two boundaries\footnote{In most models of Dilaton gravity the Dilaton is monotonic and renders the solution effectively singular on one of the two boundaries (region of infinite string coupling from a string theory perspective). See though~\cite{Maldacena:2018lmt} for two boundary $AdS_2$ wormholes supported by additional matter fields, without a singular Dilaton. Our model contains even more non-trivial wormhole geometries, such as ``wineglass'' type of wormholes, that we analyze in section~\ref{ComplexLiouville}.}.

Let us also adopt the other option commonly used in the literature, that is fixing a frame where $\hat{g}_{\m \n} = \delta_{\m \n}$, so that $\hat{R} = 0$. The metric then that becomes $(A)dS$ is the original
metric $g_{\m \n} = e^{2 \phi} \hat{g}_{\m \n}$, since $R(e^{2 \phi} \hat{g}_{\m \n}) = - \mu/2$. We find therefore that these two choices of frame ultimately describe the same physics. In section~\ref{Gaussian}, we shall show how the two options for the sign of $\mu$ corresponding to (A)dS spacetimes, are imprinted the properties of the associated Wheeler-DeWitt (WDW) wavefunction.

\subsubsection{Simple solutions for the gravitational Sine-Gordon theory}

The SG matter equation admits two types of simple solutions localised at critical points of the cosine potential, when $X$ is also a constant. The two types of critical points are maxima/minima of the cosine potential and labeled by an even/odd integer $m$
\be\label{criticala}
X^{(n)}_m =   m \pi \frac{R}{n} \, ,  \qquad m \in \mathbb{Z} \, ,
\ee
We observe from the rest of the EOMs that exchanging the even/odd type of solutions/critical points amounts to effectively changing the sign $t \rightarrow - t$. For example, for the odd solutions, the $\phi$ equation reduces to
\bea\label{EOMsconformal2a}
- \partial^2 \phi - 2 \partial^2 \sigma + e^{2 \sigma} \left(  \mu e^{2 \phi}  - t \xi_n e^{\xi_n \phi}  \right) = 0 \, , 
\eea
and the even type of solutions are obtained upon changing the sign of $t$.

\paragraph{The Sine-Liouville limit -} 
Let us first consider the case $\mu \rightarrow 0$ (which is sometimes called the Sine-Liouville limit, see~\cite{Betzios:2022pji} for its T-dual description and analysis). In this case once again the simplest types of solutions are constant $\phi = \phi_{(m)}$ solutions for which
\be
\hat{R} = - 2 e^{- 2 \sigma} \partial^2 \sigma = \pm t \xi_n e^{\xi_n \phi_{(m)}} \, .
\ee
We observe that due to the cosine potential odd/even solutions correspond to dS and AdS respectively for $t>0$ (and vice versa had we assumed $t<0$). This model therefore correponds to a well defined CFT with a potential that is oscillatory but bounded from below, while at the same time admitting saddles with either positive or negative curvature at its two types of critical points.
Once again we could also adopt the perspective/frame $\hat{g}_{\m \n} = \delta_{\m \n}$ ($\sigma = 0$) and then find that the metric 
$g_{\m \n} =  e^{\xi_n \phi} \delta_{\m \n}$ is a space of constant positive/negative curvature
\be
R(e^{\xi_n \phi} \delta_{\m \n}) = - \xi_n e^{- \xi_n \phi} \partial^2 \phi = \pm t \, .
\ee

\paragraph{General parameters ($\mu \, , \, t \neq 0$) -} 
In the generic SG case, we find that once again constant $\phi$ solutions are either (A)dS and the sign of the curvature is now dependent on both parameters
\be
\hat{R} = - 2 e^{- 2 \sigma} \partial^2 \sigma = - \mu e^{2 \phi_{(m)}} \pm t \xi_n e^{\xi_n \phi_{(m)}} \, .
\ee
This more general case admits both (A)dS solutions depending also on the relation between between the values
of $\mu, t$ (for example if $t$ is very small the dS solution can cease to exist).

\paragraph{The on-shell action for the simplest solutions -} 
Since the simplest solutions have both $X, \, \phi_{(m)} $ and the curvature $\hat R$ to be constants, the on-shell action is simply proportional to the Euler characteristic of the surface (when discussing the AdS saddles, we consider the surface boundary to be a compact $S^1$ and thus the topology is a disk) 
\be
S_{\text{on-shell}}  \sim \phi_{(m)} \chi_{\mathcal{M}} \, .    
\ee
This means that the contribution to the path integral from the $EdS_2$
saddle is at the saddle point level $Z_{SG.} \sim e^{- S_{\text{on-shell}}} \sim g_s^{-\chi_{\mathcal{M}}} \sim g_s^{- 2} \sim \mathcal{F}_{target}$, where we defined the ``string coupling'' $g_s \sim e^{\phi_{(m)}}$. This qualitative behaviour of the on-shell action is expected on general grounds from a string theoretic perspective of the gravitational SG model, see~\cite{Betzios:2022pji} and references within for a more detailed discussion. We did not keep track of the numerical prefactors though, since as we argued at $c=1$ the path integral over the fluctuations of the Liouville field is important and will lead to quantitative changes of such numbers. The exact result, using CFT techniques is provided in the next section, exhibiting a scaling behavior in accordance with the $EdS_2 = S^2$ saddle results, in the regime where the latter is expected to dominate the physics. 

\subsubsection{The exact genus zero free energy}\label{genuszerofree}

The exact genus zero free energy, when the first deformation is turned on ($t_1  = t \, , \, \xi_n = 2 - 1/R$) was computed first in~\cite{Moore:1992ga} and then later on in~\cite{Kazakov:2000pm} and~\cite{Betzios:2022pji} for the T-dual model, using CFT, integrability and matrix model techniques. The simplest derivation uses the fact that 
\be
\mathcal{Z}_{SG} = \langle e^{ t \,\mathcal{T}^{-}_{1/R} + {\bar t}\,\mathcal{T}^-_{-1/R}} \rangle
=\sum_{n=0}^\infty \sum_{m=0}^\infty \frac{t^n}{n!}
\frac{{\bar t}^m}{m!}
\langle (\mathcal{T}^-_{1/R})^n(\mathcal{T}^-_{-1/R})^m\rangle =
\sum_{n=0}^\infty \frac{|t|^{2n}}{(n!)^2}
\langle (\mathcal{T}^-_{1/R}\mathcal{T}_{-1/R})^n\rangle \, ,
\ee
the last equality following from momentum conservation. The form of the correlators at genus zero is
\be
\langle (\mathcal{T}^-_{1/R}\mathcal{T}^-_{-1/R})^n\rangle^{(0)} = \frac{n!}{R^{2 n}} \mu^2 \left( (1-1/R) \mu^{1/R-2}\right)^{n} \frac{\Gamma \left(n(2-1/R) - 2 \right)}{\Gamma \left(n(1-1/R) +1 \right)} \, , \qquad n \geq 1 \, .
\ee
In appendix~\ref{exactPartition} we provide the all-genus exact answer for these correlators using the dual matrix model. Unfortunately it is not very convenient to work with beyond genus zero, unless one invokes a combinatorial large ``representation limit'' discussed in~\cite{Betzios:2022pji}.

One can then compute the SG partition function/target space genus zero free energy in an expansion at large $t$ and small $\mu$, with either integrability~\cite{Kazakov:2000pm,Alexandrov:2002pz} or matrix model techniques~\cite{Betzios:2022pji} to find
\be
\mathcal{Z}_{SG.}^{(0)} = \mathcal{F}^{(0)}_{target} \simeq - A(R) t^{4 R/(2R-1)} - B(R)  \mu t^{2R/(2R-1)} + O(\mu^2) \, , 
\ee
with $A(R),B(R)$ coefficients that vanish at $R=1/2$\footnote{This is the T-dual point of the Kosterlitz-Thouless point ($R=2$) of vortex liberation studied in~\cite{Kazakov:2000pm} and~\cite{Betzios:2022pji}.}. Remembering the KPZ/DDK scaling in eqn.~\ref{KPZDDK}, we observe the aforementioned genus zero scaling $\mathcal{Z}_{SG.}^{(0)} = \mathcal{F}^{(0)}_{target} \sim {1}/{g_s^2}$. 

The non-trivial $R$ dependence of the result is of utmost importance - in the opposite regime of small $t$ and large $\mu$, the leading piece in the free energy is found to scale linearly with $R$~\cite{Kazakov:2000pm,Alexandrov:2002pz,Betzios:2022pji}. When this happens, the leading classical entropy computed from the formula  $S^{(0)} = (- R \partial_R  + 1) \log \mathcal{Z}^{(0)}_{M.M} = (R \partial_R  - 1) \mathcal{F}^{(0)}_{target}$, vanishes identically. This exact genus-zero result from the matrix model is also consistent with our semi-classical analysis in the previous section, which shows that in this latter regime only negatively curved semiclassical saddles contribute in the path integral, leading to higher genus (quantum) contributions to the entropy\footnote{Let us emphasize once more that we are discussing a target space notion of entropy. This means that we consider compact manifolds, without a boundary to compute it. This is distinct from any potential notion of entropy for the putative dual that lives on the $S^1$ boundary created by loop operators on the surfaces, a perspective more akin to recent analyses of JT and Dilaton gravity models, that we describe in section~\ref{LoopMacro}.}. Going back to the regime of large $t$ and small $\mu$, since the genus zero (classical) entropy is non-zero, we are led to an object in target space with the properties of a gravitating string condensate or a black hole (for the T-dual model~\cite{Kazakov:2000pm,Betzios:2022pji}), see also~\cite{Alexandrov:2002pz} for further discussion in the present case of (target space) Tachyon deformations\footnote{Phase transitions between the two regimes where studied in~\cite{Betzios:2022pji}, but only for the T-dual model of winding/vortex deformations. It would be interesting to revisit this study for the present gravitational SG model with Tachyon/vertex deformations.}. .

From the perspective of two dimensional quantum gravity (the gravitating SG model), this also means that the two dimensional $EdS_2 \equiv S^2$ becomes an on-shell saddle of the effective action in the large $t$ small $\mu$ parameter regime, in accordance to what we found in the previous section - our saddle point analysis resulted into the same behavior $\mathcal{Z}_{SG.}^{(0)} = \mathcal{F}^{(0)}_{target} \sim {1}/{g_s^2} = e^{- 2 \phi_{(m)}}$ (up to coefficients that are affected by the strong quantum effects of the fluctuations of the Liouville mode at $c=1$, that the saddle point analysis cannot capture correctly). This computation shows that the presence of target space entropy (that is associated to a thermodynamic object such as a black hole) is directly related to the presence of a de-Sitter saddle on the two dimensional quantum gravity theory. Finally, we should mention that this result is also consistent with the physical picture proposed in~\cite{Susskind:1994sm}.

\subsection{A complex Liouville field and ``wineglass'' wormholes}\label{ComplexLiouville}

\begin{figure}[t]
\begin{center}
\includegraphics[width=0.4\textwidth]{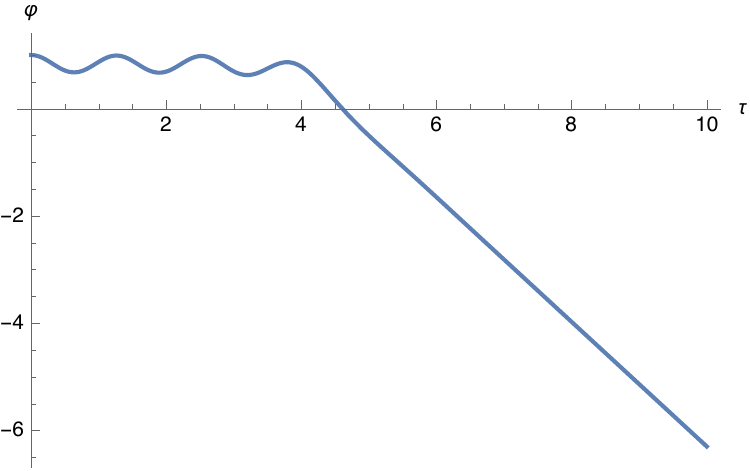}
\hspace{0.2em}
\includegraphics[width=0.4\textwidth]{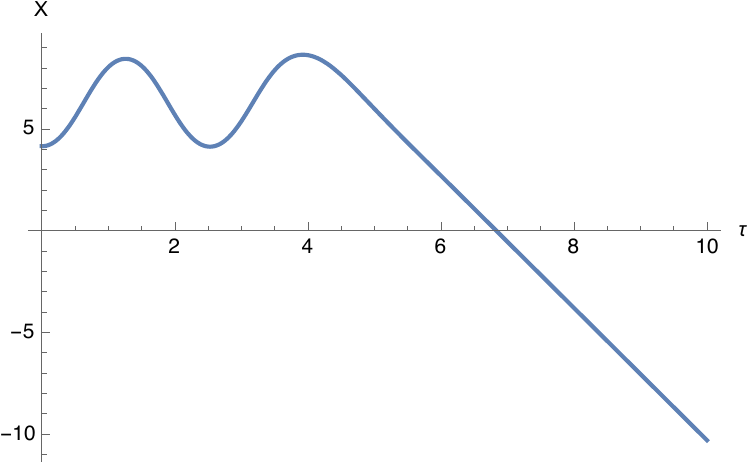}
\end{center}
\caption{Euclidean geometries (for $\xi_n = 1$), that oscillate and (generically) end in a singularity ($e^{\phi} \sim e^{- c \tau} \rightarrow 0$). The field $X$ initially oscillates, but never reaches any critical point of the potential (runaway). Forming a periodic de Sitter wormhole such as the ones found in~\cite{Halliwell:1989pu,Rey:1989th,Aguilar-Gutierrez:2023ril,Blommaert:2025bgd}, is an extremely fine-tuned unstable case.}
\label{fig:singular}
\end{figure}

One observes that the saddle point equations simplify considerably in  the case where $\xi_n = n/R$ and $\mu = 0$
. We then form the combination $\Phi = \phi + i X$ that obeys
\be\label{ComplexLiouvilleEOM}
- \nabla^2 \Phi + \hat{R} + t \xi_n e^{\xi_n \Phi^*} = 0 \, , \qquad - \nabla^2 \Phi^* + \hat{R} + t \xi_n e^{\xi_n \Phi} \, ,
\ee
so that the problem splits into two coupled Liouville equations for the complex conjugate parts (it is convenient to set $\hat{g}_{\m \n} = \delta_{\m \n} \, , \, \sigma = 0$ so that the covariant derivatives reduce to simple derivatives). This set of equations arises from the following complex Liouville field action to which the gravitational SG model action reduces
\be\label{ComplexLiouvilleaction}
S_{SG} \, \xrightarrow[\xi_n = n/R]{\mu = 0} \, S_{CL}[\Phi, \Phi^*] = \frac{1}{4 \pi} \int d^2 x \sqrt{\hat{g}} \left(\partial_\mu \Phi \partial^\mu \Phi^* + (\Phi + \Phi^*) \hat{R} + t (e^{\xi_n \Phi} + e^{\xi_n \Phi^*} )  \right) \, .
\ee
We notice that this action is reminiscent of the complex Liouville string~\cite{Collier:2024kmo,Collier:2025pbm}, with a crucial difference in that the kinetic term here consists
on the product of complex conjugate fields, instead of the sum of their squares\footnote{If the field $X$ was non-compact and timelike, we would have acquired the kinetic term of the complex Liouville string (but then one faces the problem of defining a cosine potential for a timelike field). Another option is to consider $\phi$ to be timelike, but then one needs additional matter fields to describe ``timelike'' Liouville theory.}.

Unfortunately we have not been able to integrate~\eqref{ComplexLiouvilleEOM}, or to use a method such as the Backlund-transform to reduce the problem to a linear equation. We shall therefore resort to a numerical analysis. Let us moreover assume that the functions depend only on a single coordinate, so that equations~\eqref{ComplexLiouvilleEOM} simplify to
\be
 - \Phi''  + t \xi_n e^{\xi_n \Phi^*} = 0 \, , \qquad - {\Phi^*}''  + t \xi_n e^{\xi_n \Phi} = 0 \, ,
\ee
in Euclidean signature ($\tau$ is our Euclidean coordinate) and to
\be
 \ddot{\Phi}  + t \xi_n e^{\xi_n \Phi^*} = 0 \, , \qquad  \ddot{\Phi}^*  + t \xi_n e^{\xi_n \Phi} = 0 \, ,
\ee
in Lorentzian signature ($T = i \tau$). This set of coupled ODEs is invariant under Euclidean/Lorentzian time-reversal symmetry $\tau \leftrightarrow - \tau$ or $T \leftrightarrow - T$. This means that we can seek $Z_2$ symmetric solutions if we impose initial conditions with a certain value for $\Phi(0)$ and also demand that $\Phi'(0) = \dot{\Phi}(0) = 0$. We can also glue the Euclidean to the Lorentzian solutions at the point of time reversal symmetry
as in the semi-classical descriptions of the Hartle-Hawking~\cite{Hartle:1983ai} or ``wineglass'' wormhole proposals~\cite{Betzios:2024oli}, to describe the nucleation of a two dimensional universe from nothing or from a patch in the interior of an asymptotically AdS space respectively.

\begin{figure}[t]
\begin{center}
\includegraphics[width=0.4\textwidth]{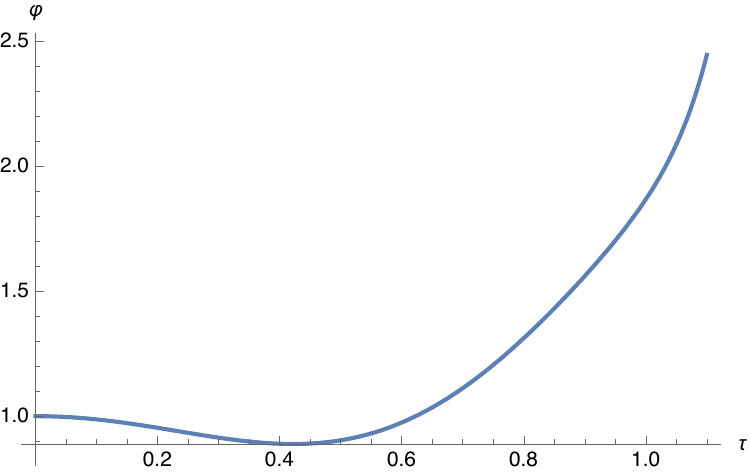}
\hspace{0.2em}
\includegraphics[width=0.4\textwidth]{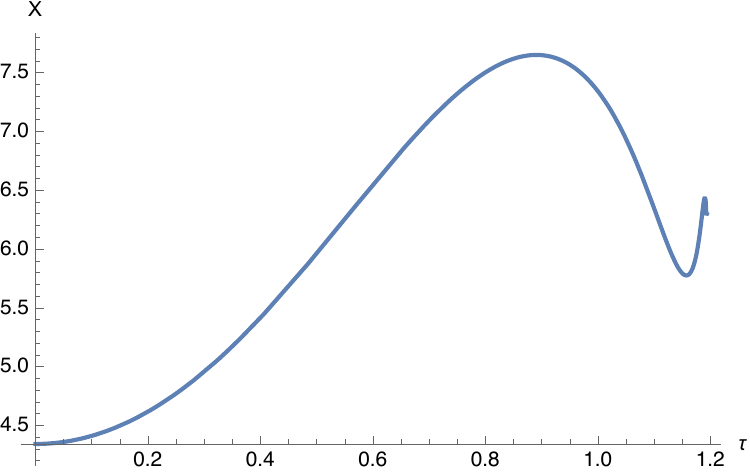}
\vspace{0.2em}
\includegraphics[width=0.4\textwidth]{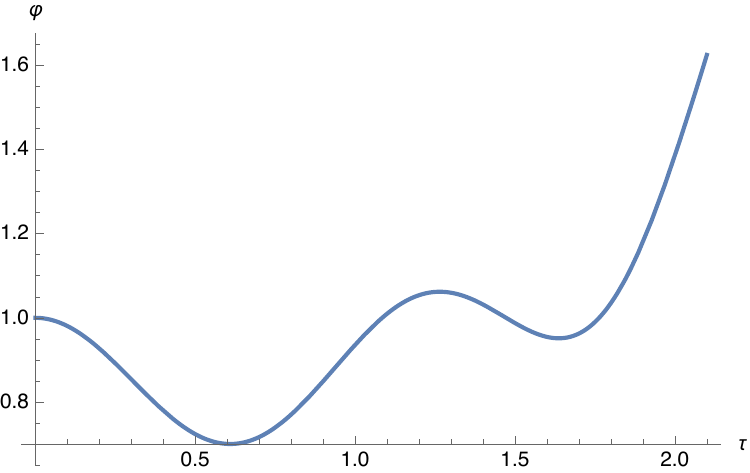}
\hspace{0.2em}
\includegraphics[width=0.4\textwidth]{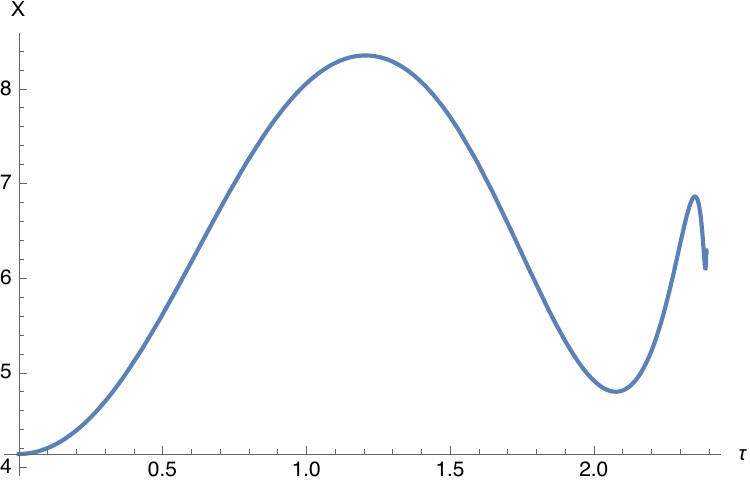}
\end{center}
\caption{Top ($\xi_n = 1$): Euclidean ``wineglass'' $AdS$ wormhole~\cite{Betzios:2024oli,Betzios:2024iay}. Bottom ($\xi_n = 1$): Bouncing $EAdS$ wormhole~\cite{Ghodsi:2024jxe}. The scalar field $X$ asymptotes to an attractor $EAdS$ critical point (Here $X_{crit.} = 2 \pi$). Both solutions are $Z_2$ symmetric under $\tau \leftrightarrow - \tau$, and satisfy $\phi'(0) = X'(0) = 0$.}
\label{fig:bouncing}
\end{figure}

Apart from the trivial solutions that we analyzed in the previous sections (where $X = \Im \Phi$ is localized at the critical points $X_m^{(n)} = \pi m /\xi_n $), the solution space is quite rich. In fig.~\ref{fig:singular} we depict bouncing Euclidean geometries that begin or end in a (Euclidean) singularity\footnote{This is the generic behaviour. For extremely tuned parameters/conditions, they could resemble the ``periodic de-Sitter wormholes''~\cite{Halliwell:1989pu}, recently studied in~\cite{Aguilar-Gutierrez:2023ril,Blommaert:2025bgd}. The paradox of an infinitely negative Euclidean action for such configurations, is evaded here due to their instability towards forming a singularity or an $EAdS$ boundary. This offers a complementary mechanism to evade the paradox, compared to the one proposed in~\cite{Halliwell:1989pu}.} and where the field $X$ exhibits a runaway behaviour. At the top of fig.~\ref{fig:bouncing} we depict the ``wineglass'' type of asymptotically $EAdS_2$ wormholes, that are two dimensional analogues of the ones discussed in~\cite{Betzios:2024oli,Betzios:2024iay}. At the bottom of the same figure, we depict more general ``bouncing'' versions of such two boundary wormholes~\cite{Ghodsi:2024jxe}. In these examples, the scalar field $X$ gets eventually ``trapped'' in an $AdS$ critical point (minimum) of the cosine potential.

\begin{figure}[t]
\begin{center}
\includegraphics[width=0.4\textwidth]{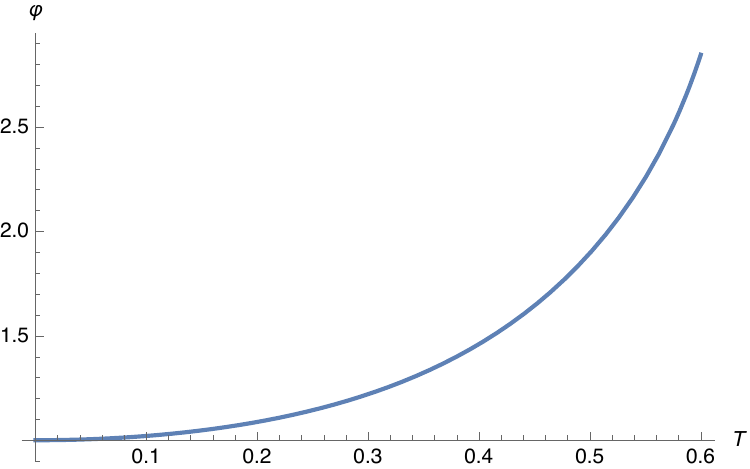}
\hspace{0.2em}
\includegraphics[width=0.4\textwidth]{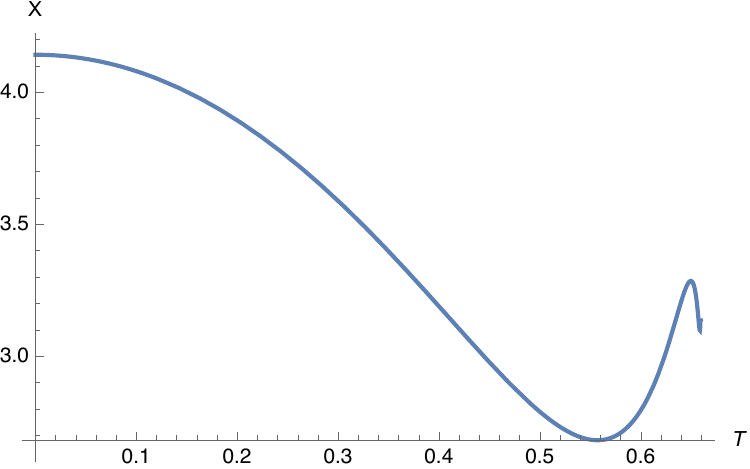}
\end{center}
\caption{The Lorentzian continuation of the ``wineglass'' wormhole (top fig.~\ref{fig:bouncing}). The Lorentzian Universe is expanding and the field $X(T)$, reaches a de-Sitter critical point (Here $X_{crit.} = \pi$).}
\label{fig:expanding}
\end{figure}

All these solutions as long as they exhibit a local maximum of $\phi$ at the point of reflection symmetry, so that $\phi''(0) < 0$, can be glued to a Lorentzian expanding Universe ($\ddot{\phi}(0) > 0$), that we depict in fig.~\ref{fig:expanding} in the case of a simple ``wineglass'' wormhole. See~\cite{Betzios:2024oli,Betzios:2024iay}, for more details on the cosmological importance of such solutions in higher dimensions, that lead to a natural inflationary phase for the universe even if the global minimum of the potential corresponds to an AdS space\footnote{Similarly to the current two dimensional case, the effective potential in higher dimensions, contains both regions where it is positive and regions where it is negative.}.

More complicated solutions without a $Z_2$ symmetry can also be constructed numerically, resembling the ``centaur'' geometries of~\cite{Anninos:2017hhn,Anninos:2022hqo}, having a single asymptotic AdS boundary connected to a piece of the Euclidean sphere (dS) in the interior. These solutions do not admit a natural $\tau = - i T$ continuation and gluing to Lorentzian signature, though.

In light of these interesting results, we believe that an analysis of the coupled equations~\eqref{ComplexLiouvilleEOM} deserves further study\footnote{The analogous equations for the complex-Liouville string or Sine-Dilaton gravity, decouple and can be solved exactly, and it would be interesting to study the potential presence of more general solutions interpolating between dS and AdS, such as the ones we find here.}.

\subsection{The minisuperspace Wheeler-DeWitt wavefunction}\label{MinisuperspaceWDW}

Further insight in the physics of such geometries, is provided via the solution of the \emph{Wheeler-DeWitt} (WDW) equation.
A general issue with this approach, is that the WDW functional equation is not well defined (especially in higher dimensions). A common approximation scheme is the so-called minisuperspace approximation, which suppresses the non-zero mode fluctuations of the fields. While perhaps not so well justified in higher dimensions, in the present two dimensional context, it has been argued and even proven in some cases, to provide exact results for Liouville theory at genus zero~\cite{Fateev:2000ik,Moore:1991sf,Moore:1991ag,Mertens:2020hbs,Betzios:2020nry}.

The minisuperspace WDW equation for the gravitational SG model reads
\be\label{minisuperspaceSG}
\left(- \frac{\partial^2}{\partial \phi_0^2} - \frac{\partial^2}{\partial X_0^2} + \mu e^{2 \phi_0} + t e^{\xi_n \phi_0} \cos \frac{n}{R} X_0 \right) \Psi(\phi_0, X_0) = 0 \, .
\ee
If we also impose periodicity $X_0 \sim X_0 + 2 \pi R$ and expand 
\be
\Psi(\phi_0, X_0) = \sum_{m=-\infty}^\infty  e^{i m X_0/R} \psi_m(\phi_0) \, ,
\ee
we can also recast it into a difference-differential equation 
\be\label{minisuperspaceSGperiodic}
-\psi_m''(\phi_0) + \left( \mu e^{2 \phi_0} + \frac{m^2}{R^2} \right) \psi_m(\phi_0) + t e^{\xi_n \phi_0} \left(\psi_{m+n}(\phi_0) + \psi_{m-n}(\phi_0) \right) = 0  \, .
\ee
In the undeformed $c=1$ case where $t=0$, the momentum $q$ conjugate to $X_0$ is a good quantum number (in the compact case $q=m/R \, , \, m  \in \mathbb{Z}$), leading to solutions that are Bessel functions, which in the case $\mu > 0$ read
\be\label{minisuperspaceBesselK}
\left(- \frac{\partial^2}{\partial \phi_0^2} + \mu e^{2 \phi_0} + q^2 \right) \Psi(\phi_0, q) = 0 \, , \qquad \Psi(\phi_0, q) = A_q K_q(\sqrt{\mu} \ell) + B_q I_q(\sqrt{\mu} \ell) \, , \quad \ell = e^{\phi_0} \, .
\ee
The length $\ell = \oint \sqrt{\hat{\gamma}} e^{\phi} = e^{\phi_0}$ measures the size of the boundary (we assume it has an $S^1$ topology) on which we define the WDW wavefunction. One of these solutions ($K_q(\sqrt{\mu} \ell)$) blows up for small $\ell$ while the other ($I_q(\sqrt{\mu} \ell)$) for large $\ell$\footnote{Usually in Liouville theory coupled to $c \leq 1$ matter one chooses the $K_q(\sqrt{\mu} \ell)$ wavefunction to describe microscopic local operators that create punctures on the two dimensional surface.}.
These wavefunctions describe $AdS_2$ geometries (with appropriate operator insertions that carry the momentum $q$). These two types of wavefunctions also appear in the large-N limit of the undeformed matrix model dual, see section~\ref{GaussianLoop} and~\ref{noncompactloop}. 

On the other hand when $\mu < 0$, the solutions become oscillatory
\be\label{minisuperspaceBesselJ}
\Psi(\ell = e^{\phi_0}, q) = A_q J_q(\sqrt{|\mu|} \ell) + B_q Y_q(\sqrt{|\mu|} \ell) \, ,
\ee
and are better suited to describe cosmological type of geometries
consistent with the fact that the $dS_2$ saddle of the quantum effective action exists for  $\mu < 0$. More details on various types of wavefunctions and their physical applications can be found in~\cite{CarneirodaCunha:2003mxy}. In~\cite{Betzios:2020nry} the all-genus/non perturbative version of these wavefunctions for the undeformed $c=1$ model (when $X$ is non-compact) is analyzed in detail and their physical properties are contrasted with respect to the genus zero results\footnote{For example even for $\mu > 0$ the exact non-perturbative wavefunction exhibits oscillations that are non existent at any finite order of the genus expansion.}. Different asymptotic expansions of the exact wavefunction, lead to the various genus zero wavefunctions. 

Currently it seems impossible to analytically solve~\eqref{minisuperspaceSG} in full generality. Further progress can be achieved by defining
\be
u = e^{n \phi_0 /R} \cos \frac{n}{R} X_0 \, , \quad v = e^{n \phi_0/R} \sin \frac{n}{R} X_0 \, , \qquad u^2 + v^2 = e^{2 n \phi_0/R} \, ,
\ee
so that the WDW equation~\eqref{minisuperspaceSG} becomes ($\xi_n = 2 - n/R$)
\be
\left[- \frac{n^2}{R^2} \left(\frac{\partial^2}{\partial u^2} + \frac{\partial^2}{\partial v^2}  \right) + \mu (u^2 + v^2)^{(2 R - n)/(2 n)} + t u (u^2 + v^2)^{(R - n)/n} \right] \Psi(u,v) = 0 \, .
\ee
This equation becomes separable, as long as the powers in the exponents are either $0, 1$. The simplest such option corresponds to the case analyzed in section~\ref{ComplexLiouville}, that is $\xi_n = n/R $, or in other words $R = n$, with $\mu = 0$. In such a case we can expand in momenta conjugate to $v$ i.e. $\Psi(u,v) = e^{i k v} \psi_k(u)$ to find
\be
- \psi_k''(u) + (t u + k^2) \psi_k(u) = 0 \, , \quad \psi_k(u)  =  A_k \, \text{Ai}\left(\frac{t u + k^2}{t^{2/3}} \right) + B_k \, \text{Bi}\left(\frac{t u + k^2}{t^{2/3}} \right)  \, . 
\ee
The solutions are Airy functions. Their behavior(s) closely resemble the one(s) found in the saddle point solutions of section~\ref{ComplexLiouville}. In particular they exhibit both oscillatory as well as exponentially increasing/decaying behavior depending on the value of their argument $(t u + k^2)/t^{2/3}$. For example (assuming $t>0$ for definiteness) near the (AdS/even) critical points $X_0 = 2 m \pi $, $\Re u > 0$, so that the wavefunction is either exponentially increasing ($ \text{Bi}$-branch) or decays to zero ($ \text{Ai}$-branch). Asymptotically, they are very similar to the Bessel-I and Bessel-K wavefunctions of~\eqref{minisuperspaceBesselK} and describe a geometry that approaches $EAdS$. On the other hand near the odd critical points, when $X_0 = (2 m +1) \pi \, , \, $ $\Re u < 0$, the wavefunction becomes oscillatory for $ t u + k^2 < 0$ describing a Lorentzian universe, as the wavefunctions in eqn.~\eqref{minisuperspaceBesselJ}. Finally, there is a lower bound on $u = e^{\phi_0} \cos X_0 > - k^2/t$ for a Euclidean geometry, translating to the minimum size of the wormhole (correlated with the value of $X_0$). In general this leads to the oscillatory wormhole solutions, shown in fig.~\ref{fig:bouncing}. It is hard to progress much beyond these qualitative remarks though, since the variables $u, v$ correlate $X_0, \phi_0$ in a non-trivial manner.

This concludes our minisuperspace analysis of the WDW equation from the gravitational SG model action. In section~\ref{LoopMacro} we discuss how the WDW wavefunctions can be obtained from the (large-N) spectral curve of the dual matrix model and in section~\ref{LoopMicro} we discuss and contrast them with certain exact all orders results (derived once more using the matrix model).

\section{The dual Matrix Models}\label{MMduals}

\subsection{$c=1$ Matrix Quantum Mechanics in the chiral formalism}\label{sec:Chiral}

The $c=1$ Liouville theory is known to have a dual description in terms of Matrix Quantum Mechanics (MQM) in an inverted oscillator potential~\cite{Klebanov:1991qa,Boulatov:1991xz,Ginsparg:1993is,Nakayama:2004vk,Alexandrov:2003ut,Martinec:2004td}. The singlet sector is described by a gauged MQM model defined by
\begin{equation}\label{gaugedMQM}
S_{MQM}=\int dt \, \tr \left(\frac{1}{2}\left(D_{t}M\right)^{2} + \half  M^2  \right),
\end{equation}
where the covariant derivative with respect to the gauge group is $D_{t}M={\partial}_{t}M-i\left[A,M\right]\,$, and where we have simplified the action using the double scaling limit of MQM that focuses in the unstable inverted harmonic oscillator potential~\cite{Klebanov:1991qa,Boulatov:1991xz,Ginsparg:1993is,Nakayama:2004vk,Alexandrov:2003ut,Martinec:2004td}.

It turns out that one can simplify the description of MQM and its vertex operator deformations by passing to the so-called chiral variables introduced in~\cite{Alexandrov:2002fh}. We define
\be\label{chiral}
\hat{X}_\pm = \frac{\hat{M} \pm \hat{P}}{\sqrt{2}} \, ,
\ee
having commutation relations $[ (\hat{X}_+)_{i j}, \,  (\hat{X}_-)_{k l} ]  = - i \delta_{i l} \delta_{j k}  $.
The Hamiltonian of the model is then
\be\label{chiralH}
\hat{H}= \half \tr \le(\hat{P}^2- \hat{M}^2 \ri) = - \half \tr \le(\hat{X}_+ \hat{X}_- + \hat{X}_- \hat{X}_+ \ri) \, ,
\ee
and the action can be written as
\be\label{chiralaction}
S_{MQM}^{ch}= \int dt \,  \tr \le(i X_+ D_t X_- \ri) - \half \tr X_+ X_- \, .
\ee
At finite temperature (Euclidean) we rotate $t = - i \tau$ and compactify  $\tau \sim \tau +  2 \pi R$, making the matrices $X_\pm(\tau)$ periodic (up to a unitary twist that we need to integrate over in the path integral).

Adding now the vertex operator deformations in order to describe the generalised gravitating Sine-Gordon model amounts to deforming the action by
\be
S_{def} = \int d \tau \sum_{n>0} \left( t_n \tr X_+^{|n|/R} + t_{-n} \tr X_-^{|n|/R} \right) \, .
\ee
It is then possible to use results of integrability to study the partition function of the deformed model (which is a $\tau$-function of the Toda hierarchy), see~\cite{Kostov:2002tk,Alexandrov:2003ut} for reviews and~\cite{Alexandrov:2023fvb} for recent non-perturbative results. At the semiclassical level (dispersionless limit of the hierarchy), the profile of the fermi sea is given by a curve
\be\label{chiralfermisea}
x_+ x_- = \mu + \sum_{n>0} \left( n t_{\pm n} x_\pm^{n/R} \, + \, n v_{\pm n} x_\pm^{-n/R} \right) \, ,
\ee
where the coefficients $v_{\pm n}$ can be determined solving the Lax-equations (zero curvature conditions) of the Toda hierarchy. For the undeformed model this gives rise to hyperbolae $x_+ x_- = \mu$, describing the semiclassical Tachyon scattering.

While this formalism is interesting in its own right and can be applied also in the case of (target space) Lorentzian signature when $X_\pm$ are non-compact, to compute scattering amplitudes (even in the non-singlet sector~\cite{Kostov:2006dy}), it has not lead to an easy computation of more complicated observables of the gravitational SG model (such as insertions of loop operators the create macroscopic boundaries) and neither clarifies an underlying simplicity and finiteness present when the $c=1$ boson $X(z, \bar{z})$ is compact. For these reasons, we now turn to a Normal Matrix Model (NMM) description of the compact SG model, which will define for us its non-perturbative completion and will be shown to connect more naturally to cosmological/de-Sitter type of physics.

\subsection{The Normal Matrix Model and its large-N limits}\label{sec:Normal}

In this section we shall propose another, manifestly finite, formulation of the gravitational (compact) Sine-Gordon/Matrix model duality that involves a \emph{Normal Matrix Model} (NMM)\footnote{The reader can consult appendix~\ref{NormalMatrix} for various technical details on NMMs.}. Apart from offering various conceptual advantages that we shall soon describe, the NMM is also very effective as a tool for computing various physical quantities (see~\cite{Alexandrov:2003qk} and~\cite{Mukherjee:2006hz,Mukherjee:2005aq}, for an analysis of the NMM in the present context). 

In this form of the duality, first described in~\cite{Alexandrov:2003qk}, the generalised SG model and its dual chiral MQM description, is related to the following normal matrix model\footnote{In our work we use slightly different conventions and identifications from the ones in~\cite{Alexandrov:2003qk,Mukherjee:2006hz,Mukherjee:2005aq}, since they clarify better the appropriate ``macroscopic'' and ``microscopic'' large-N limits of the model. With these identifications we are also able to match both the expectation value of loop operators at the disk topology level with Liouville as well as the closed surface perturbative genus expansion, for the undeformed model.}
\be\label{NMM}
{\cal Z}^{(N)}_{NMM}
= \int D Z DZ^\dagger  e^{{ \tr}\left( - \frac{1}{\lambda}
(ZZ^\dag)^R + \left(R\n + \frac{R-1}{2}\right){\rm\,log}\,ZZ^\dag + 
\sum_{n=1}^{\infty}(t_n^{NMM} Z^n + \overline{t}_n^{NMM} {Z^\dag}^n)\right)} \, .
\ee
In this expression $Z$, $Z^\dag$ are Normal $N\times N$ matrices that satisfy
\be
[Z, \, Z^\dag] = 0 \, .
\ee 
Since the matrix $Z$ commutes with its adjoint, both of them can be
simultaneously diagonalised, so that the NMM reduces to an eigenvalue model, the later occupying a finite region of the two dimensional complex plane. The parameters $R,\nu$ correspond to the compactification radius of $X \sim X + 2 \pi R$ and the cosmological constant $\mu$ via the relation $\nu = - i \mu$.
The parameters $t_n^{NMM}$,
$\overline{t}_k^{NMM}$ are (in general complex) couplings to the gauge-invariant operators $\tr
Z^n$, $\tr {Z^\dag}^n$. These operators are identified with the vertex/Tachyon operators $\mathcal{T}^-_{n/R}, \mathcal{T}^-_{-n/R}$~\eqref{Liouvilleoperators} of Liouville theory and hence eqn.~\eqref{NMM} provides a microscopic description of the generalised gravitational SG model, incorporating arbitrary perturbations. Finally the parameter $\lambda$ plays the role of a 't Hooft coupling, and in a general circumstance it can be absorbed and set to one by a rescaling  of the matrices $Z, Z^\dagger$ and the couplings $t_n^{NMM}$, $\overline{t}_n^{NMM}$. Nevertheless we chose to keep it explicit, since it will elucidate the nature of the different large-N limits one can consider and make clear that in the undeformed case where all $t_n^{NMM} = \bar t_n^{NMM} = 0$, the only scale present is provided by $\lambda = \mu$, with $\mu$ corresponding to the Liouville cosmological constant coupling, see section~\ref{MacrolargeNlimit} and eqns.~\eqref{densityfiniteR},~\eqref{radiusfiniteR}
in the Appendix. More generally the various terms in the action of the NMM scale as $ Z , Z^\dagger \sim \mu^{1/2 R}$, so that $t_n^{NMM} \sim \mu^{-n/2R} \sim t_n / \mu $ in this parametrization. 

In~\cite{Alexandrov:2003qk}, two slightly different versions of the normal matrix model were investigated, and we shall adopt the first version which states that the exact (closed surface) partition function of the gravitating Sine-Gordon model is described by the normal matrix model in eqn.~\eqref{NMM}, after performing a certain version of the large-N limit and a matching of parameters on the two sides of the duality
\be
{\cal Z}_{SG}(\mu,R  ;  t_n,\overline{t}_n) 
= \lim_{N \rightarrow \infty} \log {\cal Z}^{(N)}_{NMM}\left(\lambda = \mu = i \nu ,  R \,;\, t_n^{NMM} = - i \frac{t_n}{\mu} ,  \overline{t}_n^{NMM} = - i \frac{\bar{t}_n}{\mu}\right) 
\ee
We observe a form of Wick rotation in the relation of parameters between the two sides, owed to the fact that the Lax operators and the resulting string equations and Toda structure of the normal and the chiral matrix models are exactly the same up to these factors of $i$~\cite{Alexandrov:2003qk}\footnote{In fact, since the NMM is perfectly well defined even without this analytic continuation, we shall also briefly discuss the physics when the NMM parameters are real.}. 
 Quite remarkably, it was subsequently proven by a brute force explicit computation in~\cite{Mukherjee:2006hz}, that \emph{one can actually
consider $N$ to have a finite value} and yet obtain the correct exact result for certain correlators of the model as non-trivial functions of the couplings $\mu, \, R, \, t$. The precise finite value of $N$ is determined by the specific correlator/observable one wishes to compute and in particular by the total (discrete) momentum $P_{d}$ that flows within the correlator (i.e. $\tr Z^n$ carries discrete momentum $P_d = n$). A simple intuitive explanation of this remarkable fact, is that when computing a given correlator of total momentum $P_d$, and after dividing by the partition function to properly normalize it, infinitely many terms in the numerator and the denominator are exactly the same and cancel out in the ratio formed. The remaining (finite) terms which do contribute to the correlator of interest, are
the same as those found when considering a finite value of $N$, as long as $N \geq P_d$. This means that the observables that do actually require taking a large-N limit is the partition function itself and expectation values that involve an infinite number of local single trace operators (such as loop operators, that we discuss in sections~\ref{LoopMacro} and~\ref{LoopMicro}).
A more mathematically refined explanation involves the relation between the NMM and the triangulation of the moduli space of Riemann surfaces, and is briefly alluded to in the discussion section.

It is important to emphasize though, that even for the observables that do require one to take a large-N limit, the latter is neither a 't Hooft type of limit, nor the usual double scaling limit~\cite{Ginsparg:1993is,Moore:1991sf}  employed in studies of two dimensional models of quantum gravity, since the NMM is already expressed in terms of the physical parameters of the gravitating SG model. To clarify the nature of the present form of the limit, in appendix~\ref{OrthoNormalMatrix}, we introduce a family of orthogonal polynomials in the complex plane appropriate for the class of normal matrix models studied in this work. We then show that the large-N undeformed partition function of the NMM (all $t_n^{NMM} = 0$), corresponds to a certain non-perturbative completion of that of the undeformed $c=1$ model. This involves taking a direct large-N limit, without any further scaling of physical parameters such as $\mu, R$. This result can be extended to the case of any deformations $t_n$, using integrability techniques following~\cite{Alexandrov:2003qk}.

An alternative and more useful description of this exact ``microscopic'' large-N limit can be provided via the use of the general orthogonal polynomials in the complex plane for the deformed NMM measure. The mathematical theory of such polynomials is developed in~\cite{Totik}. We provide a brief discussion relevant for our examples in appendices~\ref{OrthoNormalMatrix} and~\ref{CDKernel}. 
In particular let us consider the set of orthonormal functions with respect to the matrix model measure and the associated
Christoffel-Darboux (CD) kernel (see appendix~\ref{CDKernel} for definitions and more details\footnote{In the context of two dimensional JT gravity the CD-kernel was studied in~\cite{Johnson:2022wsr} in relation to non-perturbative physics.})
\bea
\psi_n (z)=\frac{1}{\sqrt{h_{n-1}}}\, P_{n-1}(z) e^{-W(z, \bar z)/2} \, , \quad 
\int_{\mathbb{C}} \, \frac{d^2 z}{2 \pi i} \, \psi_n (z){\psi_m (\bar{z})}  = \delta_{mn} \,  , \qquad n = 1, 2, ... \, ,  \nn \\
e^{-W(z, \bar{z})} \, = \, 
e^{- (z\bar{z})^R/\lambda - V(z) - \overline{{V}}(\bar{z}) } \, 
(z\bar{z})^{(R-1)/2 + R \nu } \, , \nn \\
K_N^{(CD)}(z, \bar w) = \sum_{n=1}^{N}
\psi_n (z) {\psi_n (\bar{w})}  \, . \qquad \qquad \qquad \qquad \, \, \, \,
\eea
 One can then define the appropriate ``microscopic'' large-N limit,
that reproduces any non-perturbative observable,
directly using the orthogonal polynomials and the CD-kernel. This ``microscopic'' limit, does not require any scaling of any physical parameters in the NMM, that are to be kept fixed. In particular it is defined by
\be\label{microlimit}
N \rightarrow \infty \, , \qquad \text{with} \quad \lambda = \mu = i \nu , \, \,  R, \,  \, t_n , \, \, \bar{t}_n  \quad \text{fixed} \, . 
\ee
In this limit the orthonormal polynomials $\psi_n(z)$ also remain \emph{unscaled}, the only effect being that their indices can now run up to infinity. 

Let us now contrast this with the other types of large-N limits encountered in the literature. The global spectral statistics such as the global ``macroscopic'' eigenvalue density, are concerned with correlations between eigenvalues that have many other eigenvalues (a finite fraction) in between them\footnote{This is also the limit recovered by solving loop-equations at leading order in the large-N expansion.}. In this
limit the fluctuations of the eigenvalues on a local scale (of a few eigenvalues) are averaged out. Another option is to magnify fluctuations among eigenvalues at a distance of $1/N^\delta$ near particular regions of the spectrum - this defines the so-called ``microscopic double scaled limits'' (see~\cite{Ginsparg:1993is,Eynard:2015aea} for a general discussion and~\cite{Deift2014} for the case of eigenvalues on the complex plane). The value of $\delta$ and the form of the limiting CD-kernel (and the associated scaled orthogonal polynomials) depend on the location in the spectrum: In the bulk of the spectrum the Sine-kernel is typically found, while at the so-called soft edges one typically finds the Airy-kernel~\cite{Tracy:1992rf}. Another common option for cases that exhibit a hard edge ($\delta = 1/2$) is the limiting Bessel kernel~\cite{Tracy:1993xj}. In contrast to these cases, the limit in eqn.~\eqref{microlimit} keeps the original unscaled polynomials intact, while large-N replaces various summations or products appearing in computations of observables by infinite ones (so that the matrix model spectrum now contains an infinite number of discrete eigenvalues). 

In the following section, we analyze the exactly solvable case of the self-dual radius $R=1$. In this case the orthogonal polynomials become of the Laguerre type and we can explicitly compute various observables in closed form and compare and contrast the ``macroscopic'' and ``microscopic'' large-N limits in detail.

\section{Exact solvability at the self-dual radius ($R=1$)}\label{sec:selfdual}

The case where we specialize the compactification radius to $R=1$, deserves a special attention, being the case for which we can provide completely explicit non-perturbative results for various observables. This case corresponds to the self-dual radius (from a target space string theory perspective), and the matrix model simplifies considerably (for example in appendix~\ref{Laguerreexamples} we show how the orthogonal polynomials correspond in this case to the well studied classical Laguerre polynomials). 

Moreover when $R=1$, the normal matrix model is related to the Imbimbo-Mukhi-Kontsevich-Penner or $W_\infty$ matrix model and exhibits various further connections with topological strings and the $1/2$-BPS sector of four-dimensional $\mathcal{N}=4$ Super-Yang-Mills (SYM), see~\cite{Imbimbo:1995np,Ghoshal:1995wm,Mukhi:2003sz} and~\cite{Gopakumar:2022djw,Gopakumar:2024jfq} for some recent works describing the relevant web of dualities. Very similar normal matrix models have also been proposed to model properties of QCD (the spectrum of the Dirac operator) at a finite baryon chemical potential~\cite{Stephanov:1996ki,Osborn:2004rf}.
We therefore expect our analysis to also be relevant for these different physical examples.

\subsection{``Macroscopic'' large-N analysis}\label{MacrolargeNlimit}

Let us now describe the ``macroscopic'' large-N analysis of the normal matrix model for $R=1$, for more details see appendix~\ref{largeNmacro} and the works/reviews~\cite{Wiegmann:2003xf,Wiegmann:2005eh,Sommers:1988zn,Eynard:2015aea}. In this limit the eigenvalues of the NMM form a finite area compact fermi surface on the 2d complex plane (a ``droplet'' ${D}$).

The potential simplifies into
\begin{align}
W(z,\bar{z}) = \frac{1}{\lambda} z \bar{z} + \nu \log (z \bar z)  +  V(z) +  \overline{V}(\bar{z}) \, , 
\end{align}
where we parametrized the Gaussian term in terms of a constant coefficient ('t Hooft coupling) $\lambda$, in order to compare our results with those of the usual 't Hooft limit. The logarithmic term and the terms $ V(z)+  \overline{V}(\bar{z})$, do not contribute to the large-N density, that is computed via (see appendix ~\ref{largeNmacro} for more details)
\begin{align}
  \partial_z \partial_{\bar z} W(z,\bar{z}) = \frac{1}{\lambda}  \, \qquad \rho(z, \bar z) = \frac{1}{\pi }\partial_z \partial_{\bar z} W(z,\bar{z}) = \frac{1}{\pi \lambda} \, .
 \label{eq:laplacianv}
\end{align}
The only non-trivial task is to determine the support of the density itself, in other words the domain/droplet $D$.  Since the density is constant and normalized $\int_{D} \rho = 1$, the support must be a domain/droplet of area $\pi \lambda$. We then parametrize the domain by the shape of its boundary curve $\partial D$, with an equation 
\begin{equation}
\bar z = S(z) \, .
\end{equation}
The analytic function $S(z)$ obeys an involution property
\begin{align}
 \overline S \circ S = I \, ,
 \end{align}
and is called the \emph{Schwarz function} of the curve $\partial D$, see appendix~\ref{Schwarzproperties}.
Then the ``Stieltjes'' transform of the density (or the resolvent)
\begin{equation}
{\omega}(z) = \int_{\mathbb{C}}\frac{ \rho(z',\bar z')}{z-z'}d^2z' 
\, , 
\end{equation}
can be rewritten in terms of the function $S(z)$ using Stokes' theorem \footnote{This can be proven using $\frac{1}{z-z'} d^2z'= \frac{1}{2i} \frac{1}{z-z'} dz' \wedge d\bar{z}' = \frac{1}{2i} d\left(\frac{\bar{z}'}{z-z'} dz'\right)$ .}
\begin{equation}
{\omega}(z) 
=  \frac{1}{2\pi i \lambda} \int_{\partial D}\frac{S(z')}{z-z'}dz' 
 \, ,
\end{equation}
and combining this with the saddle-point equation at large-N
\begin{equation}
\forall z\in D \, ,
\qquad
\frac{1}{\lambda} \bar z + \partial_z V(z) =  \omega(z)\ , 
\end{equation}
leads to
\begin{equation}\label{affineintegralSchwarz}
\forall z \in \partial D \, ,
\qquad
\frac{1}{\lambda} S(z) + \partial_z V(z)
=  \frac{1}{2\pi i \lambda} \int_{\partial D}\frac{S(z')dz' }{z-z'}
 \, .
\end{equation}
This (affine) integral equation must actually be valid even outside $D$ by analyticity.

We shall now solve this equation, together with the analogous equation for the function $\overline{S}$, in simple examples of interest. The more general theory of the Schwarz function in relation to the solution of such problems is discussed in appendix~\ref{Schwarzproperties}.

 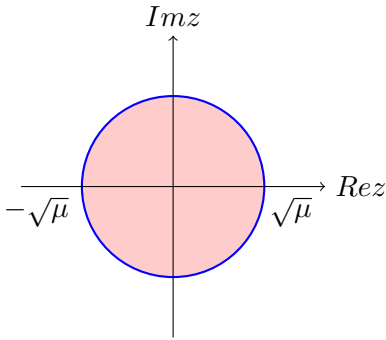
\begin{figure}
\begin{center}
 \begin{tikzpicture}

  \filldraw [fill=red,fill opacity=0.2,thick,draw=blue] 
  (0,0) circle [radius=1.2]; 
  
  \draw [->] (-2,0) -- (2,0) node [right] {$Re z$};
  \draw [->] (0,-2) -- (0,2) node [above] {$Im  z$};

  \node at (1.55,0) [below] {$\sqrt{\mu}$};
  \node at (-1.8,0) [below] {$- \sqrt{\mu}$};

 \end{tikzpicture}
 \caption{The Gaussian NMM eigenvalue distribution in the complex plane. It forms a disk droplet of area $\pi \mu$ (upon identifying the 't Hooft coupling $\lambda \equiv \mu$).}
 \label{fig:Gaussian}
\end{center}
\end{figure}

\subsubsection{Gaussian Normal Matrix model (and its relation to the conifold)}\label{Gaussian}

We first consider the undeformed case $V(z)=\overline V(z)=0$. From the integral equation \eqref{affineintegralSchwarz} and the behaviour of $\omega(z)$ at infinity, we must have $S(z) \underset{z\to\infty}{=} {\lambda}/{z} + O({1}/{z^2})$. Repeating the analysis for $\overline{S}(z)$ one finds the same asymptotic behaviour for $\overline S(z)$. 
One then easily observes that
\begin{align}
 S(z) = \frac{\lambda}{z} \, , \qquad  \overline S(\bar z) = \frac{\lambda}{\bar z} \, ,
\end{align}
provide a solution of the saddle point integral equations. Therefore, in the macroscopic large $N$ limit, the eigenvalues of the undeformed Gaussian normal matrix model fill a disk of radius $\sqrt{\lambda}$, depicted in fig.~\ref{fig:Gaussian}. The boundary of the support of the equilibrium density is defined by $\bar z= S(z) = {\lambda}/{ z}$ (see appendix~\ref{Schwarzproperties}), and corresponds to a circle of radius $\sqrt{\lambda}$. Therefore the spectral curve of the model is given by $z \bar z = \lambda$ and corresponds to a certain projection (a certain cycle) of a more general spectral curve on $\mathbb{C}^2$~\cite{Alexandrov:2003qk}. The dual cycle is described by the non-compact chiral matrix model spectral curve $x_+ x_- = \mu$, see eqn.~\eqref{chiralfermisea}. The relation between the two dual cycles then identifies the 't Hooft coupling $\lambda \equiv \mu$, as discussed in the previous section. For the self-dual radius the unified description/curve on $\mathbb{C}^2$ describes topological strings (the so called ``B-model'') on the conifold, see~\cite{Ghoshal:1995wm} and followups for more details\footnote{It would be interesting to understand what the radius deformation away from $R=1$ means from the topological string perspective.}.

\paragraph{From the Gaussian NMM to the Wigner semicircle -}

One can also relate the two dimensional circular droplet distribution to the one dimensional Wigner semicircle of the Hermitian Gaussian matrix model. To do so, we define a one dimensional projection
of the droplet density ($z = x + i y$)
\be
\rho(x) = \int_{\text{supp.} (y)} dy \, \rho(z, \bar z) = \frac{2}{\pi \lambda} \int^{\sqrt{\lambda - x^2}}_0 d y =  \frac{2 \sqrt{\lambda - x^2}}{\pi \lambda} \, , \qquad \lambda \equiv \mu \, ,
\ee
proving that the one dimensional projection of the constant circular droplet density is indeed the Wigner semicircle. This also means that the expectation value of (single trace) observables that depend only on $x = (z + \bar z)/2$ is the same as those computed using a Gaussian Hermitian matrix model in the macroscopic ('t Hooft) large-N limit.

\subsubsection{Exactly solvable deformations}

Let us now deform the model with the terms $V(z)=   t_1 z + t_2 z^2/2  \, , $ and $\overline V(\bar z)=  \bar t_1 \bar z + \bar{t}_2 \bar{z}^2/2$\footnote{In this section we drop the upperscript NMM in $t_n^{NMM}$ in order not to clutter the notation.}. These deformations do not spoil the exact solvability of the Gaussian NMM. In particular the linear deformations can be reabsorbed in the Gaussian potential as a shift in the coordinates $z, \, \bar z$ i.e. $z \rightarrow z + \lambda \bar{t}_1, \, \bar{z} \rightarrow \bar{z} + \lambda {t}_1 $. This has the mere effect of shifting the origin/location of the large-N droplet
in the complex $z$ plane. We therefore focus only in the non-trivial quadratic deformations that change the shape of the droplet in the rest of this section.

Studying the integral equation, we find asymptotically
\begin{equation}
S(z) \underset{z\to\infty}{=}  - \lambda{t_2} z  + \frac{\lambda}{z} + O\left(\frac{1}{z^2}\right)
\, , \quad
\overline S(\bar z) \underset{\bar{z}\to\infty}{=} -\lambda \bar{t}_2 \bar z   + \frac{\lambda}{\bar z} + O\left(\frac{1}{\bar z^2} \right) \ .
\end{equation}

The inverse functions therefore behave as
\begin{equation}
S^{-1}(z) \underset{z \to\infty}{=} -\frac{1}{\lambda t_2} z  - \frac{\lambda }{z} + O\left(\frac{1}{z^2}\right)
\, , \quad
\overline S^{-1}(\bar z) \underset{\bar z \to\infty}{=} -\frac{1}{\lambda \bar{t}_2} \bar z  - \frac{\lambda }{\bar z} + O\left(\frac{1}{\bar z^2}\right)
\, .
\end{equation}

\begin{figure}
\begin{center}
 \begin{tikzpicture}

  \filldraw [fill=red,fill opacity=0.2,thick,draw=blue,rotate=30] 
  (0,0) circle [x radius=1.5, y radius = 2]; 
  
  \draw [->] (-3.,0) -- (3.,0) node [right] {$Re z$};
  \draw [->] (0,-2.5) -- (0,2.5) node [above] {$Im z$};

  \draw [->, rotate=30] (0,0) -- (2.5,0) node [above right]{$\sqrt{\frac{\mu-\mu^2 |t_2|}{(1+\mu |t_2|)}}$};
  \draw [->, rotate=30] (0,0) -- (0,2.5) node [above left]{$\sqrt{\frac{\mu+\mu^2|t_2|}{(1-\mu|t_2|)}}$};

 \end{tikzpicture}
\end{center}
\caption{An elliptically shaped droplet, for the deformed normal matrix model with $V(z)= N t_2 z^2/2  \, , $ $\overline V(\bar z)= N  \bar{t}_2 \bar{z}^2/2$. We also depict the size of the two axes of the ellipse. One should also perform the replacement $t_2 \rightarrow - i t/\mu$ to connect with the gravitational Sine-Gordon model.}
 \label{fig:Elliptic}
\end{figure}
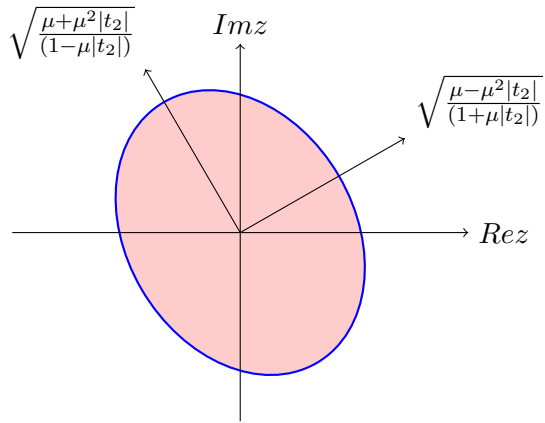

Since $\overline S \circ S = I$ or in other words $\overline S (S(z)) = z $, and moreover $V(z)$ is a polynomial of degree two, this means that for a solution being consistent with the asymptotics, $S(z)$ has to live on a double cover of the complex $z$-plane, see appendix~\ref{Schwarzproperties} for a general discussion. If we label the two sheets by $\pm$, we find the asymptotic behaviours
\be
 S_+(z) \underset{z\to\infty}{\sim}  - \lambda{t_2} z  + \frac{\lambda}{z} + O\left(\frac{1}{z^2}\right)
\, , \quad S_-(z) \underset{ z \to\infty}{=} -\frac{1}{\lambda \bar{t}_2}  z  - \frac{\lambda }{ z} + O\left(\frac{1}{ z^2}\right)
\, .
\ee
Moreover the two branch structure means that $S(z)$ obeys a second order algebraic equation i.e. $S^2 - S \, (S_+ + S_-) \, + \, S_+ S_- \,= \, 0$, with the two expected solutions $S(z) = S_\pm(z)$. Assuming that the coefficients of the second order equation are polynomials (analytic and single valued), we find in consistency with the asymptotic expansions that
\begin{align}
 S_+(z) \, S_-(z) 
 & \,=\, \frac{t_2}{\bar{t}_2}z^2 + \lambda^2 {t_2} - \frac{1}{\bar{t}_2} \,, \nn \\
 S_+(z) +  S_-(z) 
 & \,=\, -\left(\lambda {t_2} + \frac{1}{\lambda \bar{t}_2} \right) z  \, .
\end{align}
From this we can determine the boundary of the support of the equilibrium density using $\bar z = S(z)$. It corresponds to an ellipse defined via the equation
\be
t_2 z^2 + \bar{t}_2 \bar z^2 + \left(\lambda t_2 \bar{t}_2  + \frac{1}{\lambda} \right) z \bar z \,=\, 1 - \lambda^2 t_2\bar{t}_2 \, ,
\ee
see fig.~\ref{fig:Elliptic} for a depiction of the elliptic droplet.
Once more the appropriate identifications are $\lambda \equiv \mu$ in order to match the undeformed model in the corresponding limit, and $t_2 , \bar t_2 \rightarrow  - i t_2/\mu  $
in order to match with the gravitational SG parameters.

When including higher deformations $t_n , \, n > 2$, the droplet starts developing a complicated ``fingering'' (and eventually fractal) pattern and provides the solution to the so called \emph{Hele-Shaw} problem, whereby the deformations $t_n$ capture the moments of the distribution of a two dimensional viscous fluid~\cite{Mineev-Weinstein:2000jhx,Teodorescu:2004qm}. From a 2d quantum gravity/Liouville perspective this is related to the fact that the Tachyon/vertex operators $t_n e^{(2 - n)\phi} e^{i n X}$ are not well defined for $n > 2$ and lead to an instability of the vacuum/background (they heavily backreact in the UV ($\phi \rightarrow  - \infty$) of the geometry). What happens geometrically in these cases is not understood and is a very interesting future problem.

\subsection{Expectation values of loop operators in the ``macroscopic'' limit}\label{LoopMacro}

We can now use our results for the NMM density of eigenvalues and its support, in order to compute the expectation value of loop operators in the macroscopic large-N limit (this keeps only the leading contribution in the genus expansion of these observables). From a two dimensional quantum gravity perspective the expectation value of fixed length loop operators can also be interpreted either as describing the Wheeler-DeWitt (WDW) wavefunction $\Psi(\ell) \equiv \langle W(\ell) \rangle$ of the two dimensional geometries, or as a dual (coarse grained/averaged) partition function $\langle W(\ell) \rangle = \langle \tr e^{- \ell M} \rangle \equiv \langle Z(\beta)\rangle$ with $\beta \equiv \ell$ and $M$ the (Hermitian) matrix, see~\cite{Betzios:2020nry}. In the NMM there is a further complication in that the model is essentially a two matrix model and one can in principle define two types of loop and determinant operators related to $Z$ or $Z^\dagger$, see appendix~\ref{sec:determinantoperators} for more details. We can in principle compute the expectation value of any combination of such operators, and it will generically lead to a complex result. A real result can be given once a ``projection'' is chosen. In particular the usual loop operator insertion in the NMM eigenvalue basis is expressed as $W(\ell) = \sum_i e^{- \ell(z_i + \bar z_i)/2}$~\cite{Okuyama:2006jc}, corresponding to the analogue of the loop operator in the Hermitian matrix model.

\subsubsection{The Gaussian Normal Matrix Model}\label{GaussianLoop}

Let us first discuss the expectation value of a loop operator creating a boundary with length $\ell$ in the undeformed case when all $t_n =0 $ (Gaussian NMM).  It can be computed using the density of eigenvalues and its support (see appendix~\ref{Observables}) 
\be
\langle W(\ell) \rangle = \int_{\text{supp.}} d^2 z \, \rho(z, \bar z) \, e^{-\ell(z+\bar{z})/2} \, ,
\ee
and is expressed in terms of a Bessel function integral
\bea
\langle W(\ell) \rangle = \frac{1}{2\pi \mu} \int_0^{\sqrt{\mu}} d r r \int_0^{2\pi} d \theta e^{- \ell r \cos \theta } = \nn \\
= \frac{1}{\mu} \int_0^{\sqrt{\m}} d r r   I_0(\ell r) =  \frac{1}{\sqrt{\mu} \ell} I_1(\sqrt{\mu} \ell) \, .
\eea
We therefore find that the wavefunction has most of its support at large boundary length $\ell$, where it diverges.

This result corresponds also to the expectation value of a circular $1/2$-BPS Wilson loop in four dimensional $\mathcal{N}=4$ SYM~\cite{Okuyama:2006jc,Pestun:2007rz} and can be written equivalently in terms of the Wigner semicircle density integral
\be\label{I1loop}
\langle W(\ell) \rangle = \frac{1}{\pi \mu} \int_{-\sqrt{\mu}}^{\sqrt{\mu}} d x \sqrt{\mu - x^2} e^{- \ell x} =  \frac{1}{\sqrt{\mu} \ell} I_1(\sqrt{\mu} \ell) \, , 
\ee
that is the projection of the circular droplet density into the one dimensional $x$-axis as we discussed in~\ref{Gaussian}. Similarly to what happens here, in the context of $\mathcal{N}=4$ SYM, this loop operator creates an $AdS_2$ geometry inside $AdS_5 \times S^5$ whose boundary is the loop operator. In the $\mathcal{N}=4$ SYM context $AdS_2$ corresponds to the geometry of the two dimensional (open) string worldsheet with a loop boundary.

\subsubsection{Comparison with the non-compact spectral curve and $c=1$ Liouville}\label{noncompactloop}

Let us now compare and contrast the loop operator/WDW wavefunction between the cases of the compact vs. the non-compact spectral curve.
In the non-compact case one can again project the spectral curve $x_+ x_- = \mu = x^2 - p^2$ on the real $x$-axis that leads to an integral over the complementary region of the $x$-axis~\cite{Betzios:2020nry}
\bea\label{K1loop}
\langle W(\ell) \rangle^{(n.c.)} \, &=& \, \int_{supp.} d x_+ d x_- \rho(x_+, x_-) e^{- \ell (x_+ + x_-)/2}\, = \, \nn \\ 
\, &=& \, \frac{1}{\mu} \int_{\sqrt{\mu}}^\infty d x \sqrt{x^2-\mu} e^{- \ell x} \, = \,  \frac{1}{\sqrt{\mu} \ell} K_1(\sqrt{\mu} \ell) \nn \\
\, &=& \, \frac{1}{\mu} \int_{\sqrt{\mu}}^\infty d r r K_0(\ell r) = \frac{1}{\mu} \int_{\sqrt{\mu}}^\infty d r r  \int_{0}^\infty d t e^{- \ell r \cosh t} \, .
\eea
The first thing to notice is that both the Bessel-K and Bessel-I wavefunctions exactly appear in the minisuperspace/zero mode treatment of the undeformed $c=1$ Liouville model, see eqn.~\ref{minisuperspaceBesselK}. The specific index is related to the presence of the cosmological constant operator from the two dimensional quantum gravity perspective~\cite{Betzios:2020nry}. More general operator insertions would lead to more general indices for these Bessel functions, as seen by inserting higher powers of $x$ in the integral expressions~\eqref{I1loop} and~\eqref{K1loop}.

We then observe that we can solve the constraint of the spectral curve using the auxiliary parametrization $x = \sqrt{\mu} \cosh 2 \pi s$ and $p = \sqrt{\mu} \sinh 2 \pi s$, to recover an alternative integral expression
\be
\langle W(\ell) \rangle^{(n.c.)} = \frac{1}{\sqrt{\mu} \ell} K_1(\sqrt{\mu} \ell) = 2 \pi \int_0^\infty d s e^{- \ell \sqrt{\mu} \cosh{2\pi s}} \sinh^2 (2 \pi s) \, .
\ee
This form makes manifest the fact that $E \equiv x = \sqrt{\mu} \cosh 2 \pi s$ can be interpreted as the energy of the disk boundary Hamiltonian and $\rho(s) = \sinh 2 \pi s = \sinh \cosh^{-1}(E/\sqrt{\mu})$ as a density of states~\cite{Mertens:2020hbs}.

On the other hand, by resolving the constraint in the compact case $x= \sqrt{\mu} \cos 2 \pi s$ and $y = \sqrt{\mu} \sin 2 \pi s$, we recover the analogous expression
\be
\langle W(\ell) \rangle^{(c.)} = \frac{1}{\sqrt{\mu} \ell} I_1(\sqrt{\mu} \ell) = 2 \pi \int_0^{1/2} d s e^{- \ell \sqrt{\mu} \cos{2\pi s}} \sin^2 (2 \pi s) \, ,
\ee
which upon identifying $E = \sqrt{\mu} \cos 2 \pi s \, , \,$ $\rho(s) = \sin 2 \pi s = \sin \cos^{-1} E/\sqrt{\mu} $ leads to a compact support for the density of states, that bears a resemblance to models of Sine-Dilaton gravity~\cite{Blommaert:2024whf}\footnote{With our choice of definition of energy, we have placed the boundary energy spectrum to be symmetric with respect to $E=0$, it is possible at this stage to perform shifts of this origin.}, being well defined only when restricted in a region (half period) of $\sin 2 \pi s$, where the density is positive definite. A similar issue of potential non positivity will also appear later for the non-perturbative density of states in the ``microscopic'' limit. 

As we mentioned previously, insertions of local (puncture) operators, shift the arguments of the Bessel functions. The differences between the compact vs. the non-compact spectral curves and the different Bessel functions (i.e. $I_\nu(\sqrt{\mu} \ell)$ vs. $K_\nu(\sqrt{\mu} \ell)$, see also section~\ref{MinisuperspaceWDW}) that describe the expectation values of Loop and microscopic operators in $c=1$, is analogous to the differences found in the models of Sine vs. Sinh Dilaton gravity~\cite{Blommaert:2025avl,Okuyama:2023byh}.

\subsubsection{Continuation to the cosmological regime }

So far our discussion described the creation of a macroscopic (loop) boundary in the purely Euclidean regime corresponding to the WDW wavefunction of the cosmological constant operator with $\mu > 0$, that corresponds to an $EAdS_2$ geometry with disk topology, see sections~\ref{undeformedLiouville} and~\ref{MinisuperspaceWDW}. Depending on the choice of compact vs. non-compact spectral curve, we found that the wavefunction can have support on large vs. small boundary lengths $\ell$. This discrepancy is similar to what is found in the Hartle-Hawking vs. Tunneling proposals in the Euclidean regime.

If we wish to pass to the cosmological/dS regime of the undeformed model, we have to perform a continuation to negative $\mu < 0$, as discussed in~\ref{undeformedLiouville}\footnote{Notice also that at the level that we are currently working (leading in the genus expansion) sending $\mu \rightarrow - \mu$ is equivalent to rotating $\ell \rightarrow i \ell$, making contact with the ``$-AdS_2$'' proposal analyzed in~\cite{Maldacena:2019cbz,Betzios:2020nry}.}. In the compact spectral curve case, the wavefunction then becomes a Bessel-J function i.e.
\bea
\langle W(\ell) \rangle^{(c.)} =  \frac{1}{\sqrt{|\mu|} \ell} J_1(\sqrt{|\mu|} \ell) \, .
\eea
This is a delta function normalized oscillatory wavefunction with a finite maximum at $\ell = 0$, that corresponds to a well defined WDW wavefunction for the corresponding two dimensional universe\footnote{Further insertions of local operators would change the index of the Bessel function, see for example~\cite{Betzios:2020nry}.}. Due to its normalizability we expect this to properly describe the two dimensional smooth de-Sitter like cosmologies at the semiclassical (minisuperspace) level. 

On the other hand in the case of the non-compact spectral curve, we find the cosmological ($\mu < 0$) wavefunction becoming a Hankel function~\cite{Betzios:2020nry} that is non-normalizable
\bea
\langle W(\ell) \rangle^{(n.c.)} =  \frac{i}{\sqrt{|\mu|} \ell} H_1^{(1)}(\sqrt{|\mu|} \ell) \, ,
\eea
again bearing some similarities with the tunneling proposal~\cite{Betzios:2020nry}. Since this wavefunction blows up at small $\ell$, we expect it to describe cosmologies with bang/crunch singularities. This discussion can be generalised to other wavefunctions, whereby (microscopic) local operators are turned on the two-dimensional geometry, see~\cite{CarneirodaCunha:2003mxy} for an extended analysis.

We therefore conclude that the NMM and its compact spectral curve lead to a better behaved (normalizable in the cosmological regime of interest) genus zero WDW cosmological wavefunction, in contrast with the chiral non compact spectral curve. This is another indicator that the former can serve as a better microscopic model of two dimensional cosmological spacetimes.

\subsection{``Microscopic'' large-N analysis}\label{MicroLargeN}

In this section, we discuss and compare the ``microscopic'' large-N analysis of some exactly solvable examples of the NMM, and contrast them with the previous ``macroscopic'' large-N results (that hold to leading order in the genus expansion). The basic object to determine is the Christoffel-Darboux (CD) kernel, which can be used to compute most of the physical observables (see the discussion in appendix~\ref{Observables} and~\ref{CDKernel}). In particular we use it in section~\ref{LoopMicro},~\ref{dosrealmu} and~\ref{nurealproperties} to compute the exact one point functions of Loop operators and the dual density of states. This will allow us to understand, the extend in which the all orders results for the wavefunctions resemble or differ from the leading asymptotic approximation. We should also mention at this point that the CD-kernels that we find upon the continuation $\nu = - i \mu$ etc. are naturally complex, since the matrix model measure $e^{-W(z,\bar z)}$ is complex. The effect of this is that the expectation values of various observables are also a priori complex. Some further discussion is provided in appendix~\ref{CDkernelimaginary}. A practical cure to this issue regarding the expectation values of local operators is to take the real part of the result\footnote{A similar approach was also implemented in~\cite{Mukherjee:2005aq} and shown to be correct for the expectation values of local operators.} . In section~\ref{nurealproperties} we consider also the case that $\nu$ is real, leading to expectation values that are manifestly real, and discuss how this possibility can be realized from the dual quantum gravity system.

\subsubsection{The undeformed model for arbitrary $R$}

In the case that all the $t_n = 0 = \bar t_n$ the orthogonal polynomials are simply $P_n(z) \sim z^n $, and it is possible to compute exactly the Christoffel-Darboux (CD) kernel, see eqn.~\eqref{CDkernelAxial} and~\eqref{CDWright}. The undeformed CD-kernel is expressed in terms of the generalised Mittag-Leffler function $E_{a,b}(x)$ (for arbitrary $R$ and $\nu = - i \mu$)
\bea\label{CDundeformed1}
  K_\infty( z , \bar w  ) 
=  e^{- \frac{1}{2 \mu} (z\bar{z})^R  - \frac{1}{2 \mu} (w\bar{w})^R}
(|z| |w|)^{(R-1)/2 - i R \mu } E_{1/R , (R+1)/2 R - i \mu }(z \bar w) \, ,
\eea
see appendix~\ref{Formulae} for more details on the properties of the Mittag-Leffler function\footnote{The Mittag-Leffler and the related Wright functions are also used as a probability measures in studies of partitions and stochastic processes.}. For $R=1$, using eqn.~\eqref{R=1Mittag} and the properties of the lower incomplete gamma function, one can explicitly check that
the leading term in its $1/\mu$ asymptotic expansion is
\be
 K_\infty( z , \bar w  ) \underset{\mu \rightarrow \infty}{\simeq}  \, \frac{(|z| |w|)^{- i  \mu }}{
(z\bar{w})^{-i \mu }} \, ,
\ee
so that the genus zero diagonal form of the kernel $K_\infty(z, \bar z)$ (the density of eigenvalues) reduces to a constant, in accordance with the macroscopic large-N analysis, of section~\ref{MacrolargeNlimit}. In general since $\nu = - i \mu$ is complex, the CD kernel is also complex.

\subsubsection{The deformed model at $R=1$ and Laguerre polynomials}

The other exactly solvable example in the microscopic limit, that we are most interested in, corresponds to the self dual radius case $R=1$ when only the first ($t_1, \bar t_1$) deformations are turned on. As shown in appendix~\ref{Laguerreexamples}
the relevant orthogonal polynomials in this case are Laguerre polynomials. A detailed technical analysis is provided in appendix~\ref{Laguerreexamples}. When $t_1^{NMM} = \bar{t}_1^{NMM} = - i t_1/\mu$ (since this is the case we can compare with the results of section~\ref{SolutionsEOM}), we find that eqn.~\eqref{largeNkernelt1} is expressed as
\bea\label{largeNkernelt1main}
K_\infty^{(- i \mu)}(z,\bar{w}) &=& 
\frac{\left(1 - \frac{t^2_1}{\mu^2}\right)^{1 - i \mu/2} }{
\mu^{- i \mu + 1}\pi  \,
_1F_1 \left(1 - i \mu , 1 ,  -  t^2_1/\mu    \right)
}\, \frac{|z \bar w|^{- i\mu}}{(z \bar w)^{- i \mu/2}} 
 \nn \\
&\times& \exp\left[
- \frac{1}{2 \mu}  (|z|^2 + |w|^2)
- i  \frac{t_1}{2\mu} (\bar z - z + w
- \bar w)
\right]
I_{- i \mu}\left(2 \sqrt{1- \frac{t^2_1}{\mu^2}} \, \sqrt{{z \bar w}}\right) \, . \nn \\
\eea
A good check of this expression, is to take the limit $\mu \rightarrow \infty$ on its diagonal part. Using the properties of the modified Bessel functions eqn.~\ref{BesselAsymptotics1}, one finds that it reduces to a constant as expected\footnote{This can also be used to fix its overall normalization such that it conforms with the genus zero result in this limit.}. In the rest we shall use the kernel~\eqref{largeNkernelt1main} in order to compute loop observables (exactly) and then to compare them with their leading (minisuperspace) expressions.

\subsection{The WDW wavefunction/loop operator in the ``microscopic'' large-N limit}\label{LoopMicro}

Using the eigenvalue density, that is equivalent to the diagonal part of the Christoffel-Darboux (CD) kernel, see eqn.~\eqref{eigenvaluechristoffel},  one can compute the expectation value of loop operators, that create a macroscopic boundary of length $\ell$ on the two dimensional surfaces. As mentioned before, the corresponding operator insertion in the normal matrix eigenvalue basis is\footnote{One can define more refined loop operators that distinguish $Z$ with $Z^\dagger$, using the determinant operators in appendix~\ref{sec:determinantoperators}. These are useful if one wants to introduce dependence both on the zero mode $\ell = e^{\phi_0}$ as well as on the zero mode $X_0$, since $\tr Z^n \leftrightarrow e^{(2-n) \phi_0 + i n X_0}$ according to the dictionary.} $W(\ell) = \sum_i e^{- \ell(z_i + \bar z_i)/2}$~\cite{Okuyama:2006jc}. From a 2d quantum gravity perspective its expectation value corresponds the WDW wavefunction of the 2d geometry with boundary length $\ell$ i.e.
$\Psi(\ell) = \langle W(\ell) \rangle$~\cite{Betzios:2020nry}. As mentioned previously the results for the expectation values of loop operators will turn out to be complex when $\nu = - i \mu$ and at the end of the computation one has to take their real part, if one additionally wishes to interpret them as an averaged partition function of a (loop) boundary dual i.e. $\langle W(\ell)\rangle = \langle Z(\ell)\rangle \, , \ell \equiv \beta$.

For the undeformed model we can obtain a closed form integral expression for arbitrary radius $R$, using eqn.~\eqref{CDundeformed1} involving
the Mittag-Leffler function
\be
\langle W(\ell , \mu ; R) \rangle =  \int_0^{\infty} d r  r^{R- 2 i R \mu}  e^{-  r^{2R}/\mu } I_0( \ell r )  E_{1/R, (R+1)/R -i \mu}\left( r^2 \right) \, .
\ee
We are not aware a closed form description of this integral in terms of special functions.

Here we are mostly interested in the $R=1$ case when only $t_1^{NMM} = - i t/\mu$ is turned on (which we also studied in sections~\ref{ComplexLiouville} and~\ref{MinisuperspaceWDW} from a saddle point and WDW perspective). Using the large-N diagonal form of the CD-kernel~\eqref{largeNkernelt1main} (and by rescaling $|z| = r \rightarrow \sqrt{\mu} r$) we find the expectation value\footnote{In the rest we drop any irrelevant overall normalization factor, in order not to clutter the notation. It can be easily reinstated using our exact formulae.}
\be\label{Loopt1main}
\langle W(\ell , \mu, t_1^{NMM} = - i t_1/\mu = - i \zeta/\sqrt{\mu}) \rangle =  \int_0^{\infty} d r  r^{1-i \mu} e^{-  r^2 } I_0(\sqrt{\mu} \ell r )  I_{-i \mu}\left(2 \sqrt{1- \zeta^2 } \, r \right) \, ,
\ee
where we expressed the result in term of the dimensionless variable $\zeta = t_1/\sqrt{\mu} $, according to the KPZ-DDK scalling analysis in eqn.~\eqref{KPZDDK}. In general the result is complex, due to the complex measure that involves $\nu = -i\mu$\footnote{In the next section, we discuss the case where we take $\nu$ to be real, that leads to real results both for the loop operator and the dual density of states.}. The asymptotic $1/\mu$ expansion is very similar to the undeformed case (this is the limit of large $\mu$ with respect to $t_1$ or equivalently small $\zeta$), giving rise to the disk $EAdS_2$ result of~\eqref{I1loop} to leading order (with corrections that can be computed order by order, by expanding the second Bessel function both at large $\mu$ and small $\zeta$, using higher order terms in eqn.~\eqref{BesselAsymptotics1}).

\begin{figure}[t]
\begin{center}
\includegraphics[width=0.4\textwidth]{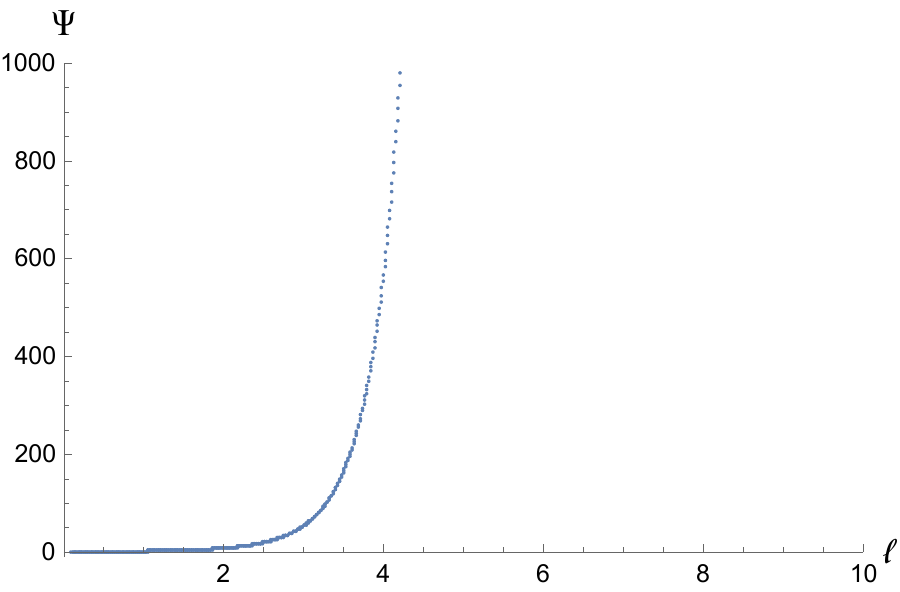}
\hspace{0.2em}
\includegraphics[width=0.4\textwidth]{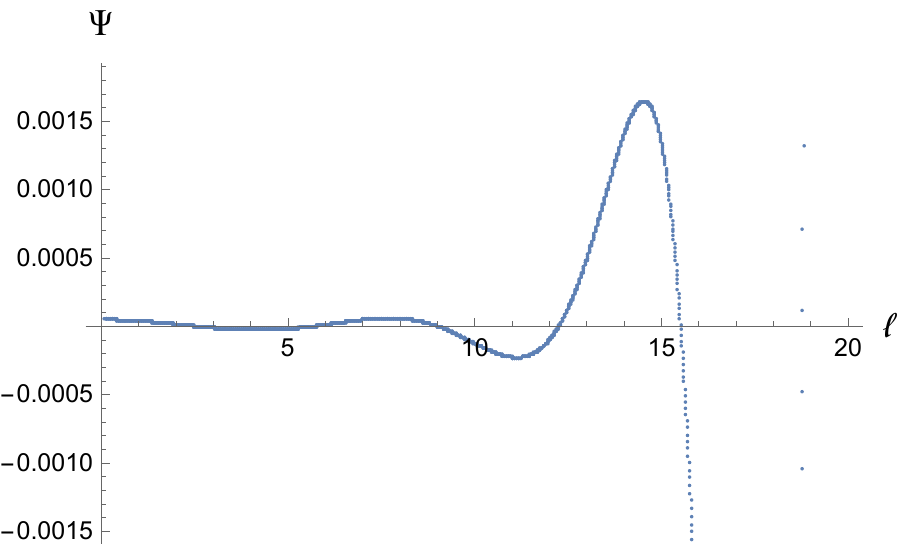}
\end{center}
\caption{Two typical behaviours of the real part of the (non-perturbative) WDW wavefunction for $\zeta < 1$ (left) and $\zeta > 1$ (right) for real $\mu = 0.1$. We observe that in one phase the wavefunction resembles the one for $AdS_2$ (Bessel-I), while in the other it is oscillatory (and finite at $\ell = 0$) similarly to the Bessel-J case corresponding to $dS_2$. It is difficult to study the large $\ell$ behavior numerically, and we do so using analytic techniques in eqn.~\eqref{largeloopexpansion}. The resulting power law decay at large $\ell$ indicates that the non-perturbative wavefunction is normalizable for $\zeta > 1$.}
\label{fig:wavefunctionnumerics}
\end{figure}

We are mostly interested in the opposite limit of small $\mu$ and large $t$ or large $\zeta$, where we expect the wavefunction to acquire cosmological characteristics. We observe that the expression~\eqref{Loopt1main}, changes behaviour at $\zeta = 1$, above which the Bessel-I is replaced by a Bessel-J function i.e.
\be\label{Loopt1mainJ}
\langle W(\ell , \mu,  \zeta > 1) \rangle =  \int_0^{\infty} d r  r^{1-i \mu} e^{-  r^2 } I_0(\sqrt{\mu} \ell r )  J_{-i \mu}\left(2 \sqrt{\zeta^2-1} \, r \right) \, ,
\ee
indicating a form of phase transition in the behaviour of this observable. This is evident from the numerical plots in fig.~\ref{fig:wavefunctionnumerics}. In particular in one phase we describe a wavefunction that is exponentially growing with $\ell$, while in the other phase it is oscillatory and bounded (so that the role of changing the sign of $\mu$ in the undeformed model, is now played by changing the value of the dimensionless parameter $\zeta$).

It is actually possible to re-express these integrals analytically, using eqns.~\eqref{I1} and~\eqref{I2} and then~\eqref{HypertoJacobi}, as a series of orthogonal Jacobi polynomials (again up to overall normalization)
\bea
\langle W(\ell , \mu, \zeta < 1) \rangle &=& \frac{(1-\zeta^2)^{-i \mu/2}}{2 \Gamma(1-i\mu)} \sum_{k=0}^\infty \frac{\Gamma(k+1-i\mu)}{\Gamma(k+1)}  \frac{ \left(\mu \ell^2\right)^k}{4^k k!}  \, _2 F_1\left(-k,  - k,  1-i\mu ; \frac{4(1-\zeta^2) }{\mu \ell^2} \right)  \nn \\
&=& \frac{(1-\zeta^2)^{-i \mu/2}}{2} \sum_{k=0}^\infty \frac{\left(\frac{\mu \ell^2}{4}- (1-\zeta^2) \right)^k}{\Gamma(k+1)}   \, P_k^{(-i \mu , 0)} \left(\frac{\mu \ell^2 + 4 (1-\zeta^2)}{\mu \ell^2 - 4 (1-\zeta^2)} \right) \, , \nn \\
\eea
and similarly for $\zeta > 1$. This expression, shows that the WDW wavefunction can be written as an infinite convergent series over normalizable wavefunctions (the Jacobi polynomials)\footnote{This series can also be performed in terms of the generalized hypergeometric functions $F_4$ or $\Psi_2$~\cite{Prudnikov2,Prudnikov3}, but this does not seem to offer any further physical insights.}, which can be thought of as some kind of ``normalizable microstates''. It would be interesting to uncover their geometric interpretation if any.

Simpler expressions can be found in various limiting cases.
For example in the strict $\mu \rightarrow 0$ limit\footnote{Notice that $\sqrt{\mu} \ell$ can still be non-zero for large $\ell$.}, the Jacobi polynomials reduce to Legendre polynomials to zeroth order in the small $\mu$ expansion and the summation can be performed using eqn.~\eqref{LLegendresum1} 
\be
\langle W(\ell , \mu, \zeta > 1) \rangle \underset{\mu \rightarrow 0}{=} \half e^{\mu \ell^2 + 4(1-\zeta^2)} J_0\left(\sqrt{(\zeta^2 - 1) \mu} \ell \right) \, ,
\ee
the Bessel-$J_0$ being replaced by the Bessel-$I_0$ when $\zeta < 1$. This limit again exhibits the afforementioned difference between the $AdS_2$ and the normalizable cosmological ($dS_2$) wavefunction. 

Another interesting case to study, is the limit near the phase transition i.e. $\zeta \rightarrow 1^\pm$. In this case we can use the expansion~\eqref{BesselAsymptotics1} to find at leading order
\be\label{Loopt1mainJ3}
\langle W(\ell , \mu,  \zeta ) \rangle \underset{\zeta  \rightarrow 1^\pm}{\simeq} \frac{1}{\Gamma(1-i \mu)} \int_0^{\infty} d r  r^{1-2 i \mu} e^{-  r^2 } I_0(\sqrt{\mu} \ell r )  = \half L_{i \mu -1}\left(\frac{\mu \ell^2}{4} \right) \, .
\ee
in terms of Laguerre functions (or equivalently the confluent hypergeometric functions $_1F_1$). The subleading terms start to differ on whether we approach the limit from above or below.

The final limiting case that we shall study is the large loop length limit, $\sqrt{\mu} \ell \rightarrow \infty$, which is the hardest limit to access numerically. Assuming that we are also in the regime of large $\zeta$, we can simplify the integral expression into
\be\label{Loopt1mainJ2}
\langle W(\ell , \mu,  \zeta > 1) \rangle \underset{\mu \ell^2  \rightarrow \infty}{=}  \frac{1}{(\sqrt{\mu} \ell)^{1- i \mu}}  \int_0^{\infty} d x  x^{1-i \mu}  I_0(x)  J_{-i \mu}\left(\frac{2 \zeta}{\sqrt{\mu} \ell} \, x \right) \, ,
\ee
and perform the resulting integral using eqn.~\eqref{I5} to yield
\be\label{largeloopexpansion}
\langle W(\ell , \mu, \zeta > 1) \rangle \underset{\mu \ell^2  \rightarrow \infty}{\simeq}   \left( \frac{2 \zeta}{\sqrt{\mu} \ell} \right)^{- i \mu } \left(\sqrt{\mu} \ell - 2 \zeta \right)^{i \mu - 1} \, , \qquad \sqrt{\mu} \ell \geq 2 \zeta \, ,
\ee
where we again dropped the overall normalization. This result shows that the wavefunction decays with a power law having persistent oscillations with a period governed by the value of $\mu$. This behaviour is reminiscent of the behaviour of the non-perturbative WDW wavefunction in the undeformed uncompactified $c=1$ model~\cite{Betzios:2020nry} at large loop lengths.

We conclude by summarizing that both our analytic and numerical results show that similarly to the case of $\mu >0$ vs. $\mu<0$ in the undeformed model, we obtain a divergent vs a normalizable oscillatory wavefunction depending on whether $\zeta < 1$ or $\zeta > 1$ respectively\footnote{In contrast though with the $\mu < 0$ case, the gravitational SG model is well defined with a potential that is bounded from below for all values of $\zeta$.}.  Once again, this is the (non-perturbative) WDW wavefunction counterpart of the transition between the dominant $AdS_2$ and $dS_2$ saddles we found in section~\ref{sec:sols1}. It would be interesting to study the expectation value of loop operators in various other regimes, additionally separating the $Z$ and $Z^\dagger$ contributions, so that we can freely tune the zero modes of both $\phi_0$ and $X_0$. According to the minisuperspace results of section~\ref{MinisuperspaceWDW}, we expect this analysis to capture the physics of the more general interpolating ``wineglass wormholes''. 

\subsection{The non-perturbative density of states for $\nu = - i \mu$}
\label{dosrealmu}

\begin{figure}[t]
\begin{center}
\includegraphics[width=0.4\textwidth]{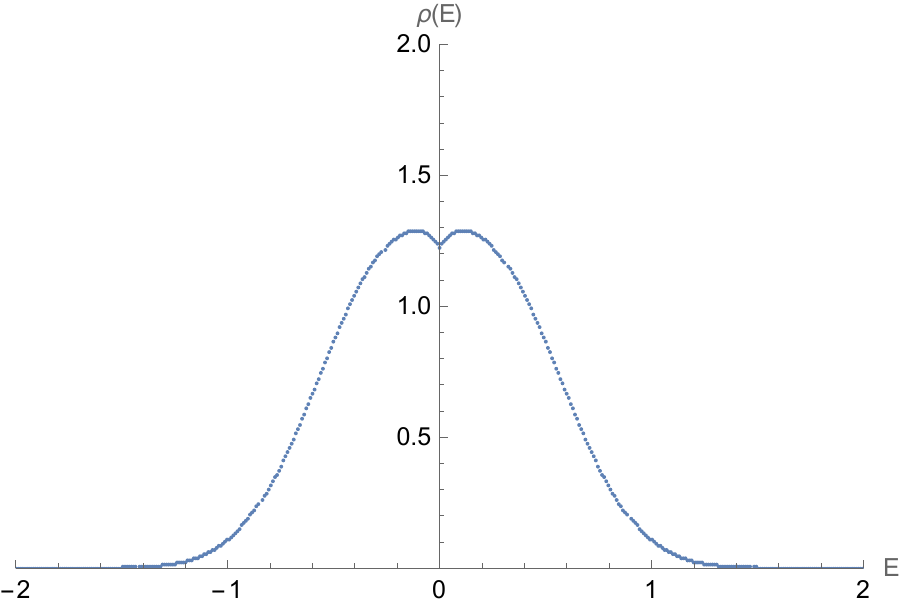}
\hspace{0.2em}
\includegraphics[width=0.4\textwidth]{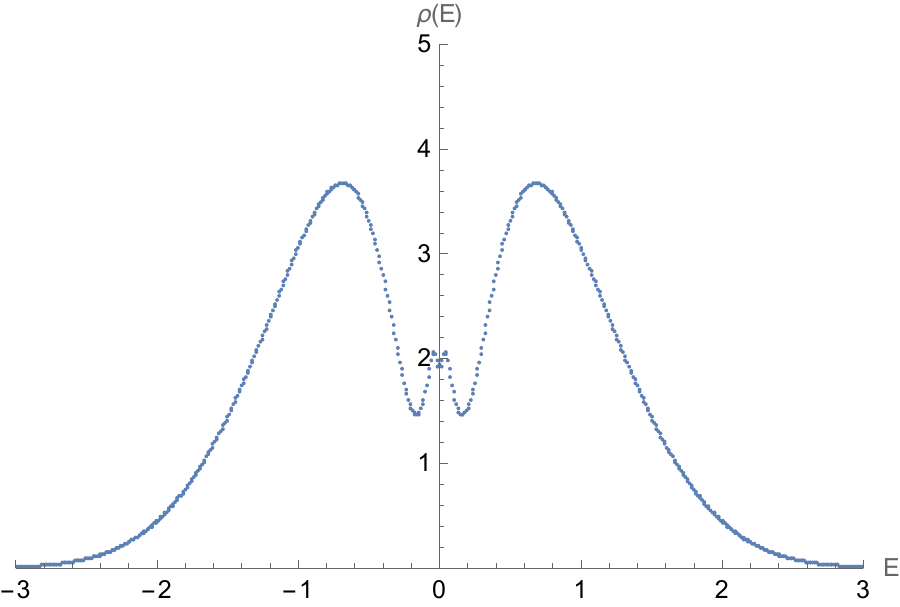}
\end{center}
\caption{The non-perturbative density of states $\langle \rho(E) \rangle$ of a single (loop) boundary dual for $\zeta < 1$ (AdS regime) and $\mu =0.1, 1.1$ respectively. We observe that as we increase $\mu$ it exhibits larger oscillations. After a certain critical value $\mu_c$, it becomes negative. For large $\mu \rightarrow \infty$ it becomes once more positive and approaches the Wigner semi-circle asymptotically.}
\label{fig:dosnumericssmallzeta}
\end{figure}

\begin{figure}[t]
\begin{center}
\includegraphics[width=0.4\textwidth]{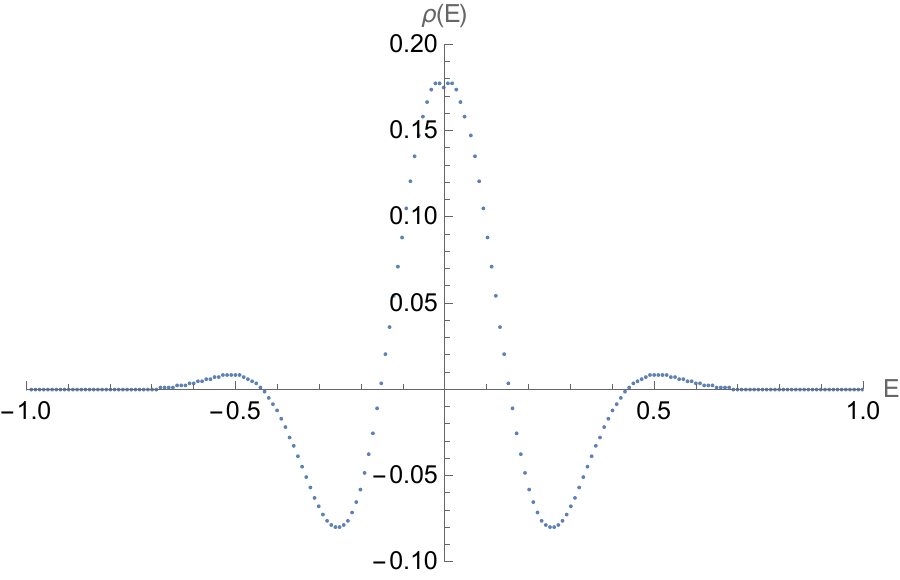}
\hspace{0.2em}
\includegraphics[width=0.4\textwidth]{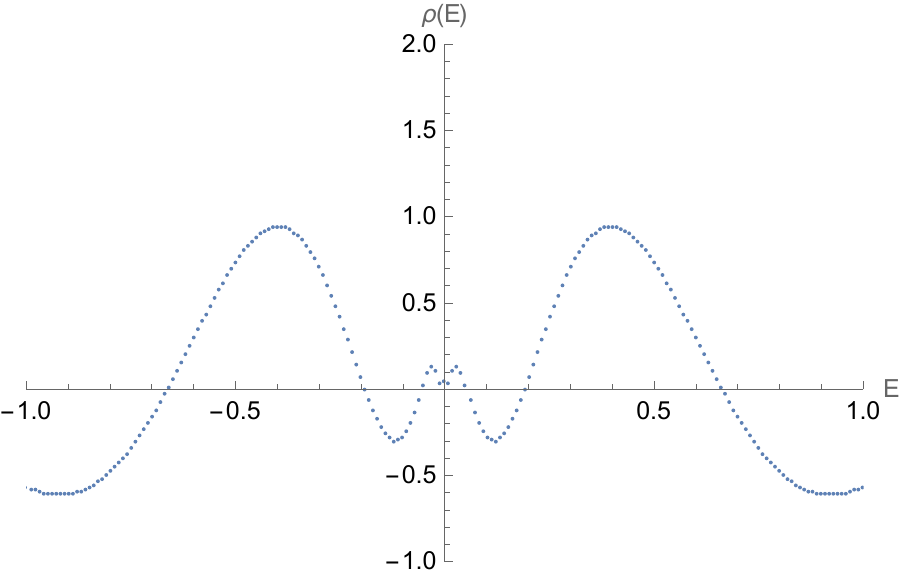}
\end{center}
\caption{The non-perturbative density of states $\langle \rho(E) \rangle$ for $\zeta > 1$ and $\mu =0.1, 1.1$ respectively (cosmological regime). We observe that it always contains large oscillations and regions where it becomes negative. This behaviour seems to be in conflict with the presence of a boundary dual on a single loop boundary (even if the NMM is a perfectly well defined non-perturbative description of the physics).}
\label{fig:dosnumericsslargezeta}
\end{figure}

We now adopt the second interpretation of the loop observable as providing the averaged partition function of the dual theory on the $S^1$ boundary i.e. $\langle W(\ell) \rangle = \langle Z(\beta) \rangle  , \, \beta \equiv \ell$. Motivated by the macroscopic large-N results of section~\ref{LoopMacro}, we rewrite the expectation value for the loop operator using the variables $z =  x + i y $, expecting that $  x \equiv E$ will play the role of an energy for the boundary dual. We find
\be\label{dosfromloop1}
\langle W(\beta , \mu, \zeta < 1 \rangle =  \int_{-\infty}^\infty d x \int_{-\infty}^\infty dy \,   (x^2+y^2)^{-i \frac{\mu}{2}} e^{- \frac{x^2 + y^2}{\mu}} e^{- \beta  x }  I_{-i \mu}\left(2 \sqrt{\frac{(1- \zeta^2)(x^2+y^2)}{\mu} } \,  \right) \, .
\ee
From this expression it is obvious that $  x \equiv E$ as expected and that the non-perturbative density of states (DOS), after we take its real part reads
\be\label{dos1}
\langle \rho(E) \rangle^{(\zeta < 1)} =  \Re \int_{-\infty}^\infty dy \,   \left(E^2+y^2\right)^{-i \mu/2} e^{- (E^2+ y^2)/\mu }   I_{-i \mu}\left(2 \sqrt{(1- \zeta^2)(E^2 + y^2)/\mu } \,  \right) \, ,
\ee
with a similar expression that replaces $I_{- i \mu} \rightarrow J_{-i\mu}$ for $\zeta > 1$. We observe that the density of states is $Z_2$ symmetric under $E \rightarrow - E$. There is a large exponential suppression factor $e^{- E^2/\mu}$ that quickly drives it to zero at large energies, so it has an effective compact support, with exponentially small tails. An obvious issue with this expression is that even if we adopt the prescription of taking the real part of the quantities we obtain with the complex measure, it can still become negative. In  fact in the small $\mu$ limit, when $\zeta < 1$, the density of states has a compact support and is positive definite. As $\mu$ grows, it starts developing oscillations, up to a point where it acquires also negative values, see fig.~\ref{fig:dosnumericssmallzeta} for more details. The large $\mu$ result coincides with and was treated in the macroscopic analysis, giving rise to an asymptotic Wigner semi-circle as $\mu \rightarrow \infty$\footnote{The exponential tails due  to $e^{-E^2/\mu}$ are replaced by power law square root edges as $\mu \rightarrow \infty$.}. The cosmological case $\zeta > 1 $, is always pathological exhibiting regions where the density of states becomes negative, see fig.~\ref{fig:dosnumericsslargezeta}. This leads to an obstruction for interpreting the non-perturbative result in terms of a boundary dual on a single loop and giving a true meaning to a density of states $\langle \rho(E) \rangle$.

Let us summarize the findings of this section. Assuming that $\nu = - i \mu$ is imaginary, there are three possible explanations for these peculiar results for the density of states. One is that an averaged boundary dual exists only for some range of parameters of the model (and in particular only in the AdS regime), but this begs the question on what physical principle constrains these parameters\footnote{For example from the string theory perspective, the gravitating SG model is well defined for any value of $\mu, \zeta$.}. A second possibility is to keep only the real part of the DOS and in addition exclude any regions where it becomes negative. A similar manipulation is relevant in the case of the complex Liouville string~\cite{Collier:2024kmo,Collier:2025pbm,Collier:2024lys} (in that case the result for the density of states is real but non positive definite). We believe this option/manipulation to be rather ad-hoc without further justification or indications on its validity and necessity\footnote{Some arguments were provided in the case of the complex Liouville string, see~\cite{Collier:2024kmo,Collier:2025pbm,Collier:2024lys} for details.}.
A final possibility (that we believe is the most conservative) is that in general there is no single (loop) boundary dual to the bulk quantum gravity theory. This means that it is a form of a ``lucky coincidence'' that there exist some very simplified models of low dimensional quantum gravity and in a certain regime of parameters, that are consistent with such an interpretation. In our present example this can happen only in the parameter regime when the bulk theory admits asymptotically AdS saddles and not in the regime where it also admits dS, or other cosmological types of saddles without an asymptotic boundary. This is precisely because in these latter cases any (loop) boundary is always a gravitating Cauchy slice in the bulk, in contrast with the AdS case where the fluctuations become frozen near the actual physical asymptotic boundary (or in simple two dimensional models are trivial in the bulk apart from boundary modes governed by the simple Schwarzian action).  On the other hand, the matrix model is directly definining a non-perturbative sum over the two dimensional geometries
even when highly fluctuating\footnote{This is as long as they do not ``crease/tear apart or resemble branched polymers''. In such cases the matrix model can potentially still describe the physics, but the notion of two-dimensional (smooth) geometry is lost.} - and holography works according to the original 't Hooft type of paradigm that relates matrix models with the triangulations of the dual graph, or according to the Gopakumar et.al. paradigm that relates them with a discretization of the moduli space of Riemann surfaces~\cite{Gopakumar:2022djw,Gopakumar:2024jfq}. This last explanation also means that generically, non-perturbative two dimensional quantum gravity with dynamical matter contains more degrees of freedom and richer physics than those existing in any possible single one-dimensional dual theory that can live on a macroscopic loop boundary, as first indicated in~\cite{Betzios:2020nry}. At this point though, we cannot exclude some mutual overlap between these options depending on the perspective/context one wishes to adopt and study.

\begin{figure}[t]
\begin{center}
\includegraphics[width=0.4\textwidth]{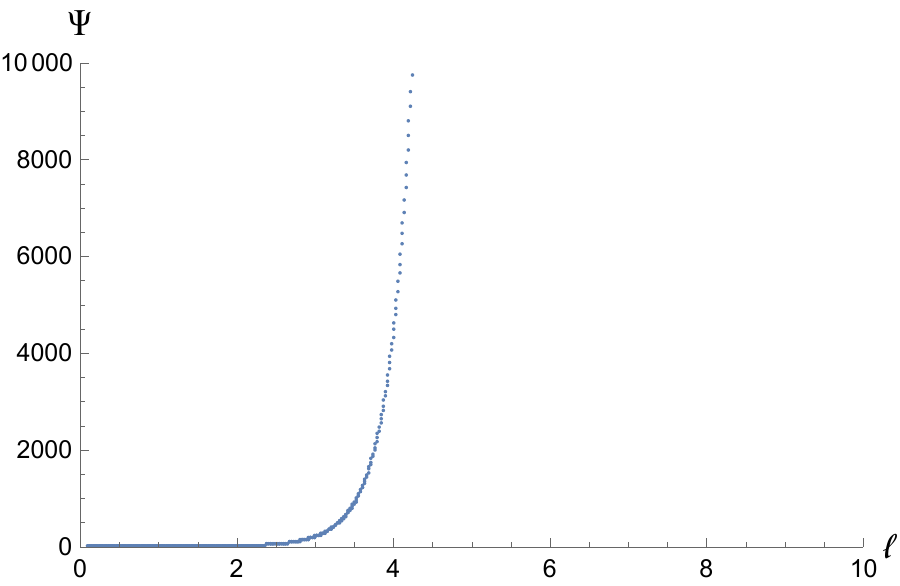}
\hspace{0.2em}
\includegraphics[width=0.4\textwidth]{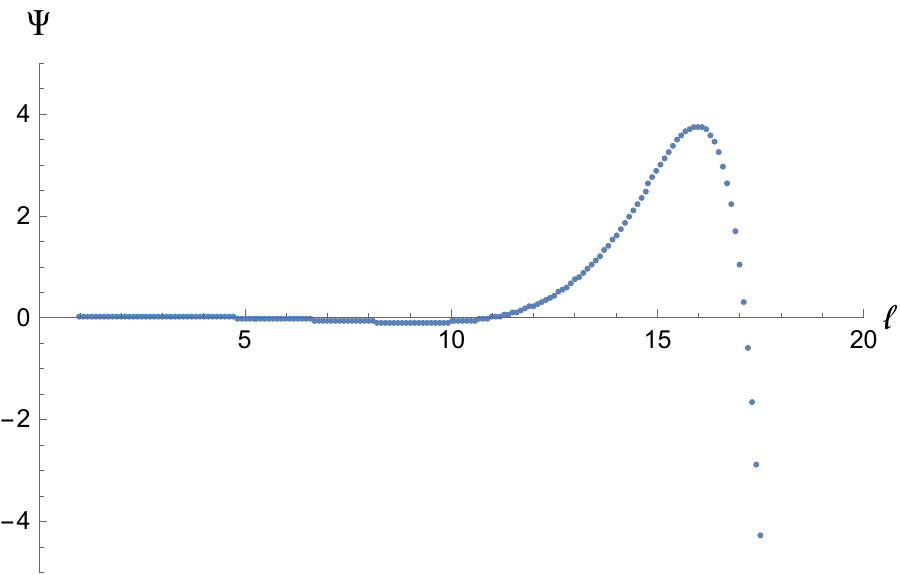}
\end{center}
\caption{The WDW wavefunction for $\zeta = 0.1, 2.1$ and $\nu = \mu = 0.1, 1.1$ respectively. It's qualitative behaviour in the two phases ($\zeta \lessgtr 1$) is similar to the previous case ($\nu = - i \mu$) in fig.~\ref{fig:wavefunctionnumerics}, but since $\nu$ is real the wavefunction now is also manifestly real.}
\label{fig:WDWnumericsreal}
\end{figure}

\subsection{The density of states and WDW wavefunction for $\nu = \mu$ real}\label{nurealproperties}

In this section we consider one last possibility: the case where $\nu$ is a real number, so that the $1/\nu$ expansion of the undeformed model becomes an alternating series corresponding to an imaginary coupling for the genus expansion for the undeformed model $g_{st} \sim 1/i \nu$. This can be shown in detail using an integral representation for the Digamma function, since~\cite{Betzios:2022pji}
\be
\frac{\partial \mathcal{F}_{undef.}}{\partial \nu} = \sum_{n \geq 0} \psi\left(\half + \nu + \frac{n+1/2}{R} \right) \, ,
\ee
and then forming an asymptotic genus expansion at large $\nu$.
More generally, when $\nu$ is real the diagonal part of the CD-kernel becomes real, and we obtain real expectation values for real single trace operators. This choice might sound ad-hoc at first glance, but has in fact been motivated in a similar context, in studies of 2d de-Sitter JT gravity~\cite{Cotler:2024xzz} (see also the discussion section~\ref{sec:discussion}). This complex string coupling for the undeformed model can also be obtained by shifting the contour of integration of the Liouville field in the path integral i.e. $\phi \rightarrow \phi + i \pi/2$.

\begin{figure}[t]
\begin{center}
\includegraphics[width=0.4\textwidth]{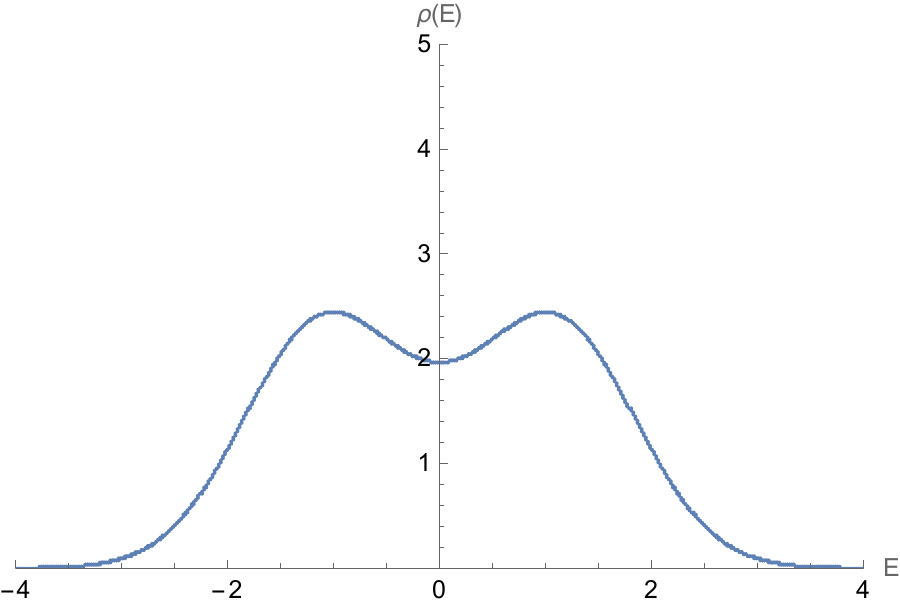}
\hspace{0.2em}
\includegraphics[width=0.4\textwidth]{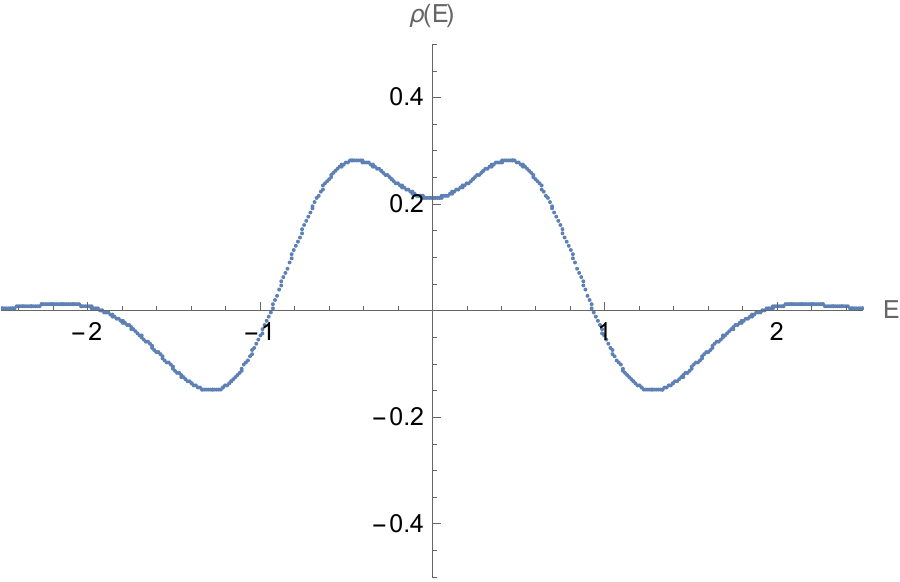}
\end{center}
\caption{The non-perturbative density of states $\langle \rho(E) \rangle$ for  $\nu=\mu = 1.1$ and $\zeta =  0.1, 2.1$ respectively. It is a manifestly real density of states that is always positive in the (AdS) regime $\zeta < 1$, while it can acquire negative values in the (cosmological) $\zeta > 1$ regime.}
\label{fig:Dosnumericsreal}
\end{figure}

The analysis of the previous section follows in the same fashion with minor changes. We summarize our results with two characteristic plots for the WDW wavefunction in fig.~\ref{fig:WDWnumericsreal} and another two plots for the non-perturbative density of states
in fig.~\ref{fig:Dosnumericsreal}. The wavefunction has a qualitatively similar behaviour both for real and imaginary $\nu$ (with the crucial difference that in the first case it is a manifestly real function, while in the second we had to consider its real part). The density of states for $\nu$ real and $\zeta<1$ is always a positive bound function (it resembles a semicircle envelope which large amplitude oscillations superimposed). On the other hand the pathology in the cosmological regime $\zeta > 1$, is cured only partially, and while the DOS is manifestly real, it is again not positive definite (this is more similar to the case of the complex Liouville string). This is again a strong indication against the existence of a boundary dual on a single macroscopic loop boundary, for the two dimensional cosmologies.

\section{Discussion}\label{sec:discussion}

In this last discussion section, we mention some future directions to explore. Finally, we provide some further comments in relation to various other microscopic matrix models that have appeared in the literature that have either been proposed to describe the physics of de-Sitter space or they have a close connection with the NMM.

\subsection*{Exact loop correlators - The WDW propagator/spectral form factor}\label{LoopcorrelatorMicro}

Another very interesting observable that one can study, is the loop correlator corresponding to the bulk WDW propagator (or the spectral form factor from a boundary perspective - in the cases this perspective makes sense). This observable gives information about the spectrum of the bulk quantum gravity path integral and the transition amplitudes between different geometries. It has been possible to non-perturbatively compute it only for the undeformed (uncompactified) $c=1$ model in~\cite{Betzios:2020nry}, while usually one has to resort to a perturbative evaluation together with applying the  topological recursion to recover higher genus results. In the present normal matrix model, there are various options for this correlator, depending on whether the loop operators involve $Z$ or $Z^\dagger$, or a combination of them. Given an explicit form of the CD-kernel 
$K_\infty( z , \bar w  )$, the simplest real loop operators $W(\ell) = \sum_i e^{- \ell(z_i + \bar z_i)/2}$, lead to the following expression for the connected part of the correlator (see appendix~\ref{Observables} and~\ref{CDKernel})
\bea
\langle W(\ell_1) \, W(\ell_2) \rangle_c &=& \int {d^2 z_1} \int {d^2 z_2}  e^{- \ell_1(z_1+\bar{z}_1)/2 } K_\infty( z_1 , \bar z_2 ) e^{- \ell_2(z_2+\bar{z}_2)/2} K_\infty( z_2 , \bar z_1  ) \, . \quad
\eea
Due to the nature of such correlated observables it can be advantageous to use the expression of the CD-kernel as an infinite sum over orthogonal polynomials (see appendix~\ref{CDKernel}), in order to separately perform the two integrals. In the case of the $R=1$ model with the first deformation, after writing $z_{1,2} = r_{1,2} e^{i \theta_{1,2}}$ we can perform the $r_{1,2}$ integrals in terms of the generalized Hypergeometric $_3F_3$, but the result is once more rather opaque and a further explicit integration over the angles is not evidently possible. We therefore leave its analysis in various physical regimes for the future.

\subsection*{UV complete models of de-Sitter space}

The present work, as well as the recent works related to the complex Liouville string, seem to indicate that non-perturbative - UV complete models of de-Sitter space can exist only when they admit the simultaneous presence of both anti-de Sitter and de-Sitter solutions. 
Perhaps this coexistence of AdS and dS saddles is a necessary feature in order for any quantum gravity model containing de-Sitter space to have a UV (non-perturbative) completion. Something similar could potentially happen in higher dimensional models, such as those admitting ``wineglass'' wormhole saddles~\cite{Betzios:2024oli,Betzios:2024iay}. Our results regarding the expectation value of loop operators that create macroscopic boundaries in these two different cases, also show that a boundary dual can only be assigned in the case of asymptotically AdS saddles (even if this dual is an averaged one). It would be interesting to explore this idea further, also in conjuction with various swampland conjectures that also seem to rule out the existence of pure (``isolated'') de-Sitter saddles in UV complete models of quantum gravity~\cite{Obied:2018sgi}.

\subsection*{Alternating genus expansion for $\nu$ real}

In~\cite{Cotler:2024xzz} it was argued that the correct genus expansion for dS-JT gravity, forms an alternating series with better convergence (Borel-summation) properties. This has also been encountered before in the genus expansion of Fermionic matrix models, see~\cite{Semenoff:1996vm}. In fact, as we found in section~\ref{nurealproperties}, in our compact gravitating SG model the analogue of the alternating genus expansion appears when we use its dual Normal matrix model description and define the parameter $\nu$ to be real, indicating the presence of an imaginary cosmological constant for the two dimensional quantum gravity (Liouville) action. One can absorb such a factor of $i$ through a redefinition of the Liouville field by an imaginary shift i.e. $\phi + i \pi/2$. This corresponds to picking a different contour prescription for it, when performing the path integral. While from a Liouville/two dimensional quantum gravity perspective the complete ramifications of such a choice are far from clear (whether the model is free from any pathologies), the NMM is perfectly well defined in this regime, having a well behaved and bounded real measure, without the need to consider taking the real part in various (a priori) complex results. Nevertheless, we also found that even in this case the averaged density of states $\langle \rho(E) \rangle$ of any putative dual on the (loop) boundary is still not positive definite in the cosmological $\zeta > 1$ regime. This is similar to what happens in the complex-Liouville string~\cite{Collier:2024kmo,Collier:2025pbm,Collier:2024lys} pointing to an inherent difficulty in assigning (even an averaged) boundary dual as in the asymptotically AdS examples. This seems to point towards the direction that one should really treat the matrices as the true fundamental microscopic degrees of freedom in cosmological models of two dimensional quantum gravity, since intuitively there is no ``cold'' non fluctuating asymptotic boundary on which to place a non-gravitating dual\footnote{The only other possibility is a microscopic description tied to each observer that explore the cosmological spacetime around them (a ``Matrix Quantum Mechanics model for Observers''). It is not clear how to make very precise such a solipsistic perspective.} and the gravitational degrees of freedom remain everywhere active.

\subsection*{Imbimbo-Mukhi-Kontsevich-Penner matrix model and topological strings}

In the case of the self dual radius $R=1$, $c=1$ strings are also described in terms of the Imbimbo-Mukhi or $W_\infty$ matrix model~\cite{Imbimbo:1995np,Ghoshal:1995wm,Mukhi:2003sz}
\be
Z_{IM}(\mu ; t_k , \bar{t}_k) = \det X^{- i \mu} \int D A e^{i \mu \sum_{k=1}^\infty t_k \tr(A^k) + i \mu \tr (A X) - (i \mu+Q) \tr \log A }
\ee
where $A$ are $Q\times Q$ matrices and $\bar t_k = \tr_Q X^{-k} / k$ a Miwa parametrization of the $\bar t_k$\footnote{The matrix size $Q$ here is related to how many of the $t_k$'s are independent.}. This can also be thought of as a generalized Kontsevich-Penner type of model. In~\cite{Mukherjee:2005aq} it was shown by explicit manipulation of the integral above that it is in fact equivalent to the NMM integral~\eqref{NMM} at $R=1$ (up to an overall numerical prefactor). From the perspective of the Kontsevich-Penner type of  model, the matrices are parametrizing the moduli space of the Riemann surfaces $\mathcal{M}_{g,n}$. This relation can be understood a Strebel parametrization of the moduli space and explains why we can recover exact results for certain correlators maniplating the NMM even at finite-N (as long as N is larger than the momentum that flows within the correlator), see~\cite{Gopakumar:2022djw,Gopakumar:2024jfq} and in particular the discussion near fig. (21) of~\cite{Gopakumar:2024jfq}\footnote{The ``KM'' model of~\cite{Gopakumar:2024jfq}, is in a sense a ``rotated version'' of the NMM when $R=1$.}. For the self-dual radius there exists also a unified description/spectral curve on $\mathbb{C}^2$ that describes both the chiral and the normal matrix model, whose spectral curves are certain lower dimensional projections. In fact the same spectral curve also describes topological strings (the so called ``B-model'') on the conifold, see~\cite{Ghoshal:1995wm} and followup work.
Finally, the Gaussian NMM also describes the physical properties of $1/2$ BPS states in $\mathcal{N}=4$ super Yang Mills (SYM) and their holographic dual geometries (see for example~\cite{Okuyama:2006jc} and references within).  It is not clear what the radius $R$ deformation in the NMM corresponds to from the point of view of  $\mathcal{N}=4$ SYM, or the conifold and it would be interesting to explore this further.

\subsection*{Fermionic (Grassmann) matrix models}

There exists also another class of manifestly finite matrix models~\cite{Semenoff:1996vm}, that have been proposed to be related to the physics of de-Sitter space~\cite{Anninos:2015eji,Anninos:2016klf}. These are matrix models, where the fundamental degrees of freedom (matrices) are Grassmann variables. A fermionic matrix model appears when exponentiating the determinant operators of the NMM (see appendix~\ref{sec:determinantoperators} for further details on such operators)
\bea
\langle\prod_{i=1}^n\det(a_i-Z)
\prod_{j=1}^m\det(b_j-Z^\dag)\rangle_{t_k,\bar t_k}^{NMM}
 = \frac{\mathcal{Z}_{NMM}(t_k-t^0_k,\bar t_k-\bar t^0_k)}{
\mathcal{Z}_{NMM}(t_k,\bar t_k)}\,(\det A\det B)^N  \nn \\
= \int D \chi^\dagger D \chi \,  D \psi^\dagger D \psi  \, D Z^\dagger D Z \, e^{-S_{NMM}(Z,Z^\dagger)} \,  e^{\tr \chi^\dagger (A \otimes I - I \otimes Z ) \chi + \tr \psi^\dagger (B \otimes I - I \otimes Z^\dagger ) \psi} \, , \nn \\
t^0_k = -\frac{1}{ k}\,\tr A^{-k} \, , \qquad \bar t^0_k = -\frac{1}{ k}\,\tr B^{-k} \, ,
\eea
the first line showing that the parameters $a_i, b_j$ play equivalently the role of ``time'' deformations. Hence the deformed model can be understood in terms of the undeformed model with particular determinant insertions. We can then exponentiate them as in the second line above and further integrate out the normal matrices $Z, Z^\dagger$ with the Gaussian measure to remain with a fermionic matrix integral (and the fixed auxiliary parameters). Similar manipulations of integrating in/out are also used to prove the web of dualities in~\cite{Gopakumar:2022djw,Gopakumar:2024jfq}. If we embed these $a_i, b_j$ parameters into $n \times n$ and $m\times m$ - ($A, B$) matrices as above, we can also imagine that they are dynamical variables as well and integrate over them in the path integral. From this last perspective the resulting Fermionic matrix models are related to a collection (an average) of NMMs over different couplings. It is not clear though what the averaging over the parameters/couplings corresponds to in the two dimensional quantum gravity theory (an average over the SG couplings) and whether the resulting model is still a CFT or an averaged version thereof.

We close this discussion section by mentioning that it would be interesting to explore further our ``microscopic'' version of the large-N limit in other matrix models and beyond, since it is manifestly different from the usual double scaling or 't Hooft limits, and can perhaps teach us valuable lessons in several other models of lower dimensional quantum gravity, or in cases where supersymmetric localization reduces complicated path integrals into a matrix model.

\acknowledgments

We wish to thank a plethora of colleagues for discussions and exchanges related to microscopic models with cosmological characteristics, throughout the years. We are especially grateful to Olga Papadoulaki for a large-N number of discussions on matrix models and Liouville theory and Andreas Blommaert, Thomas Mertens, Jacopo Papalini and the rest of the string theory group at Ghent University for several illuminating discussions related to Sine-Dilaton and JT gravity. We would also like to express our gratitude to Peter J. Forrester for informing us on some errors in the appendices of the first version of this article.
\\
\\
We are deeply saddened by the loss of our dear friend and mentor Umut G\"ursoy who introduced us to the world of matrix quantum mechanics and two dimensional quantum gravity.
\\
\\
\noindent P.B. acknowledges financial support from the European Research Council (grant BHHQG-101040024), funded by the European Union. Views and opinions expressed are those of the author only and do not necessarily reflect those of the European Union or the European Research Council. Neither the European Union nor the granting authority can be held responsible for them.

\appendix

\section{Normal matrix models (NMM)}\label{NormalMatrix}

In this Appendix we review some facts about Normal Matrix Models (NMM)'s and focus in the particular example of the models dual to the $c=1$ generalised SG models compactified at radius $R$. The NMM when written in terms of the matrix eigenvalues, can be thought of as a fermionic Dyson gas, with particles living on a two dimensional complex plane in the presence of a magnetic field (this lends to a direct relation between NMM and the quantum Hall effect, see). At the large-N semiclassical level, they occupy some finite region of the two dimensional plane. Some further reviews can be found in~\cite{Wiegmann:2003xf,Wiegmann:2005eh,Sommers:1988zn,Eynard:2015aea}. A recent review and survey of the subject is~\cite{Forrester}.

\subsection{Orthogonal polynomials for the normal matrix model}\label{OrthoNormalMatrix}

Let us start by writing the normal matrix integral as an integral over the eigenvalues of the matrix $Z$:
$z_1, ... , \, z_N$ 
\be
\mathcal{Z}_{NMM} \, = \, \frac{1}{ N!} \int_{\mathcal{C}}
\prod_{k=1}^N \frac{d^2 z_k}{2\pi i} \, e^{- W(z_k,\bar{z}_k)} \, \Delta(z)\Delta(\bar{z}) \, .
\ee
The specific measure of interest in our case reads
\be 
e^{-W(z, \bar{z})} \, = \, 
e^{-  \left[(z\bar{z})^R/\lambda - V(z) - \overline{{V}(z)} \right]}
(z\bar{z})^{(R-1)/2 + R \nu } \, .
\ee
It is useful to introduce an appropriate set of orthogonal polynomials with respect to this measure. In general the orthogonal polynomials are introduced as mean values of the characteristic polynomials of some random matrix $M$
\be\
P_n(\lambda )=
\langle \det (\lambda - M )\rangle_n \, ,
\ee
where $P_n$ are polynomials in $\lambda$ of the form
$P_n(\lambda )=\lambda^n +\, \mbox{lower degrees}$. In the present normal matrix case where the matrices are complex there exist two types of determinant operators, see appendix~\ref{sec:determinantoperators}. The resulting orthogonal polynomials are defined via their orthogonality relation in the complex plane $\mathbb{C}$
\be\label{orthog}
\int_\mathbb{C} \frac{d^2 z}{2 \pi i} \,  P_n (z)\overline{P_m(z)}e^{-W(z, \bar z)} =
h_n \, \delta_{mn}
\ee
The square of their norm $h_n =||P_n ||^2$ is directly related to the partition function via the relations
\be
h_n =\frac{1}{n\! +\! 1} \, \frac{\mathcal{Z}_{n+1}}{\mathcal{Z}_n}\,,
\qquad
\mathcal{Z}_N = N! \prod_{n=0}^{N-1}h_n
\ee
In particular for the undeformed model, we set all the ``time'' variables to zero, $z^n$ and $\bar{z}^n$ are the orthogonal polynomials for the axially symmetric
$W(|z|)$, and we find 
\be\label{undeformednorms1} 
h_n(\nu, R; 0)={1\over 2\pi i}
 \int_{\mathbb{C}} d^2 z \,
e^{-  (z\bar z)^R/\lambda}  (z \bar z)^{(R-1)/2+ R \nu + n} \, .
\ee
Since the measure is rotationally invariant we can use $z = r e^{i \theta}$ and perform the integral to obtain (see appendix~\ref{Formulae})
\be\label{undeformednorms} 
h_n (\nu, R ; 0) = \frac{\lambda^{1/2 + \nu + (n+1/2)/R }}{2 R}
\Gamma\left(\nu + \frac{n+\half}{R}+\half\right) \, \simeq \, e^{i \Phi_0 \left(E = i \nu + i  \frac{n+1/2}{R}\right)}  \, .
\ee
We find that upon setting $\lambda = \mu = i \nu $, the result coincides with the reflection coefficient for $c=1$ MQM (up to non-perturbative and irrelevant constant normalization terms~\cite{Betzios:2022pji}). Therefore we obtain
\be
 \mathcal{F} = \lim\limits_{N\to \infty}  \log \mathcal{Z}_{NMM} (\lambda = \mu = i \nu  , R ; t_k = \bar{t}_k = 0) 
=i  \sum\limits_{n=0}^{\infty}
\Phi_0\left( i  \frac{n+\half }{ R} + \mu\right) \, . 
\ee
This corresponds to the unperturbed free energy for finite temperature MQM when all the time variables are zero $t_{\pm k}=0$. Since, both deformed models are $\tau$ functions of the same Toda hierarchy (upon relating $t_k^{NMM} = - i t_k^{MQM}$ and $\nu^{NMM} = - i \mu^{MQM}$), this extends the correspondence of the NMM with the finite temperature MQM for any deformations (at any order of perturbation theory)~\cite{Alexandrov:2003qk}. Non-perturbatively the models do differ though. In our work we consider the NMM and its compact spectral curve as providing the correct non-perturbative definition of the (compact) gravitational SG model, when the $t_k$'s are turned on.

\subsection{Observables}\label{Observables}

The density of eigenvalues is defined as usual
\be
\rho(z, \bar z) = \frac{1}{N}  \sum_{i=1}^N \delta^{(2)}(z-z_i)
\ee
Simple observables arising from matrix traces can be evaluated using the density of eigenvalues and its connected correlators. For example for functions $f_i(z, \bar{z})$
\bea
\langle \tr f \rangle = \int \frac{d^2 z}{2 \pi i} f(z, \bar z) \langle \rho(z, \bar z) \rangle \, , \nn \\
\langle \tr f_1 \, \tr f_2 \rangle_c = \int \frac{d^2 z_1}{2 \pi i} \int \frac{d^2 z_2}{2 \pi i} f_1(z_1, \bar z_1) f_2(z_2, \bar z_2) \langle \rho(z_1, \bar z_1) \rho(z_2, \bar z_2) \rangle_c \, ,
\eea
and so forth. We shall now pass to a description of the eigenvalue density correlators using orthogonal polynomials and the associated Christoffel-Darboux kernel.

\subsection{The Christoffel-Darboux (CD) kernel}\label{CDKernel}

Using the orthogonal polynomials it is also possible to define a most basic object in random matrix theory: The Christoffel-Darboux (CD) kernel, which is useful to compute the observables in the previous section. A review with more details can be found in~\cite{Eynard:2015aea} and its application to two dimensional JT gravity in~\cite{Johnson:2022wsr}.

First we rescale the polynomials to define the functions\footnote{Usually one considers weight functions for which $\overline{W(z, \bar z)} = W(\bar{z}, z)$. More general cases are understood via analytic continuation in the parameters. There do exist though potential technical issues, when discussing orthogonal polynomials with complex weights~\cite{Stahl}, but the orthogonal polynomials we discuss here have a well defined continuation of their parameters in the complex domain. This is true when all the moments $\int d^2 z e^{- W(z,\bar z)} |z|^{2n} < \infty$ exist, for example eqns.~\eqref{undeformednorms1} and~\eqref{undeformednorms} are well defined even after replacing $\nu = - i \mu$ that leads to a complex weight function, see also appendix~\ref{CDkernelimaginary} for further discussion in the present context.}
\be
\psi_n (z)=\frac{1}{\sqrt{h_{n-1}}}\, P_{n-1}(z) e^{-W(z, \bar z)/2} \, , \qquad n = 1, 2, ... \, ,
\ee
that are orthonormal
\be
\int_{\mathbb{C}} \, \frac{d^2 z}{2 \pi i} \, \psi_n (z)\overline{\psi_m (z)}  = \delta_{mn} \, .
\ee
The $\psi_n$'s have also an interpretation of ``one-particle wave functions" of electrons in the presence of a magnetic field on the 2d plane. The $N$-particle wave function is then a Slater determinant
$\Psi_N (z_1 , \ldots , z_N) \sim \det [\psi_{j}(z_k)]$. The joint probability to find fermionic particles at the locations $z_1 ,
\ldots , z_{N}$ is
\be
|\Psi_N (z_1 , \ldots , z_N)|^2 =\frac{1}{N!}\,
|\det [\psi_j (z_k )] |^2 \, = \, \frac{1}{N!} \, \det
\left (
\sum_{n=1}^{N}\psi_n (z_j)\overline{\psi_n (z_k)}
\right )    \, .
\ee
The last expression inside the determinant is called the Christoffel-Darboux (CD) kernel function\footnote{In the present NMM context this is also sometimes referred to as the Bergman kernel.}
\be
K_N(z, \bar w) = \sum_{n=1}^{N}
\psi_n (z) \overline{\psi_n (w)} = e^{- W(z,\bar z)/2 - \overline{W( w, \bar w)}/2} \sum_{k=0}^{N-1}  \frac{P_{k}(z) \overline{P_{k}(w)} }{h_{k}} \, .
\ee
The main properties of the CD kernel are
\begin{itemize}
\item
Hermiticity:
$K_N (z, \bar w)=\overline{
K_N (w, \bar z)}$ 
\item
Normalization:
$\int \frac{d^2 z}{2 \pi i} K_N (z, \bar z) =N$
\item
Projection or self reproducing property:
$\int \frac{d^2 z}{2 \pi i} K_N (z_1 , \bar z) K_N (z, \bar z_2) 
= K_N (z_1 , \bar z_2)$.
\end{itemize}

The remarkable power of the CD kernel is that all density correlation functions can be expressed through it. In particular the eigenvalue density is
\be\label{eigenvaluechristoffel}
\langle \rho (z) \rangle_N =   K_N (z, \bar z) \, ,
\ee
and higher correlators of the density are given by
\be
\langle \rho (z_1 )\ldots \rho (z_n)\rangle_N
=   \det (K_N (z_i , \bar z_j))_{1\leq i,j\leq n} +
\, \mbox{contact terms} \, ,
\ee
where the contact terms vanish if all the points $z_i$
are different.

The CD Kernel for axially symmetric potentials $W(|z|^2)$, for which the orthogonal polynomials are simply $P_n(z) = z^n$ takes the simple form
\be\label{CDkernelAxial}
K_N (z, \bar w)= e^{-\frac{1}{2}(W(|z|)+\overline{W(|w|)})}
\sum_{n=0}^{N-1} \frac{(z\bar w)^n}{ h_n} \, .
\ee
In the undeformed example given by eqn.~\eqref{undeformednorms} it is possible to sum this expression even at finite-N. Calculations with the finite-N form of the kernel and density are difficult to perform, though. Sending $N \rightarrow \infty$ one can perform the summation
using the Mittag-Leffler function (see also appendix~\ref{Formulae})
\be
E_{a, b}(x) = \sum_{k=0}^\infty \frac{x^k}{\Gamma(a k + b)} \, ,
\ee
and show that in this case the CD-Kernel acquires the form
\be\label{CDWright}
K_\infty(z, \bar w) =   e^{- (z\bar{z})^R/2\lambda - (w\bar{w})^R/2\lambda}
(|z| |w|)^{(R-1)/2 + R \nu } E_{1/R , (R+1)/2 R +\nu }(z \bar w) \, .
\ee
We observe that for $R=1, \, \nu =0$ where we recover the Gaussian measure ($a=1,b=0$), so that $h_n = n!$ as expected and the Mittag-Leffler reduces to the simple exponential function $E_{1,0}(z \bar w) = e^{z \bar w}$. For $\nu \neq 0$, one finds (see also~\cite{Forrester} for similar examples)
\be\label{R=1Mittag}
E_{1, 1+ \nu}(z \bar w) = e^{z \bar w} (z \bar w)^{- \nu}  \frac{\gamma(\nu, z \bar w)}{\Gamma(\nu)}\, ,
\ee
in terms of the lower incomplete gamma function $\gamma(a,x) = \int_0^x d t e^{-t} t^{a-1}$.

\subsection{``Macroscopic'' large-N analysis of the NMM}\label{largeNmacro}

When invoking the usual ``macroscopic'' type of large-N limit, the eigenvalues of the NMM form a finite area compact fermi surface on the 2d complex plane (a ``droplet'' $\mathcal{D}$)\footnote{Depending on the model, one could also have multiple such regions/droplets that eigenvalues occupy. We shall not treat these cases in this work.}, see the works/reviews~\cite{Wiegmann:2003xf,Wiegmann:2005eh,Sommers:1988zn,Eynard:2015aea}. 
The shape of the region $\mathcal{D}$ and its boundary $\partial \mathcal{D}$ can be determined by solving the electrostatic problem
of minimising the effective action (``energy'') at large-N
\be
S_{eff.} = W(z_i, \bar z_i) - \sum_{i \neq j } \log |z_i - z_j| \, .
\ee
We recall the definition for the density of eigenvalues
\be
\rho(z, \bar z) = \frac{1}{N} \sum_{i=1}^N \delta^{(2)}(z-z_i) \, ,
\ee
and define the field/electrostatic potential
\be
\phi(z) = - \int d^2 z' \log|z-z'| \rho(z' , \bar z') \, ,
\ee
in terms of which
\be
4 \pi \rho(z, \bar z) = - \nabla^2 \phi(z, \bar z ) \, , \qquad \nabla^2 = 4 \partial_z \partial_{\bar z} \, .
\ee
The saddle point equations of the effective action are now simply given by
\be
\partial_z (\phi + W) = \partial_{\bar z} ( \phi + W) = 0 \, .
\ee
The solution to these equations as long as the density becomes a continuous distribution inside the droplet (``macroscopic large-N limit) is
\be
\rho(z, \bar z) = \frac{1}{\pi } \partial_z \partial_{\bar z} W \, .
\ee
We therefore observe that only the radially symmetric part of the potential $W$ that depends on the product $z \bar z$ determines the density. The shape of the droplet $\mathcal{D}$ and its boundary
$\partial \mathcal{D}$ do depend on all the parameters of the potential though, and a general methodology to find them is explained in the next section. 

In our present example, setting we find the following semiclassical density in the interior of the droplet (even including the ``time'' deformations)
\be\label{densityfiniteR}
\rho(z, \bar z) = \frac{1}{\pi } \partial_z \partial_{\bar z} W = \frac{R^2}{\pi \lambda} (z \bar z)^{R-1} \, .
\ee
The droplet shape though still depends on the ``time variables'' $t_n$.
For zero time variables the shape is a simple disk with a circular boundary. The normalization condition gives the radius $L_b$ of the boundary circle
\be\label{radiusfiniteR}
\int_{\mathcal{D}} d^2 z \rho(z, \bar z) = 1 \, , \qquad L_b = \left(\frac{\lambda}{R} \right)^{1/2R} \, .
\ee
We can now compute the expectation value of a loop operator creating a macroscopic boundary with length $\ell$ in the undeformed case when $t_n =0 $. The operator insertion in the eigenvalue basis is $\sum_i e^{- \ell(z_i + \bar z_i)/2}$~\cite{Okuyama:2006jc}. We find that it is expressed in terms of a Bessel function integral
\bea
\langle W(\ell) \rangle = \frac{1}{2\pi \lambda} \int_0^{L_b} d r r \int_0^{2\pi} d \theta e^{- \ell r \cos \theta } R^2 r^{2 R - 2} = \nn \\
= \frac{R^2}{\lambda} \int_0^{L_b} d r r^{2 R - 1}  I_0(\ell r) \, ,
\eea
which is equivalent to a hypergeometric function. This result 
is understood to be related to the dual Liouville theory upon setting $\lambda \equiv \mu$.

\subsubsection{The Schwarz function and conformal (univalent) maps}\label{Schwarzproperties}

To describe the boundaries and properties of the large-N solutions, we shall now start with some general definitions related to the \emph{Schwarz} function and conformal maps (see~\cite{Mineev-Weinstein:2000jhx,Teodorescu:2004qm} for more details) and then proceed to specializing to our cases of interest.

Given a closed (in general a multiply-connected) domain $\mathcal{D}$ in the plane, we define the Schwarz function $S(z)$ as an analytic function in some neighborhood of the boundary of the domain with the value 
$S(z) = \bar z$ on the boundary of the domain. The Schwarz function obeys an involution property $\bar{S}(S(z)) = z $.

Let us now define a complex curve $f(z,  w) = 0$, whose restriction $w = \bar{z}$ defines the boundary of the droplet $\partial \mathcal{D}$. The curve is a finite-sheet covering of the z-plane. The single-valued function $w(z)$ on the curve is a multivalued function on the z-plane. Making cuts, one can fix single-valued branches of this function, and in particular
$w=\bar{z}$ belongs to a particular sheet (called the physical sheet) of the covering.
The Riemann surface defined via the function $f(z,w)$ also contains an antiholomorphic involution. If the involution divides the surface into two disconnected parts, the Riemann surface is the Schottky double of one of these parts (or in other words starting from a Riemann surface with boundaries, we can add to them a ``back side'' by gluing another copy of the Riemann surface along the boundaries). 

Our models/examples fall within the case where the complement to the support of eigenvalues is a simply-connected algebraic domain in the extended (i.e. including $\infty$) complex plane. In this case, the Schottky double of the exterior of the droplet is a Riemann sphere. Then, the exterior domain can be
conformally and univalently mapped onto the exterior of the unit disc. As above the map is $w(z)$, and its inverse
$z(w)$. For algebraic domains, the inverse map is a rational function of $w$.
Choosing the normalization such that
$z(\infty )=\infty$, we represent the
inverse map
by a half-infinite Laurent polynomial
\begin{equation}\label{invconfLaurent}
z(w)=r w+\sum_{k>0}u_{k} w^{-k},\quad |w|\geq 1.
\end{equation}
Some simplifications occur for the following cases: If the vector potential $\partial_z V(z)$ is
a polynomial of degree $d-1$,  then
the inverse map $z(w)$ is a (Laurent) polynomial:
$u^{(k)}=0,\quad k>d-1$. If $\partial_z V(z)$
is a rational function with $d-1$ simple poles in the finite part of the complex plane, the map $z(w)$ is a rational function
with $d-1$ simple poles and one extra simple pole at infinity.

The Schwarz function in this case is given by
\be
S(z)=\bar z(w^{-1}(z)) \, ,
\ee
and hence Schwarz reflection of the inverse map is defined by
\begin{equation}
\bar z(w^{-1})=r w^{-1}+\sum_{k>0}{\bar u^{(k)}}{w^k},\quad |w|\geq 1.
\end{equation}
Given a point $z$, there are $d$ values of $w$ solving the
equation~\eqref{invconfLaurent}, where $d$ is the number of poles
of the rational function $z(w)$. From them the solution that corresponds to
the conformal map we seek, is the one for which $w\to z/r$ as $z\to\infty$.
The global coordinate $w$ on the Riemann sphere provides a uniformization of
the Schottky double. In this coordinate, the involution reads $w\to 1/\bar w$. If our interest is in the boundary of the domain, we can take the limit where $w$ approaches the boundary of the unit disk: $w
\rightarrow e^{i \phi}$. In this case we obtain the map, $z(w): S^1
\rightarrow \partial { \mathcal{D}}$.

Once we have the inverse conformal map we can
convert any contour integral over the boundary $\partial{\mathcal{D}}$ to a
holomorphic contour integral over $S^1$.
As some simple examples we have the area of the droplet
\bea\label{areanorm} 
N &=& \int_{\cal D} \frac{d^2z}{\pi}  = \int_{\cal D} \frac{d^2z}{\pi}
\bar  \partial \bar z = \oint_{ S^1} \frac{dw}{2\pi i} z'(w) \bar
z(w^{-1}) \, , 
\eea
and the ``moments'' defined by ($k  > 0$)
\be\label{moments} 
t_k = \frac{1}{ k} \oint_{\partial \cal D} \frac{dz}{2\pi i} \bar z z^{-k}
= \frac{1}{k} \oint_{S^{1}} \frac{dw}{2\pi i} \frac{z'(w)
\bar{z}(w^{-1})}{ z^k(w)} \, .
\ee 
In this case the moments are also related to the ``time variables'' potentials since
$V(z) = \sum_k t_k z^k$.
This procedure is useful more generally, since bulk integrals over the droplet can typically reduced to boundary integrals, which can then be evaluated once the inverse conformal map is given. For example in the main text in eqn.~\eqref{affineintegralSchwarz} we use this to write the saddle point integral equation as a contour integral involving the Schwarz function.

\subsubsection{Macroscopic large-N analysis of the CD-kernel}\label{LargeNCDkernel}

Let us now perform a macroscopic large-N analysis of the CD-kernel. We extend the weight function $W(z, \bar z)$ from $\mathbb{C}$ to a function $W(z, \bar{w})$ i.e. a function on $\mathbb{C}^2$, as in the previous section. For consistency we demand along the diagonal $W(z, \bar w = \bar z) = W(z, \bar z)$. Let us also consider a perturbation $h(z)$ of $W(z, \bar z)$, so that $W^h = W - h$, to understand its effects. One can prove that in the interior of the droplet, the perturbed CD kernel is given in the macroscopic large-N limit by the expression~\cite{Ameur1}
\be\label{LargeNCDkernel}
K_{h, \, \infty}^{int.} (z, \bar w) = \frac{1}{\pi} \partial_z \partial_{\bar w} W(z, \bar w)  \,  e^{-W(z, \bar w) + W(z, \bar z)/2 + W(w, \bar w)/2}  e^{- i \Im [ (z- w) \partial_w h(w) ]}\, ,
\ee
where we used the approximation $h(z, \bar w) = h(w) + (z-w) \partial_w h(w) + O(|z-w|^2)$\, .
The diagonal part of the CD kernel is simply given by
\be
K_{h, \, \infty}^{int.} (z, \bar z) = \frac{1}{\pi} \partial_z \partial_{\bar z} W(z, \bar z)  \, ,
\ee
verifying our previous large-N analysis for the density. As mentioned previously, the density only depends on the rotationally symmetric part of the potential and the non-rotationally symmetric parts affect the droplet shape and hence the integration region.

\section{Exactly solvable examples in terms of Laguerre polynomials}\label{Laguerreexamples}

As we argued in the main text, in the case of the self dual radius $R=1$, there exist two cases that are exactly solvable, when the potential contains up to quadratic terms 
\be
W(z, \bar z)= \frac{1}{\lambda} z \bar z - 2  \Re \, ( t_1 z +t_2 z^2 ) + \nu \log (z \bar z) \, .
\ee
Here we shall treat only the case where $t_1$ is turned on, we expect the general case to be solvable with some modifications. Similar manipulations appear in~\cite{DiFrancesco:1994fh,Akemann:2002js}.

Let us first define the normalization integral
\bea
I^{(\nu)}(t_1) &=&  \int dz  d \bar z \, |z|^{2 \nu} \exp\left[
- \left(|z|^2/\lambda - t_1 (z
+\bar z)\right)\right] \nn \\
&=& \lambda^{\nu + 1}\pi \Gamma\left(\nu + 1 \right) \,
_1F_1 \left(\nu + 1, 1 ,  \lambda t_1^2    \right)  \, ,
\eea
in terms of the confluent hypergeometric function.

We first use the scaling symmetry $z \rightarrow  e^{i \varphi} z$ to reexpress the normalization integral as
\bea
I^{(\nu)}(t) &=&\int dzdz^\ast  |z|^{2 \nu} \exp\left[
- \left(|z|^2/\lambda - t_1 ({e}^{i\varphi}z
+ {e}^{-i\varphi} \bar z)\right)\right]\nn\\
\Rightarrow \quad 1&=&
\left\langle 
\exp\left[  ( {e}^{i\varphi} - 1) t_1 z \right]
\exp\left[  ({e}^{-i\varphi} - 1) t_1 \bar z \right]
\right\rangle_c ,
\eea
In the second line, we have divided by the integral itself to define the connected average, that is to be taken with respect to the original weight $e^{- W}$.

We shall now use the generating function for the Laguerre 
polynomials ($Z$ can be complex)
\be
(1-u)^{-\nu-1} \exp\left[ \frac{u}{u-1}\,Z \right]
\ =\ \sum_{k=0}^\infty  L_k^{\nu}(Z)\ u^k \,  , \quad |u|<1 \, .
\ee
In particular, we choose $Z= { z}/t_1$ and (we work in the branch where $t>0$)
\be
\frac{u}{u-1} =  (\mbox{e}^{i\varphi} - 1) t^2_1 \, , \quad u = \frac{( \mbox{e}^{i\varphi} - 1) t^2_1}{(\mbox{e}^{i\varphi} - 1) t^2_1 - 1}  \, , \quad |u|<1 \ .
\label{udef}
\ee
Expanding the exponentials in the connected/normalised average, we obtain
\be
[(1-u)(1-\bar{u})]^{-\nu - 1}  =  \sum_{k,l=0}^\infty 
\left\langle L^{{\nu}}_k \left(  z/t_1 \right)
 L^{\nu}_l \left(  \bar{z}/t_1 \right) \right\rangle_c
u^k \bar{u}^{ l} \, .
\label{genLL}
\ee
Using the fact that 
\be
|u|^2  = \frac{4 t^4_1 \sin^2\varphi/2}{\cos^2\varphi/2+ (1+ 2 t^2_2)^2 \sin^2\varphi/2} \, , \qquad u + \bar{u} = \frac{4 t^2_1 (1+ 2 t^2_1) \sin^2\varphi/2}{\cos^2\varphi/2+ (1+ 2 t^2_1)^2 \sin^2\varphi/2}
\label{normu}
\ee
one observes that the left hand side of eq.~\eqref{genLL} only depends on the combination $u \bar{u}$ and we 
obtain 
\be
[(1-u)(1-\bar{u})]^{-\nu-1}  = \left[1- \frac{(1+t^2_1)}{t^2_1}|u|^2\right]^{-\nu -1} 
 = \sum_{k=0}^\infty \frac{\Gamma(\nu +1+k)}{\Gamma(\nu + 1)k!}\ 
\frac{(1+t^2_1)^k}{t_1^{2 k}} (u\bar{u})^k \ , 
\label{lhs}
\ee
which converges due to $(1+t_1^2)|u|^2/t^2_1<1$ from eqn.~\eqref{normu}.
Consequently also the right hand side of eq. (\ref{genLL}) depends only on the combination  
$u\bar{u}$ and thus the Laguerre polynomials have to be orthogonal. By comparing the coefficients of eqn.~\eqref{lhs} and~\eqref{genLL} we arrive at 
\be
\left\langle L^{\nu}_k \left( \frac{z}{t_1}\right)
 L^{\nu}_l \left( \frac{\bar{z}}{t_1}\right) \right\rangle_c
 =   \frac{\Gamma(\nu + 1+k)}{\Gamma(\nu+1)\,k!} \frac{(1+t_1^2)^k}{t_1^{2k}}  \delta_{kl} \, .
\label{ortho}
\ee
The orthonormal polynomials are thus given by 
\be
P_k^{(\nu)}(z)\equiv \left(
\frac{\Gamma(\nu+1+k)}{\Gamma(\nu+1)\,k!}
 \right)^{-\frac12}\,
\left( - \sqrt{\frac{t_1^2}{1+t^2_1}}\right)^k L_k^{\nu}\left( \frac{z}{t_1} \right) .
\label{Laguerreorthonormal}
\ee
We also observe that the orthogonal polynomials correctly reduce in the $t_1 \rightarrow 0$ limit to to their top component that corresponds to the orthogonal polynomial of the undeformed rotationally symmetric model (that is independent of $t_1$) i.e.
\be
P_k^{(\nu)}(z) = \left( \frac{\Gamma(\nu+1+k)}{\Gamma(\nu+1)\,k!}
\right)^{-\frac12} \, \left(  \frac{ z^{ k}}{k!}  + O(t_1^{1+}) \right) \, .
\ee
We can now compute the explicit
expression for the CD-kernel even at finite-$N$, 
\bea
K^{(\nu)}_N(z ,\bar w) =  
\frac{e^{-{W(z)}/{2} - {W(\bar{w})}/{2}}}{I^{(\nu)}(t_1)}
\sum_{k=0}^{N-1} \frac{\Gamma\left(\nu+ 1 \right) k!}{\Gamma\left(\nu+ 1 +k\right)}\, \left({\frac{t^2_1}{1+t^2_1}}\right)^k \, 
 L^{\nu}_k \left( \frac{ z}{t_1}\right)
 L^{\nu}_k \left( \frac{\bar w}{t_1} \right) \, \nn \\
 = \frac{e^{-{W(z)}/{2} - {W(\bar{w})}/{2}}}{I^{(\nu)}(t_1)}  \frac{P^{(\nu)}_{N}(z) P^{(\nu)}_{N-1}(\bar w) - P^{(\nu)}_{N}(\bar w) P^{(\nu)}_{N-1}(z) }{z - \bar w} \qquad \, , \qquad  \qquad
\label{kernelfiniteN}
\eea
Using this kernel we can compute in principle any correlation function, even at finite-N, that can be expressed in terms on the n-point function of the eigenvalue density, using the formulae in appendices~\ref{Observables} and~\ref{CDKernel}.

\subsection{The large-N limit of the CD-kernel}

There exist various options to perform the large-N limit of the CD-kernel, depending on which parameters one chooses to keep fixed in the limit. Here we shall work with an option that allows the parameters ($\nu, \mu, t$) of the model are freely tunable even after taking the limit, such that we will be able to recover the complete genus expansion of the dual quantum gravity model. In this limit, the resulting CD-kernel and correlations are still called ``microscopic'' in the literature\footnote{Perhaps a better terminology here is ``mesoscopic'' - if the truly microscopic kernels are defined to be the ones at finite-N.}, see section~\ref{sec:Normal} for more details and discussion.  In the main text we also show how the (genus zero) ``macroscopic''' large-N analysis of the previous section~\ref{largeNmacro} is appropriately recovered by performing the matrix model counterpart of the gravitational genus expansion in the ``microscopic'' results of this appendix.

As in~\eqref{microlimit},  we now consider the large-N limit of the CD-kernel defined via
\be
N \rightarrow \infty \, , \qquad \text{with} \quad \lambda = \mu = i \nu , \, \,  R, \,  \, t_n , \, \, \bar{t}_n  \quad \text{fixed} \, . 
\ee
and by keeping $z, \bar{w}$ finite. The large-N limit of the CD-kernel is defined simply as
\be
K^{(\nu)}_\infty(z,\bar{w}) \equiv  \lim_{N\to\infty}  
K^{(\nu)}_N\left(z, \bar{w} \right) \, .
\label{largeNKernel}
\ee
In this scaling limit, the arguments of the 
Laguerre polynomials in eqn. \eqref{kernelfiniteN} are kept fixed. It is convenient to also use the Mehler identity for Laguerre polynomials~\cite{Gradstein}
\be
\sum_{k=0}^\infty \frac{k! p^{2k}}{\Gamma\left(\nu+1+k\right)} 
L_k^\nu(x) L_k^\nu(y) \ =\ 
\frac{(x y p^2)^{-\frac{\nu}{2}}}{1-p^2}
\exp\left[\frac{-p^2}{1-p^2}(x+y)\right]
I_\nu\left(2\frac{\sqrt{x y p^2}}{1-p^2}\right) \, ,
\label{MehlerLaguerre}
\ee 
and obtain the explicit large-$N$ form of the CD-kernel in
eqn.~\eqref{kernelfiniteN}
\bea\label{largeNkernelt1}
K_\infty^{(\nu)}(z,\bar{w}) &=& 
\frac{(1+t^2_1)^{1+ \nu/2} }{
\lambda^{\nu + 1}\pi  \,
_1F_1 \left(\nu + 1, 1 ,  \lambda t^2_1    \right)
}\, \frac{|z \bar w|^{\nu}}{(z \bar w)^{\nu/2}} 
 \nn \\
&\times& \exp\left[
- \frac{1}{2 \lambda}  (|z|^2 + |w|^2)
+ \frac{t_1}{2} (\bar z - z + w
- \bar w)
\right]
I_{\nu}\left(2 \sqrt{1+t^2_1} \, \sqrt{{z \bar w}}\right) \, . \nn \\
\eea

The density can be read-off from the diagonal part of the kernel and takes the form
\be\label{largeNdensityt1}
\rho^{(\nu)}(z, \bar z) = \frac{(1+t^2_1)^{1+ \nu/2} }{
\lambda^{\nu + 1}\pi  \,
_1F_1 \left(\nu + 1, 1 ,  \lambda t^2_1    \right)
}\, |z|^{\nu}  \exp\left[
- \frac{1}{\lambda} |z|^2 
\right]
I_{\nu}\left(2 \sqrt{1+t^2_1} \, |z| \right) \, ,
\ee
which depends only on $|z|$.

\subsection{The orthogonal polynomials and the CD-kernel at imaginary $\nu = - i\mu$}\label{CDkernelimaginary}

Strictly speaking most of the formulae that we used so far, were derived assuming that $\nu$ is a real parameter. In this appendix we briefly describe what happens when we analytically continue $\nu = - i \mu$ to be imaginary (with $\mu$ real). We first notice that the same steps can be followed to derive the orthogonality relation eqn.~\eqref{ortho}, where now the parameter $\nu$ in the $\Gamma$ functions is imaginary.
We then observe that 
\be
 \overline{L^{\nu}_k({z})} = L^{\nu}_k(\bar{z})  \, , \qquad \overline{L^{- i \mu}_k({z})} = {L^{i \mu}_k({\bar{z}})} \, , 
\ee
that is, in the imaginary case, to get the complex conjugate of the Laguerre function, one has to take the complex conjugate both of the index $- i \mu$ and the argument of the orthogonal polynomial, as expected.  This leads to an apparent choice/freedom in the definition of the CD-kernel, expressed either as a direct analytic continuation of the real $\nu$ case
\be\label{IMCD1}
K_N^{(\nu = - i \mu) }(z,\bar w) = e^{-W(z,\bar{z})/2   - W(\bar{w}, w)/2  } \sum_{k=0}^{N-1} \frac{L_k^{-i \mu}(z) L_k^{-i \mu}(\bar{w}) }{h_k} = \sum_{k=1}^N \psi_k^{- i \mu}(z) \psi_k^{- i \mu} (\bar{w}) \, ,
\ee
or as an expression that involves the complex conjugate of the orthonormal polynomials
\be\label{IMCD2}
\tilde{K}_N^{(\mu)}(z,\bar w) = \sum_{k=1}^N \psi_k^{-i \mu}(z) \overline{\psi_k^{-i \mu} (w)}  =  e^{-W(z,\bar{z})/2   - \overline{W(\bar{w}, w)}/2  } \sum_{k=0}^{N-1} \frac{L_k^{-i \mu}(z) L_k^{i \mu}(\bar{w}) }{\sqrt{h_k \overline{h}_k}} \, ,
\ee
that leads to a manifestly real and positive result for the diagonal $z = w$. Both expressions have certain advantages, the first one satisfying the self-reproducing property and normalization of the CD-kernel with respect to what is now a complex measure\footnote{The measure becomes complex due to the term $e^{-W} \sim (z \bar z)^{- i \mu}$.}  $e^{- W}$, while the second one is manifestly Hermitian, see appendix~\ref{CDKernel} for more details on these properties.

In this work we adopt the first definition for the CD-Kernel~\eqref{IMCD1}, since it descends from the natural measure derived from the matrix model. This definition is also consistent with the structure of the Toda-hierarchy, see the original~\cite{Alexandrov:2003qk}. The disadvantage of this choice, is that the induced probability measure/density of eigenvalues from the diagonal part of the CD-kernel is not positive definite (in fact complex). Nevertheless, this form of the CD kernel retains the two most basic basic properties (self-reproduction and normalization) which in an abstract language can be written as
\be
K_N \circ K_N = K_N \, , \qquad \tr K_N = N \, .
\ee
Now these two properties are exactly the same properties that are obeyed by Wigner distributions in phase space, where in that later case the composition is replaced by the star product. The reason to point this analogy, is that similarly to what happens here, in that case the Wigner distribution is also not positive definite, yet there do exist certain quantities whose expectation value is guaranteed to be real. These quantities are written as compositions of (real) bilinears of simple operators i.e. $\langle O^* \circ O \rangle \geq 0$.

Let us finally mention that regarding the Mehler formula for the Laguerre polynomials, one can use the results of Al Salam that are of the algebraic and operatorial type and thus do not rely on the reality of the parameter $\nu$~\cite{AlSalam}. Hence the formula~\eqref{MehlerLaguerre} is correct even when $\nu = - i \mu$.


\section{Determinant operators in the normal matrix model}\label{sec:determinantoperators}

In this appendix we discuss general properties of determinant operators in the NMM.
These operators are important for two reasons. From the matrix model side, determinant operators are interesting computable observables that also allow to make manifest the relation between the NMM and the Kontsevich-Penner model of~\cite{Imbimbo:1995np,Ghoshal:1995wm,Mukhi:2003sz} at the self dual radius, as well as variants of open closed duality, recently studied in~\cite{Gopakumar:2022djw,Gopakumar:2024jfq}.
Loop operators with fixed loop length (i.e. $\tr e^{- \ell M}$) should be thought of as their formal Laplace transform and
from the two quantum gravity perspective they are important since they allow for the creation of macroscopic boundaries on the two dimensional surfaces with fixed size (the determinant operators keep the boundary cosmological constant fixed instead). From a ``holographic ensemble'' perspective these boundaries are related to observables such as $\langle \tr e^{- \beta M} \rangle \equiv \langle Z(\beta) \rangle$ and the spectral form factor $\langle \tr e^{- \beta_1 M} \, \tr e^{- \beta_2 M}   \rangle \equiv \langle Z(\beta_1) \, Z(\beta_2) \rangle$ in the case of one or two macroscopic boundaries respectively, when the ``inverse temperature'' is identified with the macroscopic loop length $\beta \equiv \ell$.

While in models of Hermitean matrices ($M^\dagger = M$) one can use determinant operators such as
\be
e^{W(x)} = \det(x-M) \, ,
\ee
in order to define the generating function of (microscopic) loop operators $\frac{1}{l} \tr M^l$ creating holes in the dual graph in the Feynman diagram expansion of the matrix model, in the NMM the basic matrix variables are complex. Hence one can introduce two types of operators
\be
\det(a- Z) \, , \qquad \det(b- Z^\dagger) \, ,
\ee
with the interpretation that they correspond to generating functions of two types of microscopic loop operators: $\frac{1}{l}\tr  Z^l$ that create a hole in a $Z$-face, or $\frac{1}{l} \tr (Z^\dagger)^l$ that create a hole in a $Z^\dagger$-face. The resulting correlators are complex, satisfying though an exchange relation for the deformations on the NMM
\be
\langle \prod_i\det(a_i-Z) \rangle_{t_k,\bar t_k} = 
\langle\prod_i\det(a_i-Z^\dag) \rangle_{\bar t_k,t_k}
\ee
where on the right hand side the role of the deformations $t_k,\bar t_k$ has been interchanged. In the case that $t_k = \bar{t}_k$, the result is the same upon using either insertion.

It is useful to exponentiate the determinant operator and show that the expectation value of a single such operator is
\be
\langle \det(a-Z) \rangle_{t_k,\bar t_k} = 
\frac{\mathcal{Z}_{NMM}(t_k-t^0_k,\bar t_k)}{
\mathcal{Z}_{NMM}(t_k,\bar t_k)}\,a^N
\ee
with $t^0_k$ given by 
\be
\label{tzero}
t^0_k = -\frac{1}{\nu k}\,a^{-k} \, .
\ee
In other words the insertion of the loop operator (that creates an open surface with a hole) has also an interpretation as deforming the background from a purely closed surface picture (this is a version of an open/closed duality from the perspective of a ``Kontsevich-Miwa'' transform). This can be easily generalized to the case of multiple insertions such as
\be
\langle \prod_{i=1}^n \det(a_i-Z)
\prod_{j=1}^m \det(b_j-Z^\dag) \rangle \, ,
\ee
Then, defining an $n\times n$ matrix $A={\rm
diag}(a_1,a_2,\ldots,a_n)$  and a $m\times m$ matrix $B={\rm
diag}(b_1,b_2,\ldots,b_m)$, and the parameters
\be
t^0_k = -\frac{1}{\nu k}\,\tr A^{-k} \, , \qquad \bar t^0_k = -\frac{1}{\nu k}\,\tr B^{-k} \, ,
\ee
we find the general result
\be
\langle\prod_{i=1}^n\det(a_i-Z)
\prod_{j=1}^m\det(b_j-Z^\dag)\rangle_{t_k,\bar t_k}
= \frac{\mathcal{Z}_{NMM}(t_k-t^0_k,\bar t_k-\bar t^0_k)}{
\mathcal{Z}_{NMM}(t_k,\bar t_k)}\,(\det A\det B)^N \, .
\ee
While this open/closed picture is conceptually appealing, it might not be the best method to practically compute the expectation values of such operators, since the deformation involves an infinite number of ``time variables'' $t_k$. It is more convenient to resort instead to an analysis exploiting the nature of determinant operator insertions, as being related to the expectation value of orthogonal polynomials.
In particular one can use the so-called ``Heine's relations''~\cite{Eynard:2015aea}
to relate the expectation values of such operators to the complex orthogonal polynomials we introduced in appendix~\ref{OrthoNormalMatrix} i.e.
\be\label{determinantfinal}
\langle \det(a-Z) \rangle_{N \times N} = P_N(a) \, , \qquad \langle \det(b-Z^\dagger) \rangle_{N \times N} = \overline{P_N(b)} \, .
\ee

\section{Useful Formulae}\label{Formulae}

In this appendix we collect some useful formulae that we use in the main text.

\begin{itemize}
    \item The first integral concerns the undeformed measure of the NMM. We find
\bea 
h(a,b ; R)={1\over 2\pi i}
 \int_{\mathbb{C}} d^2 z \,
e^{- a (z\bar z)^R}  (z \bar z)^{b} = \int_0^\infty d r r^{2 b + 1} e^{- a r^{2 R}} = \, \nn \\
= \frac{a^{-(1+b)/R}}{2 R} \Gamma \left(\frac{1+b}{R} \right)  \, , \qquad \Re a>0 \, , \, \Re b > - 1 \, .
\eea
We observe that if we set $a=N/\lambda$ and $b_n = (R-1)/2 + R\nu + n$, we obtain the orthogonal polynomial normalization factors needed to recover the partition function of compactified $c=1$ at radius $R$ (up to irrelevant and non-perturbative terms).

\item The following leading asymptotics of Bessel functions with imaginary order
\bea\label{BesselAsymptotics1}
I_{\pm i \mu}(z) \underset{\mu \to \infty}{\sim} e^{\mp i \phi(\mu)}  z^{\mp i \mu} \frac{e^{\pi \mu/2}}{\sqrt{2 \pi \mu}} \left( 1  \, + \,  O(1/\mu) \right) , \qquad \phi(\mu) = \mu \log \frac{2 \mu}{e} + \frac{\pi}{4} \, ,  \nn \\
I_{\pm i \mu}(z) \underset{z \to \infty}{\sim}  \frac{e^{z}}{\sqrt{2 \pi z}} \left( 1  \, + \,  O(1/z) \right) \, , \nn \\
I_\nu(z) \underset{z \to 0}{\sim} \left(\frac{x}{2}\right)^\nu \frac{1}{\Gamma(1+\nu)} + O(z^2) \, , \quad J_\nu(z) \underset{z \to 0}{\sim} \left(\frac{x}{2}\right)^\nu \frac{1}{\Gamma(1+\nu)} + O(z^2)
\eea

\item Some integrals of Bessel functions~\cite{Gradstein},~\cite{Prudnikov2}

\bea\label{I1}
I_1 \, &=& \, \int_0^\infty d x x^{a-1} e^{- p x^2} I_\mu(b x) I_\nu(c x)    \nn \\
I_1 \, &=& \, \frac{b^\mu c^\nu p^{-( a + \mu +\nu)/2 }}{2^{\m + \n + 1} \Gamma(\n + 1)} \sum_{k=0}^\infty \Gamma \left[ \begin{matrix}
    k + (a+\m + \n)/2 \\ \m +  k + 1
\end{matrix} 
 \right] \frac{1}{k!} \frac{b^{2 k}}{2^{2k} p^{2 k}} \, _2 F_1(-k, - \mu - k, \n + 1 ; c^2/b^2) \, \nn \\
 &\Re p& > 0 \, ,  \qquad \Re(a+\m + \n) > 0 \, \, .
\eea

\bea\label{I2}
I_2 \, &=& \, \int_0^\infty d x x^{a-1} e^{- p x^2} J_\mu(b x) I_\nu(c x)    \nn \\
I_2 \, &=& \, \frac{b^\mu c^\nu p^{-( a + \mu +\nu)/2 }}{2^{\m + \n + 1} \Gamma(\m + 1)} \sum_{k=0}^\infty \Gamma \left[ \begin{matrix}
    k + (a+\m + \n)/2 \\ \n +  k + 1
\end{matrix} 
 \right] \frac{1}{k!} \frac{c^{2 k}}{2^{2k} p^{2 k}} \, _2 F_1(-k, - \nu - k, \m + 1 ; - b^2/c^2) \, \nn \\
 &\Re p& > 0 \, ,  \qquad \Re(a+\m + \n) > 0 \, \, .
\eea

\be\label{I3}
I_3 = \int_0^\infty d x x e^{- p x^2} I_\nu(b x) I_\nu(c x) = \frac{1}{2 p} e^{(b^2+c^2)/ 4p} \, I_\nu \left( \frac{b c}{2 p} \right) \, , \qquad \Re p >0 , \quad \Re \nu > - 1 \, .
\ee

\bea\label{I4}
I_4 = \int_0^\infty d x x^{\m} e^{- a x^2} J_\n(b x) = \frac{\Gamma\left(\frac{\m + \n + 1}{2} \right)}{b a^{\m/2}\Gamma(\n+1)} e^{- b^2/8 a} M_{\mu/2, \nu/2}\left( \frac{b^2}{4 a}\right) \, , \nn \\
\Re(a) > 0 \, , \quad \Re(\m+\n+1)>0 \, .
\eea
in terms of the confluent Hypergeometric (or Whittaker-M) function.

\bea\label{I5}
I_5 = \int_0^\infty d x x^{\mu - \nu+1} J_\nu(a x) J_\mu(b x) &=& 0 \, , \qquad a < b \, , \nn \\  &=& \frac{b^\mu (a^2 - b^2)^{\nu - 1 - \mu}}{2^{\nu - 1 - \mu} a^{\nu} \Gamma(\nu - \mu)} \, ,  \quad a \geq b , \quad \Re \nu > \Re \mu > - 1 \, . \nn \\
\eea

\item A useful specialization of the Hypergeometric function in terms of Jacobi Polynomials
\bea\label{HypertoJacobi}
_2F_1(-k, b,c ; z) &=& \frac{k!}{(c)_k} P_k^{c-1, b-c-k}(1-2z) = \nn \\
&=& \frac{k! z^k}{(c)_k} P_k^{-k-b, b- c - k }\left(1-\frac{2}{z}\right) = \frac{k!(1-z)^k}{(c)_k} P_k^{(c-1, -b-k)}\left(\frac{1+z}{1-z}\right) \, , \nn \\
\eea
where $(c)_k = c (c+1) ... (c+k-1) $.

\item A summation formula of Legendre polynomials
\be\label{LLegendresum1}
\sum_{k=0}^\infty \frac{t^k}{k!} P_k(x) \,=\, e^{t x} J_0(t\sqrt{1-x^2}) \, . 
\ee

\item The generalised Mittag-Leffler function
\be
E_{a, b}(z) = \sum_{j=0}^\infty \frac{z^j}{\Gamma(a j + b)} \, .
\ee
For $b=0$, the Mittag-Leffler function interpolates between a Gaussian and a normal distribution. There exists also a useful integral
\be
\int_0^\infty d t E_{a,b}(\pm r t^a) t^{b-1} e^{- \ell t} = \frac{\ell^{a-b}}{\ell^a \mp r} \, , \qquad \Re a,b,\ell > 0 \, . 
\ee

\item The Wright function is defined by~\cite{BatemanIII}
\be
W_{a, b }(z) = \sum_{k=0}^\infty \frac{z^k}{k! \Gamma(a k + b)} \, , \qquad a > - 1 \, ,
\ee
obeys
\be
\int_0^\infty d t e^{- \ell t} W_{a, b }(t) \, = \, \frac{1}{\ell} E_{a, b}(\frac{1}{\ell}) \, , 
\ee
where $E_{a, b}(z)$ is the generalised Mittag-Leffler function.

It also obeys the relations
\bea
a z W_{a, a+ b}(z) = W_{a, b-1}(z) + (1-b)W_{a, b}(z) \, , \nn \\
\frac{d W_{a, b}(z)}{d z} = W_{a, a+ b}(z) \, , \nn \\
a z \frac{d W_{a, b}(z)}{d z} = W_{a, b-1, z} + (1-b) W_{a, b}(z) \, .
\eea
It can also be thought of as a generalization of the Bessel function, since
\be\label{WrighttoBessel}
I_{\nu}(2 z) = z^\nu W_{1, \nu + 1}(z^2)\, , \qquad J_{\nu}(2 z) = z^\nu W_{1, \nu + 1}(-z^2) \, ,
\ee
which is also what happens in our model when the radius gets deformed away from $R=1$ ($\nu = - i \mu$). The Wright function has also several important combinatorial applications and enters as as probability density in self-similar stochastic processes.

\end{itemize}

\section{Weyl/conformal frame independence of the effective action}\label{Weylindependence}

In this appendix we show the sense in which the renormalised quantum effective action of Liouville theory exhibits a form of background independence with respect to further rescalings of the Weyl/conformal factor. In particular the choice of $\hat{g}_{\m \n}$ in the effective action is arbitrary. Our approach will be slightly unconventional, but explains our definition for the renormalised stress energy tensor of the quantum effective action (and why it is traceless).

Let us consider for generality the general quantum effective action for Liouville given by
\bea\label{Liouvilleaction2}
S_{L}[\hat{g} , \phi] = \frac{1}{4 \pi} \int_{\mathcal{M}} d^2 z \sqrt{\hat{g}} \left[ \hat{g}^{ab} \partial_a \phi \partial_b \phi + Q \hat{R} \phi + \mu  e^{2 b \phi}  \right] \, . 
\eea
Under a rescaling $\hat{g}_{\m \n} = e^{2 \omega} g^0_{\m \n}$ and integration by parts we find
\bea\label{Liouvilleaction3}
S_{L}[{g}_0 e^{2 \omega} , \phi] = \frac{1}{4 \pi} \int_{\mathcal{M}} d^2 z \sqrt{{g}_0} \left[ {g}^{ab}_0 \partial_a \phi \partial_b \phi + Q {R}_0 \phi + 2 Q {g}^{ab}_0 \partial_a \phi \partial_b  \omega  + \mu  e^{2 b \phi + 2 \omega}  \right] \, . \nn \\
\eea

If we also define the shifted Liouville field $\phi_0 = \phi + Q \omega$, we find that the first three terms combine neatly to into
\be\label{Liouvilleaction3}
\frac{1}{4 \pi} \int_{\mathcal{M}} d^2 z \sqrt{{g}_0} \left[ {g}^{ab}_0 \partial_a \phi_0 \partial_b \phi_0 + Q {R}_0 \phi_0 -  Q^2( R_0 \omega + {g}^{ab}_0 \partial_a \omega \partial_b  \omega)  \right] \, . 
\ee
Now as is well known, the matter/ghost contribution in the path integral measure
is written in terms of a Liouville action 
with the same form as the last two terms in
~\eqref{Liouvilleaction3}, but with a coefficient $(c_M-26)/48 \pi$. This simply proves that if
\be
Q^2 = \frac{c_M - 26}{6} \, ,
\ee
the two contributions cancel each other rendering this part of the effective action invariant.

The final term is the cosmological constant term, that transforms into
\be
\mu \frac{1}{4 \pi} \int_{\mathcal{M}} d^2 z \sqrt{{g}_0} e^{2 b \phi_0 + 2(1 - b Q) \omega} 
\ee
Even though naively this term does not seem to be invariant, we can prove that a careful ragularization and renormalisation of the exponential operator, renders it invariant as well.

\subsection{Regularization of the exponential operator}\label{regularizationexpoperators}

Let us now describe in more detail the appropriate transformation properties of a the simplest exponential operator in Liouville theory - the cosmological operator
\be
 \int_{\mathcal{M}} d^2 z \,  \sqrt{\hat{g}} \,  e^{2 b \phi} \, ,
\ee
that show that it is a $(1,1)$ operator in the quantum theory.

We first observe that under classical Weyl rescalings $\hat{g}_{\m \n} = e^{2 \omega} g^{0}_{\m \n}$ it maps into
\be
\int_{\mathcal{M}} d^2 z \, \sqrt{{g}_0} \, e^{2 b \phi_0 + 2(1 - b Q) \omega}  \, .
\ee
This is not the complete story though. As a quantum operator it needs appropriate regularization and renormalization on a certain background metric. In fact one finds~\cite{Zamolodchikov}
\be\label{renormalisedrelation}
[ e^{2 a \phi(x)}]_{e^{2 \omega} g^0_{\m \n}} = e^{2 a^2 \omega } [ e^{2 a \phi(x)}]_{g^0_{\m \n}} \, ,
\ee
where our notation denotes the quantum operator normalized in the specific background.

This means that the cosmological constant operator actually transforms as
\be
 \int_{\mathcal{M}} d^2 z \,  \sqrt{\hat{g}} \,  [e^{2 b \phi}]_{\hat{g}_{\m \n} = e^{2 \omega} g^{0}_{\m \n}} =  \int_{\mathcal{M}} d^2 z \,  \sqrt{{g}_0} \, e^{2(1+b^2) \omega  } \,   [e^{2 b \phi}]_{g^{0}_{\m \n}} = \int_{\mathcal{M}} d^2 z \, \sqrt{{g}_0} \, e^{2 b \phi_0 + 2(1 - b Q + b^2) \omega} \, ,
\ee
so that if
\be
1 - b Q + b^2 = 0 \, \quad \Rightarrow \quad Q = b + \frac{1}{b} \, ,
\ee
it is a $(1,1)$ operator. This leads to an alternative proof that the renormalized quantum effective action of Liouville describes a CFT. A similar analysis can be performed for any such exponential operator, such as the ones used in the gravitating SG model. The result~\eqref{renormalisedrelation} also affects the stress energy tensor of the SG, so that in its renormalised form, one has to replace the classical contributions of the exponential operators with their renormalized version i.e.
\be
\mu e^{2 b \phi} \rightarrow \mu (1+b^2)  [ e^{2 b \phi}  ] \, .
\ee
This leads to the correct quantum effective stress tensor, which on a flat background is traceless.

\section{The exact partition function of the gravitational SG model for an arbitrary radius $R$}\label{exactPartition}

Let us now consider the simple gravitating SG model at an arbitrary radius $R$, but with only $t_1 = t \, , \, \bar{t}_1 = \bar{t}$ turned on. The complete (connected plus disconnected)
$2n$-point function $\langle (\mathcal{T}_{-1/R}\mathcal{T}_{1/R})^n\rangle$ for every $n$
and to all orders in perturbation theory can be computed using the normal matrix model~\cite{Mukherjee:2006hz} 
\be
\label{2npointcorrelator}
\langle (\mathcal{T}_{-1/R}\mathcal{T}_{1/R})^n\rangle = \left|\sum_{\{k_i\}}
C(\{k_i\})^2\prod_{i=1}^n \frac{\Gamma\left(\half-i\m
+(k_i-n+\half)\frac{1}{R}\right)} {\Gamma\left(\half-i\m
-(i-\half)\frac{1}{R}\right)}\right| \, ,
\ee
with $C(\{k_i\})$ defined by the following combinatorial formula
\be
C(\{k_i\}) = 
\sum_{\mathcal{P}}(-1)^\mathcal{P} 
\prod_{i=1}^n \begin{pmatrix} {n-\sum_{j=1}^{i-1}(k_j-\mathcal{P}_j)} \\ {k_i-\mathcal{P}_i} \end{pmatrix} \, .
\ee
Here the $\{k_i\}$ are strictly ordered partitions of $n(n+1)/2$, so that
\be
k_1>k_2>\cdots>k_n \, , \qquad \sum_{i=1}^n k_i = \frac{n(n+1)}{2} \, ,
\ee
and $\mathcal{P}$ denote all possible permutations of the $n$ numbers $n-1,n-2,\cdots,0$.

This result can be used to compute the partition function of the gravitating SG model using
\be
\mathcal{Z}_{SG} = \langle e^{ t \,\mathcal{T}^{-}_{1/R} + {\bar t}\,\mathcal{T}^-_{-1/R}} \rangle
=\sum_{n=0}^\infty \sum_{m=0}^\infty \frac{t^n}{n!}
\frac{{\bar t}^m}{m!}
\langle (\mathcal{T}^-_{1/R})^n(\mathcal{T}^-_{-1/R})^m\rangle =
\sum_{n=0}^\infty \frac{|t|^{2n}}{(n!)^2}
\langle (\mathcal{T}^-_{1/R}\mathcal{T}_{-1/R})^n\rangle
\ee
where the last equality follows from conservation of momenta.

The final result for the partition function is
\be
\mathcal{Z}_{SG} 
= \sum_{n=0}^\infty \frac{|t|^{2n}}{(n!)^2}
\left|\sum_{\{k_i\}}
C(\{k_i\})^2\prod_{i=1}^n \frac{\Gamma\left(\half-i\m
+(k_i-n+\half)\frac{1}{R}\right)} {\Gamma\left(\half-i\m
-(i-\half)\frac{1}{R}\right)}\right|
\ee
This expression is cumbersome to work with. An equivalent expression that admits a clear large-N limit (this differs non-perturbatively from the one above), uses the fact that the partition function is a $\tau$-function expanded in representations/partitions $\lambda$ of $GL(\infty)$~\cite{Betzios:2022pji}
\be
\mathcal{Z}_{SG} = \tau_k( t_+ , t_- ;\mu, R) = \sum_{\lambda } s_\lambda(t_+) s_\lambda(t_-) (-1)^{|\lambda|} \langle \lambda ; 0 | {\bf G}(\mu, R) | \lambda ; 0 \rangle \, ,
\ee
which acquires an explicit form in terms of a specialisation of the Schur polynomials that involves the Plancherel measure
\be
\mathfrak{M}_\lambda(\xi) = e^{- \xi^2} \xi^{2 |\lambda|} \left(\frac{\dim \lambda}{|\lambda|!}\right)^2 \, = \,  e^{- \xi^2} \sum_{n=0}^\infty \frac{\xi^{2 n}}{n!} M_n(\lambda) \delta(|\lambda| - n)\,,
\ee
where the quantity $M_n(\lambda)$ is called the ``Plancherel measure" on the partitions of $n$ 
\be
M_n(\lambda) = \frac{(\dim \lambda)^2}{n!} \, , \quad |\lambda|=n\, ,
\ee
This leads to
\be
\mathcal{Z}_{SG} = \sum_{n=0}^\infty |t|^{2 n}\sum_{\lambda \, : \, |\lambda|=n} \left( \frac{\dim \lambda}{|\lambda|!} \right)^2  \,  \langle \lambda ; 0 | {\bf G}(\mu, R) | \lambda ; 0 \rangle \, ,
\ee
with the ratio involving the dimension of the partition in the shifted highest weight coordinates $h_i = \lambda_i - i +\ell(\lambda) $
\be
\frac{\dim \lambda}{|\lambda|!} = \frac{\prod_{i<j \leq N}(h_i - h_j)}{\prod_{i\leq N} h_i !} \, ,
\ee
or in half-integer Frobenius coordinates $\lambda \equiv (p_i, q_j)\, , \, p_i,q_j \in \mathbb{Z}^{+} + \half \, , \, i,j = 1, ... d$
\be
\frac{\dim \lambda}{|\lambda|!} = \frac{1}{\prod_{i=1}^d (p_i-\half)! (q_i-\half)!} \det_{1 \leq i,j \leq d} \left[ \frac{1}{p_i + q_j} \right] \, ,
\ee
and the reflection amplitude
\be
\langle \lambda ; 0 | {\bf G}(\mu, R) | \lambda ; 0 \rangle = Z_{singlet}(\mu, R)  \prod_{j=1}^d \sqrt{ \frac{\Gamma\left(\half + i \mu  + \frac{p_j}{ R }\right)}{\Gamma\left(\half - i \mu  - \frac{ p_j}{ R} \right)}} \sqrt{ \frac{\Gamma\left(\half - i \mu  + \frac{q_j}{ R} \right)}{\Gamma\left(\half + i \mu R- \frac{q_j}{ R} \right)}} \, .
\ee
see~\cite{Betzios:2022pji} for more details and a complete analysis of the large representation limit of this expression, that reduces to the leading genus zero result for the free energy, given in section~\ref{genuszerofree}.

%




\end{document}